\documentclass[]{bytedance_seed}

\usepackage{amsmath,amssymb,amsthm}
\usepackage{bm}
\usepackage{float}
\usepackage{colortbl}
\usepackage{tabularx}
\usepackage{enumitem}
\usepackage{url}

\definecolor{trefleRow}{RGB}{233,244,235}
\theoremstyle{plain}

\theoremstyle{definition}

\newcommand{\method}{\textsc{Tr\`efle}}
\newcommand{\methodlong}{\textbf{T}ransition states via \textbf{RE}inforced \textbf{FL}ow in one \textbf{E}valuation}
\newcommand{\RR}{\mathbb{R}}
\newcommand{\EE}{\mathbb{E}}
\newcommand{\vx}{\bm{x}}
\newcommand{\vz}{\bm{z}}
\newcommand{\vv}{\bm{v}}
\newcommand{\vu}{\bm{u}}
\newcommand{\vg}{\bm{g}}
\newcommand{\vzero}{\bm{0}}
\newcommand{\Rmol}{\bm{R}}
\newcommand{\Pmol}{\bm{P}}
\newcommand{\vxs}{\bm{x}^{\star}}
\newcommand{\cond}{c}
\newcommand{\Ehat}{\widehat{E}}
\newcommand{\Ftrue}{E}
\newcommand{\policy}{\theta}
\newcommand{\reward}{r}
\newcommand{\adv}{A}
\newcommand{\angstrom}{\ifmmode\text{\normalfont\AA}\else\AA\fi}
\DeclareMathOperator{\sg}{sg}

\newcommand{\NFE}{\ensuremath{\mathrm{NFE}}}

\providecommand{\introheading}{\section{Introduction}}
\providecommand{\titlebreak}{\\}

\newcommand{\gpuname}{high-performance GPU}
\newcommand{\gpunames}{high-performance GPUs}
\newcommand{\gpushort}{high-performance GPU}
\newcommand{\gpufull}{high-performance GPU}
\newcommand{\agpufull}{a high-performance GPU}
\newcommand{\gpubare}{GPU}
\newcommand{\gpuhour}{GPU-hour}
\newcommand{\gpuhours}{GPU-hours}
\newcommand{\cpuname}{server-class CPUs}

\renewcommand{\titlebreak}{ }   %
\renewcommand{\method}{\textrm{\textsc{Tr\`efle}}}
\newcommand{\papertitle}{Reinforcement learning amortizes transition-state physics\titlebreak
into one-step flow models}

\newcommand{\abstracttext}{%
Transition-state searches remain a major bottleneck in reaction discovery, as identifying valid saddle-point structures requires numerous expensive quantum-chemical calculations. Generative models can reduce this burden by proposing candidates from reactant and product geometries, but supervised training on geometries alone does not enforce the physical conditions required of a transition state, while enforcing them during sampling is costly. We introduce \method, a one-step flow model that amortizes physics-guided generation into training: a mean-flow objective replaces multi-step sampling, a low-cost surrogate replaces density functional theory (DFT) reward evaluations, and reinforcement-learning post-training with rewards that enforce those conditions replaces inference-time guidance. \method~generates accurate transition states at $30\times$ the throughput of multi-step baselines. It transfers few-shot to ten unseen transition metals and generalizes zero-shot to molecules far larger than any in training. With DFT refinement and reaction-path validation, \method-seeded searches recover all 13 unseen $\gamma$-ketohydroperoxide channels, while post-training raises per-attempt recovery from 28\% to 50\% across 85 unseen bimolecular channels. \method~thus provides fast, reliable seeds for automated reaction discovery.
}

\title{\papertitle}

\author[1,2,\S]{Yunyang~Li}
\author[1]{Zechang~Sun}
\author[1]{Kuang~Yu}
\author[1]{Wen~Yan}
\author[2,\dagger]{Mark~Gerstein}
\author[1,\dagger]{Hung~Q.~Pham}

\affiliation[1]{ByteDance Seed}
\affiliation[2]{Yale University}

\contribution[\dagger]{Corresponding authors}
\contribution[\S]{Work done during an internship at ByteDance Seed}

\abstract{\abstracttext}

\date{\today}
\correspondence{Mark Gerstein at \email{mark@gersteinlab.org}, Hung Q. Pham at \email{hung.pham@bytedance.com}}

\begin{document}
\maketitle

\introheading
\label{sec:intro}

Predicting which reactions occur, along which pathways and at what
rates is a fundamental challenge in chemistry, with broad implications
for catalyst design, combustion, chemical synthesis and materials
discovery.  At the molecular level,
reaction pathways comprise elementary steps, each associated with a
\emph{transition state} (TS) that defines its activation barrier and,
within transition-state theory
\citep{eyring1935activated,evans1935applications}, governs its rate.
Geometrically, a TS structure is an index-one saddle point on the potential
energy surface (PES), where the energy gradient vanishes and the Hessian has
exactly one negative eigenvalue on the internal subspace.
Combined with reaction-path analysis, its geometry reveals which bonds
break and form and connects elementary steps into reaction networks
\citep{unsleber2020exploration,wen2023chemical}.  Yet the TS is never a resting
configuration.  Poised at a maximum along the reaction coordinate, it is
traversed in femtoseconds and cannot be isolated or held for
measurement.  Direct experimental characterization of transition-state regions remains
limited to special cases, including millimetre-wave spectra of isomerizing
molecules \citep{baraban2015spectroscopic} and ultrafast electron
diffraction of a photochemical ring opening \citep{wolf2019ring}, but
neither provides a general method for determining saddle-point
geometries.  Consequently, computational searches remain essential for
locating transition states.

Locating saddle points on the PES relies on a mature family of search
algorithms.  Chain-of-states methods
\citep{jonsson1998neb,henkelman2000cineb,e2002string,zimmerman2013growing},
such as the nudged elastic band, relax discretized paths between
reactants and products; local methods
\citep{henkelman1999dimer,cerjan1981finding,baker1986algorithm}, such as
the dimer method, climb uphill from an initial structure; and
exploratory schemes \citep{maeda2013exploration,shang2013stochastic},
such as artificial-force-induced reaction, uncover mechanisms without
predefined endpoints.  When coupled to density functional theory (DFT),
these methods require repeated energy and force evaluations.  Their cost
and robustness depend strongly on initialization; poor initial paths or
geometries increase the number of electronic-structure evaluations and
can steer the search to an unintended stationary point, or to no
convergence at all.  The burden compounds across the many elementary
steps of a reaction network, where the cost is dominated by repeated
electronic-structure evaluations.

Generative models reduce this burden by recasting the search as a
sampling problem: conditioned on reactant and product, a candidate TS
geometry is drawn from a learned distribution and serves as the starting
point for refinement.  Early models regressed a single TS geometry from the endpoints
\citep{pattanaik2020generating,jackson2021tsnet,makos2021gan}. A
deterministic point predictor cannot represent a multimodal conditional
distribution when several distinct saddles connect the same endpoint pair.  Later generators each
lifted a different restriction.  OA-ReactDiff \citep{duan2023oareactdiff}
replaces the point estimate with a distribution, fitting an object-aware
SE(3)-equivariant diffusion model to reactant--TS--product triples.
React-OT \citep{duan2025reactot} uses a deterministic source,
carrying the endpoint geometries themselves to the saddle along an
optimal-transport bridge.  TSDiff \citep{kim2024tsdiff} drops the
requirement of three-dimensional endpoints, generating ensembles of TS
conformers from the two-dimensional reaction graph alone.  Together
these models establish learned generators as accurate priors over TS
geometry.

Two obstacles still separate these models from routine use.  The first
is inference cost.  Diffusion and score-based bridges
\citep{ho2020denoising,song2021score} integrate a differential equation
over tens to hundreds of neural function evaluations (NFEs) per sample; even React-OT's
deterministic map is trained as a 50-step bridge.  At deployment this
cost multiplies.
Reaction-network studies can consider $10^5$--$10^6$
candidate elementary steps
\citep{unsleber2020exploration,zhao2023comprehensive}, and some
generative workflows draw or rank multiple TS candidates per step
\citep{kim2024tsdiff,duan2023oareactdiff}.  For example, an upper-end workload of $10^6$ candidate elementary
steps, 20 generated candidates per step and 50 NFEs per candidate would
require $10^9$ NFEs.   The
second
obstacle is more fundamental.  A generator
trained to imitate TS geometries is purely statistical with respect to
the PES: nothing in a supervised distribution-fitting objective requires
a generated geometry to be stationary or to have Hessian index one.
Generated geometries therefore require downstream saddle-point
validation and refinement with DFT.  The standard remedy is
\emph{guidance} \citep{dhariwal2021diffusion,ho2022classifier}, which
steers each sampling step with an energy oracle.  Gradient-based guidance
adds oracle evaluations during sampling and requires differentiable
signals. Black-box validators such as iterative saddle refinement
therefore cannot serve as guidance without an additional
gradient-estimation or search procedure.

Here we introduce \method\ (\methodlong), which
addresses both obstacles through \emph{amortization}: iterative sampling
is learned as a one-step map, while online reward preferences are
compiled into the generator parameters (Fig.~\ref{fig:overview}).
First, the sampler is amortized into a single step.  A mean flow
generator \citep{lipman2023flow,geng2025meanflow}, built on a steerable
equivariant encoder \citep{aykent2025gotennet,reschutzegger2026geodite},
maps one stochastic prior draw to a TS geometry in one forward pass.
Second, reward modelling uses fixed low-cost surrogate-PES oracles:
machine-learned force fields
\citep{batatia2022mace,batatia2023foundation} and semiempirical methods
such as g-xTB \citep{froitzheim2025gxtb} supply the energy, force and
Hessian information behind a scalar reward during online post-training,
in place of repeated DFT evaluations. Supervised pre-training continues
to use DFT-labelled TS geometries. Third, a likelihood-free
reinforcement-learning update \citep{zheng2025diffusionnft} scores each
group of sampled candidates with the oracle and refits the generator
towards the higher-scoring ones. The oracle scores completed rollouts
and is never differentiated through; at inference, sampling requires no
energy, force or Hessian oracle. Because the
reward is consumed as a black-box scalar, it can combine
non-differentiable terms with terms whose differentiation is
prohibitively expensive. One example is the index-one condition on the
oracle Hessian, whose gradient would require third derivatives of the
energy.

Our experiments show that these amortizations succeed and that each
one enables the next.  On Transition1x
\citep{schreiner2022transition1x}, one forward pass of the
 generator matches the accuracy of React-OT's fifty
steps, and on the larger Reaction-QM \citep{reactionqm2025} it
surpasses that baseline while sampling roughly thirty times faster; the
gap persists when React-OT is granted more steps.  This
speed is what makes reward-driven post-training practical, since the
model must draw and score many candidate geometries online.
To probe the choice of reward oracle, we curate ReactOracleBench, a
benchmark of transition-state structures drawn from DFT saddle-point
optimizations and intrinsic reaction coordinate (IRC) calculations and,
in part, from reinforcement-learning rollouts.
Screening seven candidate oracles against DFT places GFN2-xTB, g-xTB,
MACE-OMol and UMA-s-1.2 on the accuracy--cost Pareto frontier.
Post-training the generator with g-xTB, MACE-OMol or UMA-s-1.2 improves
the physical quality of its geometries by similar margins in every case: under independent DFT
evaluation, its geometries carry smaller residual forces and energy
errors and converge to the saddle point in fewer refinement steps, so
the framework is robust to which frontier oracle supplies the reward.
 What the generator learns also
extends beyond its training distribution. It transfers with few-shot training to reactions involving unseen main-group elements and ten transition metals, and generalizes zero-shot to reactions with unseen Si, P and S and to molecules far larger than any seen during training. Lastly, reward post-training raises intended-channel recovery across
the 13 decomposition channels of an unseen $\gamma$-ketohydroperoxide
(KHP) and 85 channels from ten bimolecular systems, from 28\% to 50\%
of attempts on the latter, while more than halving refinement steps
(Fig.~\ref{fig:epoxidation-case-study}).  The gain extends to rare
pathways: it recovers all three highest-barrier butadiene--ethene
channels, which React-OT misses entirely.  Thus post-training broadens
transition-state coverage across both unimolecular and bimolecular
reactions.

\section{Results}
\label{sec:experiments}

\subsection{A physics-reinforced one-step generator}
\label{sec:pipeline}

\begin{figure}[!tp]
\centering
\includegraphics[width=0.90\linewidth]{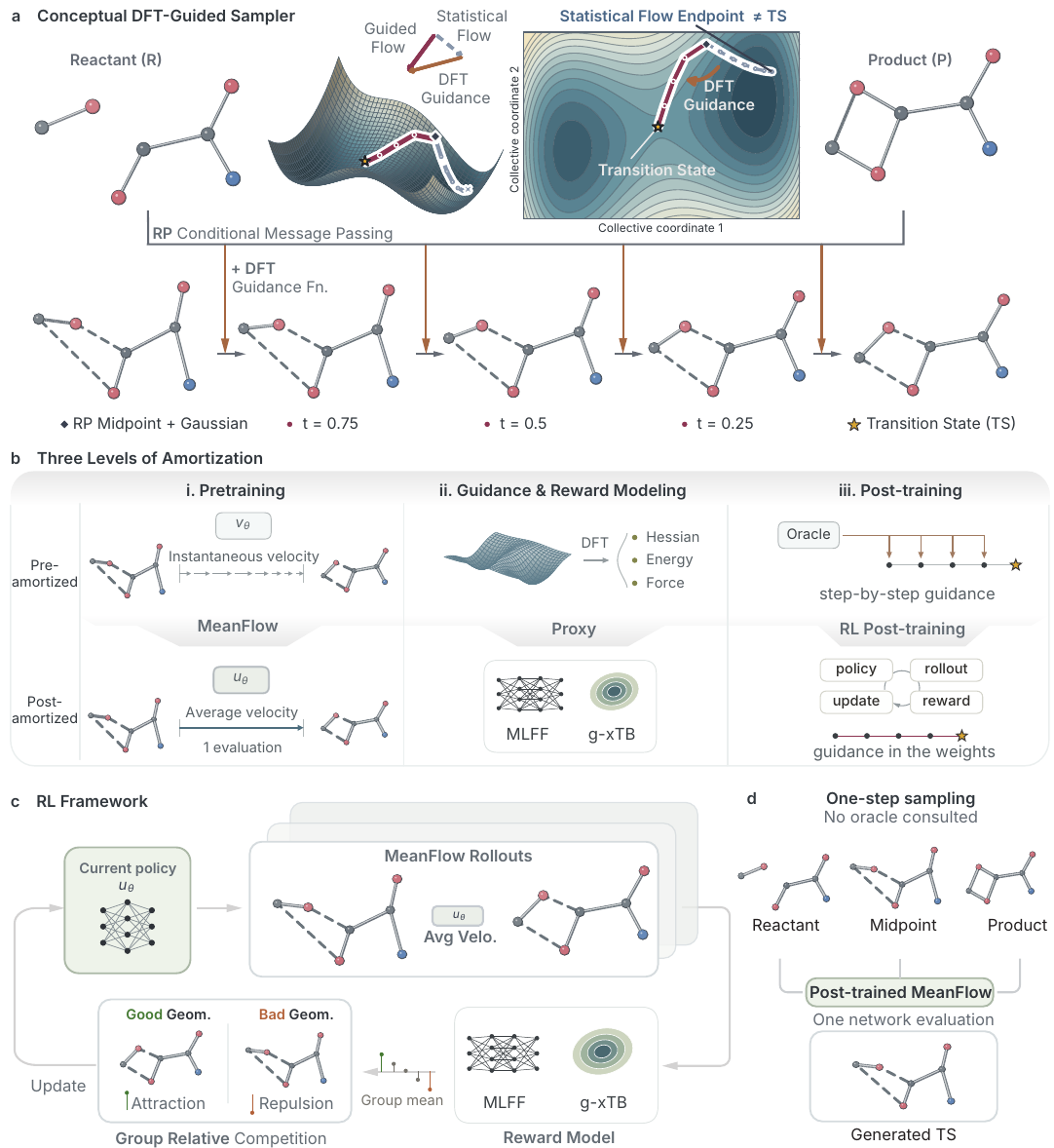}
\caption{\textbf{One-step transition-state sampling requires no energy,
force or Hessian oracle at inference.}
\textbf{a}, Conceptual DFT-guided sampler (illustrative only). Density
functional theory (DFT) guidance could in principle redirect a learned
flow from a perturbed midpoint towards a saddle-like region. We do not
quantitatively evaluate this guided sampler because of its high computational
cost.
\textbf{b}, Three stages organize the computation: MeanFlow pre-training
learns a finite-time average-velocity map from DFT-labelled TS geometries; a
fixed low-cost machine-learned force-field (MLFF) or g-xTB surrogate-PES
oracle supplies energy, force and Hessian information for reward modelling
during online post-training; and likelihood-free post-training compiles
group-relative reward preferences into the generator parameters.
\textbf{c}, During post-training, the current policy generates groups of
one-step MeanFlow rollouts; a fixed MLFF or g-xTB reward model scores them,
and group-relative comparison updates the generator towards higher-reward
samples and away from lower-reward ones.
\textbf{d}, At inference, the post-trained MeanFlow maps a
reactant--product midpoint prior directly to a transition-state candidate in
one network evaluation. Sampling requires no energy, force or Hessian oracle
at inference.}
\label{fig:overview}
\end{figure}

A useful TS candidate lies near an index-one saddle point of the
potential energy surface: vanishing forces and exactly one negative
internal Hessian eigenvalue (see \hyperref[sec:met-problem]{`Problem
formulation' section in the Methods}). 
Supervised distribution matching enforces neither condition.
A conceptual
alternative, adapted from guided generative modelling
\citep{dhariwal2021diffusion,ho2022classifier}, is to augment each
sampling step with an energy- or force-derived correction towards a
saddle-like region (schematic in Fig.~\ref{fig:overview}a).  
In our serial GPU4PySCF protocol, one joint energy--force evaluation has
a median cost of $1.60$\,s on a 32-structure cohort
(\hyperref[sec:met-dft]{Methods}). If the per-call cost remained
comparable, 10--100 serial calls would require approximately
$16$--$160$\,s per candidate.  \method{} instead separates
supervised pre-training, online reward post-training and inference across
three stages (Fig.~\ref{fig:overview}b).

In the pre-training stage, the generator learns to cover the DFT-labelled
TS distribution while amortizing iterative sampling into a single
step.  A conditional flow links the labelled geometries to a prior centred at the
reactant--product midpoint (see \hyperref[sec:met-prior]{`Centroid-projected isotropic Cartesian
prior' section in the Methods}). The MeanFlow objective \citep{geng2025meanflow} directly learns a
finite-interval average-velocity field $\vu_{\policy}$, whose boundary
map replaces numerical integration at inference. The field is
parameterized by a steerable equivariant network
\citep{aykent2025gotennet,reschutzegger2026geodite}
(see \hyperref[sec:met-meanflow]{`Pre-training: distilling transport
into one MeanFlow step' section in the Methods}). We additionally improved the MeanFlow training with the AlphaFlow curriculum~\citep{zhang2025alphaflow} (\hyperref[app:alphaflow]{Supplementary Note 1, `AlphaFlow curriculum for MeanFlow pre-training'}). A prior draw $\vx_1$
becomes the
candidate $\widehat{\vx}=\vx_1-\vu_{\policy}(\vx_1,0,1;\cond)$ for
reaction condition $\cond$.  React-OT \citep{duan2025reactot} uses a deterministic source. We retain
a stochastic prior to create within-reaction rollout groups for reward
post-training.

Reward modelling uses a fixed low-cost surrogate-PES oracle in place of
repeated DFT evaluations during online post-training. The oracle assigns a
composite reward based on force residuals, reaction progress, Hessian index,
endpoint-relative energy and atomic contacts (see
\hyperref[sec:met-reward]{`Composite heuristic reward' section in the
Methods}). Likelihood-free post-training then compiles these reward
preferences into the generator parameters, eliminating inference-time calls
to that surrogate-PES oracle.

The update maximizes the map's expected reward while regularizing the
policy towards its pre-trained weights
(see \hyperref[sec:met-post]{`Post-training: likelihood-free reward
fine-tuning' section in the Methods}). The current policy draws
per-reaction groups of one-step rollouts, which a fixed reward model
scores (Fig.~\ref{fig:overview}c). Group-relative comparison
\citep{shao2024deepseekmath} drives a likelihood-free flow-matching
update \citep{zheng2025diffusionnft} that pulls the generator towards
candidates scoring above their group mean and away from those scoring
below it; Extended Data Fig.~\ref{fig:post-training-schematic} sets out
the rollout group, the group-relative weighting and the resulting
update in detail.

At deployment, a single draw from the midpoint prior becomes a TS
candidate in one network evaluation. Sampling requires no energy, force or
Hessian oracle at inference (Fig.~\ref{fig:overview}d).

\subsection{One-step generation with reward post-training surpasses
multi-step accuracy at scale}
\label{sec:exp-accuracy}

\begin{figure}[!tp]
\centering
\includegraphics[width=\linewidth]{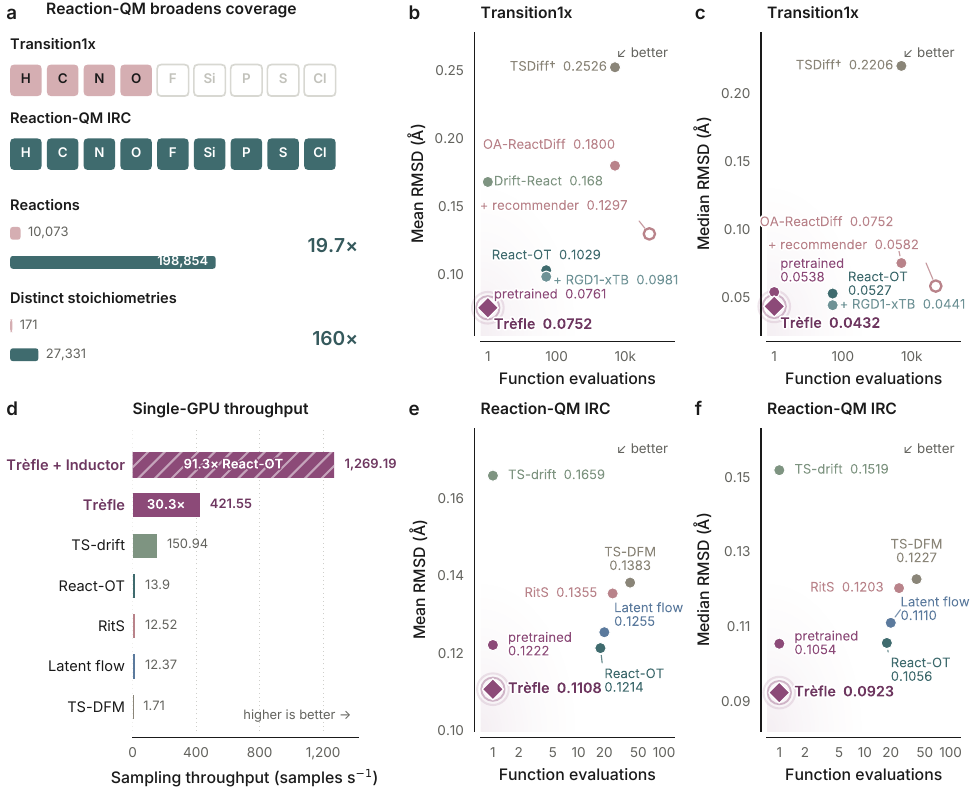}
\caption{\textbf{One-step generation surpasses multi-step accuracy on
Transition1x and Reaction-QM and samples $30\times$
faster.}
\textbf{a}, The intrinsic-reaction-coordinate (IRC) subset of
Reaction-QM broadens the evaluation beyond Transition1x: nine covered
elements versus four, $19.7\times$ the reactions ($198{,}854$ versus
$10{,}073$) and $160\times$ the distinct stoichiometries ($27{,}331$
versus $171$).
\textbf{b},\,\textbf{c}, Mean (\textbf{b}) and median (\textbf{c})
root-mean-square deviation (RMSD) between generated and labelled
transition-state geometries on the Transition1x test set ($n=1{,}073$
reactions; standard $9{,}000/1{,}073$ split) against the number of
neural function evaluations per sample (\NFE; logarithmic axis).
Published baselines are shown as reported ($\dagger$: TSDiff uses a
different split seed and conditions on the 2D reaction graph; the
OA-ReactDiff recommender ranks 40 candidates per reaction).  \method\ is
shown both after pre-training and after g-xTB reward post-training. The
pre-trained checkpoint has the lowest reported mean point estimate at
$\NFE=1$, and reward post-training gives the lowest reported mean and
median point estimates among the listed baselines, including both
50-evaluation React-OT variants.
\textbf{d}, End-to-end sampling throughput on a single \gpuname:
\method\ generates $421.6$ samples\,s$^{-1}$, $30.3\times$ the
18-evaluation React-OT baseline; the hatched bar shows throughput with
additional PyTorch Inductor compilation ($1{,}269$ samples\,s$^{-1}$,
$91.3\times$).
\textbf{e},\,\textbf{f}, Mean (\textbf{e}) and median (\textbf{f}) RMSD
on the Reaction-QM test split, where all methods are trained and
evaluated in one consistent suite (Methods); both \method\ checkpoints
are shown, and the reward-post-trained checkpoint attains the
lowest mean and median RMSD in the comparison at $\NFE=1$, improving
on the 18-evaluation React-OT by $8.7\%$ (mean) and $12.6\%$ (median).
Panels show point estimates; exact values and
evaluation protocols are given in Extended Data
Tables~\ref{tab:t1x-rmsd}--\ref{tab:throughput} and Methods.}
\label{fig:accuracy-throughput}
\end{figure}

We evaluate coordinate accuracy on Transition1x
\citep{schreiner2022transition1x} and the chemically broader
IRC test split of Reaction-QM
\citep{reactionqm2025} (Fig.~\ref{fig:accuracy-throughput}a and
Extended Data Fig.~\ref{fig:dataset-comparison}). Transition1x uses the
published standard $9{,}000/1{,}073$ split; Reaction-QM uses the
held-out test split defined in Methods ($n=19{,}883$ reactions). On both benchmarks we report
\method\ after supervised pre-training and after reward post-training.
Both post-trained checkpoints are rewarded with g-xTB; the choice of
reward oracle is examined separately in
\hyperref[sec:exp-post]{`Post-training is robust to the choice of
reward oracle'}.

On Transition1x, the pre-trained \method\ checkpoint already attains,
with a single network evaluation, a lower mean root-mean-square
deviation (RMSD; $0.0761$\,\angstrom) than every baseline; its median RMSD ($0.0538$\,\angstrom)
closely matches that of React-OT ($0.0527$\,\angstrom) and trails that
of React-OT with RGD1-xTB pre-training ($0.0441$\,\angstrom), both of
which use 50 network evaluations.  Reward post-training lowers \method\ to mean/median
$0.0752/0.0432$\,\angstrom\ with a single network evaluation, the
lowest reported point estimates among the baselines listed in Extended
Data Table~\ref{tab:t1x-rmsd}
(Fig.~\ref{fig:accuracy-throughput}b,c).
On Reaction-QM, the reward-post-trained \method\ likewise attains the
best mean and median RMSD in the comparison
($0.1108/0.0923$\,\angstrom) with a single \NFE\ per sample, whereas the pre-trained checkpoint attains
$0.1222/0.1054$\,\angstrom.  This improves on
React-OT ($0.1214/0.1056$\,\angstrom\ at $\NFE=18$), by
$8.7\%$ in mean and $12.6\%$ in median RMSD (Extended Data
Table~\ref{tab:ts-rmsd}, Fig.~\ref{fig:accuracy-throughput}e,f).  All
learned samplers surpass the geometry-only interpolants (mean RMSD
$0.322$--$0.391$\,\angstrom) by a large margin.  \method\ is also more parameter-efficient: it uses
9.38 million learned inference parameters,
12\% fewer than the matched LEFTNet React-OT and MeanFlow models and
48\% fewer than RitS (Extended Data
Fig.~\ref{fig:model-parameter-counts}a,b).

Because \method\ differs from React-OT in both its encoder (GeoditE
versus LEFTNet) and its generator (one-step MeanFlow versus a
multi-step conditional-flow bridge), we isolate each factor in a
controlled ablation on the same split.
At a matched LEFTNet encoder,
plain one-step MeanFlow raises mean RMSD from $0.1214$ to
$0.1269$\,\angstrom\ ($4.5\%$), while the AlphaFlow curriculum reduces
it to $0.1233$\,\angstrom.  A GeoditE multi-step model reaches
$0.1202$\,\angstrom, a plain GeoditE one-step model
$0.1262$\,\angstrom, and GeoditE one-step MeanFlow with the AlphaFlow
curriculum $0.1222$\,\angstrom\ before any reward post-training.
Reward post-training accounts for the remaining improvement to
$0.1108$\,\angstrom\ (Extended Data Table~\ref{tab:ablation} and
Supplementary Note~6).

We next evaluate sampling throughput. On a single \gpuname,
\method\ generates $421.6$ samples\,s$^{-1}$, $30.3\times$ the
throughput of the React-OT baseline with 18 evaluations, and reaches
$1{,}269$ samples\,s$^{-1}$ when compiled with PyTorch Inductor
(Extended Data Table~\ref{tab:throughput},
Fig.~\ref{fig:accuracy-throughput}d).

\subsection{\method\ demonstrates better \NFE\ scaling}
\label{sec:exp-nfe}
\begin{figure}[!tp]
  \centering
  \IfFileExists{figures/NFE-scaling.pdf}{%
    \includegraphics[width=\textwidth]{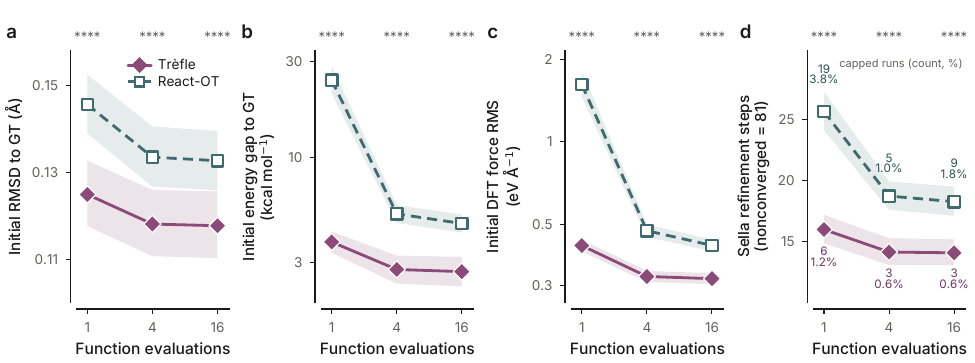}%
  }{%
    \fbox{\parbox[c][38mm][c]{0.94\textwidth}{\centering
      Upload \texttt{figures/NFE-scaling.pdf}}}%
  }
  \caption{\textbf{\method\ outperforms React-OT across the evaluated
  inference budgets.}
  The g-xTB-post-trained \method\ (purple diamonds) and React-OT (teal
  open squares) are evaluated at $\NFE\in\{1,4,16\}$ on the same
  $n=500$ reactions from the separate $19{,}897$-reaction held-out
  partition, using the DFT and Sella protocol described in Methods.
  \textbf{a}, RMSD to the reference transition state.
  \textbf{b}, Absolute DFT energy difference from the reference
  transition state.
  \textbf{c}, DFT force RMS before refinement.
  \textbf{d}, Failure-inclusive Sella optimizer-step count, with
  non-converged or errored runs assigned 81 steps.
  Labels give the corresponding failure count and rate.
  Panels \textbf{a}--\textbf{c} report pre-refinement quantities; the
  ordinates in \textbf{b} and \textbf{c} and the \NFE\ axis in every
  panel are logarithmic. Points show aggregate means across the 500
  paired reactions; shaded bands are $95\%$ percentile-bootstrap
  confidence intervals for those means, obtained from 10,000 paired
  reaction-level resamples. Asterisks compare the methods at each
  inference budget using two-sided reaction-paired Wilcoxon signed-rank
  tests with Holm correction across the twelve displayed comparisons;
  all comparisons have Holm-adjusted $P<10^{-4}$ (****). Exact adjusted
  $P$ values and resampling details are given in Methods.}
  \label{fig:nfe-scaling}
\end{figure}

We next examine how prediction quality scales with the inference budget
(NFE) and whether the predictions remain useful as TS initial guesses.
\citet{duan2025reactot} advocate this downstream view but assess it only
through the energy gap to the labelled TS; here we broaden it to DFT
forces and the cost of DFT-based Sella refinement. Because the MeanFlow
map admits multi-step integration, both methods can be compared along a
common axis.
For \method\ we apply the learned finite-interval map sequentially on a
uniform partition of $[0,1]$, so a budget of $\NFE=S$ issues exactly $S$
network calls; React-OT integrates its instantaneous velocity field with
a fixed-step explicit midpoint rule.
Grids, time orientation and update rules are given in Methods.
We evaluate the g-xTB post-trained \method\ and React-OT at
$\NFE\in\{1,4,16\}$ on a random subset of $500$ reactions from the separate
$19{,}897$-reaction held-out partition reserved for DFT evaluation
cohorts (Methods; at $\NFE=1$, where a two-evaluation midpoint step exceeds
the budget, React-OT falls back to a single explicit Euler step).
Unrefined predictions are scored by RMSD and
energy gap to the labelled TS and by DFT force root-mean-square (RMS); refined predictions
by capped Sella step count, with non-converged runs counted at the
81-step cap and success rates reported alongside
(Fig.~\ref{fig:nfe-scaling}; DFT and Sella protocol in Methods).
Across the budget range, \method\ outperforms React-OT on all four
measures at the reaction level. For both methods, nearly all of the
improvement is realized by four evaluations, and the curves are close
to flat from four to sixteen. The gap is widest at $\NFE=1$ and
persists at $\NFE=16$. Notably, \method's  single-evaluation prediction already achieves a lower initial RMSD to the labelled TS, and a comparable Sella refinement cost, than React-OT attains with sixteen evaluations.

\subsection{Benchmarking surrogate-PES reward oracles}
\label{sec:exp-oracle}

Reward modelling requires a low-cost surrogate-PES oracle that can score
every rollout during online post-training. Machine-learned force fields are
typically benchmarked on near-equilibrium structures and molecular dynamics (MD) trajectories
\citep{batatia2023foundation,levine2025omol25}, rather than the near-saddle
regions relevant to transition states. A suitable TS reward oracle must rank
candidates consistently with DFT, identify index-one Hessians and cheaply
score every rollout. 

We therefore assembled ReactOracleBench which combines path-ranking, saddle-classification and
preference-ranking tasks. The path data include $48{,}497$ observations
from $3{,}550$ complete Sella trajectories and $61{,}803$ from $4{,}415$
complete DFT IRC paths, totalling
$110{,}300$ observations of $105{,}855$ unique geometries across $69$
reaction families. Path ranking uses the complete trajectories
(Fig.~\ref{fig:mlff-benchmark}a). Saddle classification uses $3{,}550$
Hessian-labelled structures ($2{,}906$ index-one positives and $644$
negatives), together with a balanced subset of $1{,}288$ structures
($644$ per class; Fig.~\ref{fig:mlff-benchmark}b). Full-cohort
precision and recall at the natural class prevalence are reported in
Extended Data Fig.~\ref{fig:oracle-hessian-fullset-timing}a.

The preference data contain $12{,}000$ candidates, generated in groups
of eight for $500$ Reaction-QM training reactions at three noise scales.
All candidates came from the epoch-0 checkpoint of a g-xTB
reward-post-training run initialized from the supervised Reaction-QM
model. The preference task tests within-reaction energy and force ranking on
grouped generator candidates, following the same group-relative comparison
principle as post-training. The benchmark is detailed in
\hyperref[sec:met-post]{Methods} and
Fig.~\ref{fig:mlff-benchmark}c,d. Benchmark construction and oracle
selection exclude both held-out Reaction-QM partitions and the $n=200$
and $n=500$ DFT evaluation cohorts.

Seven oracles benchmarked against this reference define an
accuracy--cost Pareto frontier for machine-learned force fields in the
transition-state region, with joint energy--force throughput spanning
$38$ to $366$
structures\,s$^{-1}$ on a node with one visible \gpufull\ and
\cpuname\ (Fig.~\ref{fig:mlff-benchmark}a--d). These cost values time only the
joint energy--force evaluation. The complete finite-difference
Hessian-index operation used by the Hessian reward term is timed
separately in Extended Data Fig.~\ref{fig:oracle-hessian-fullset-timing}b.
Across the six preference-ranking accuracy--cost comparisons
(energy and force at three noise scales), GFN2-xTB and g-xTB appear on
all six Pareto frontiers, while UMA-s-1.2 appears on five, more often
than the other oracles; together with MACE-OMol XL, the strongest of the
MACE variants, these four form the overall accuracy--cost frontier. GFN2-xTB is the fastest but provides the least faithful path
and preference rankings. UMA-s-1.2 gives the most accurate index-one
classification and remains accurate on the ranking tasks, whereas
g-xTB offers consistently high fidelity at semiempirical, CPU-only
cost.  From the frontier we carry one
representative of each family forward as a candidate reward model:
g-xTB, MACE-OMol and UMA-s-1.2.  A concurrent assessment of these
potentials for automated transition-state searches likewise finds the
ranking to vary by reaction class \citep{marks2026reliable}.

\subsection{Post-training is robust to the choice of reward oracle}
\label{sec:exp-post}

Using g-xTB, MACE-OMol XL (MACE) and UMA-s-1.2 (UMA) as candidate
oracles, we post-train an otherwise identical generator. We
compare the three models with the pre-trained checkpoint on a
prespecified cohort of 200 reactions from the separate Reaction-QM
test split
(\hyperref[sec:si-test200]{Supplementary Note~9}), following the
GPU4PySCF and Sella \citep{hermes2022sella} protocol used in
Fig.~\ref{fig:nfe-scaling}.
No generator was optimized against this pipeline, though supervised
pre-training used DFT-labelled TS geometries.

\begin{figure}[!tp]
  \centering
  \includegraphics[width=1.0\textwidth]{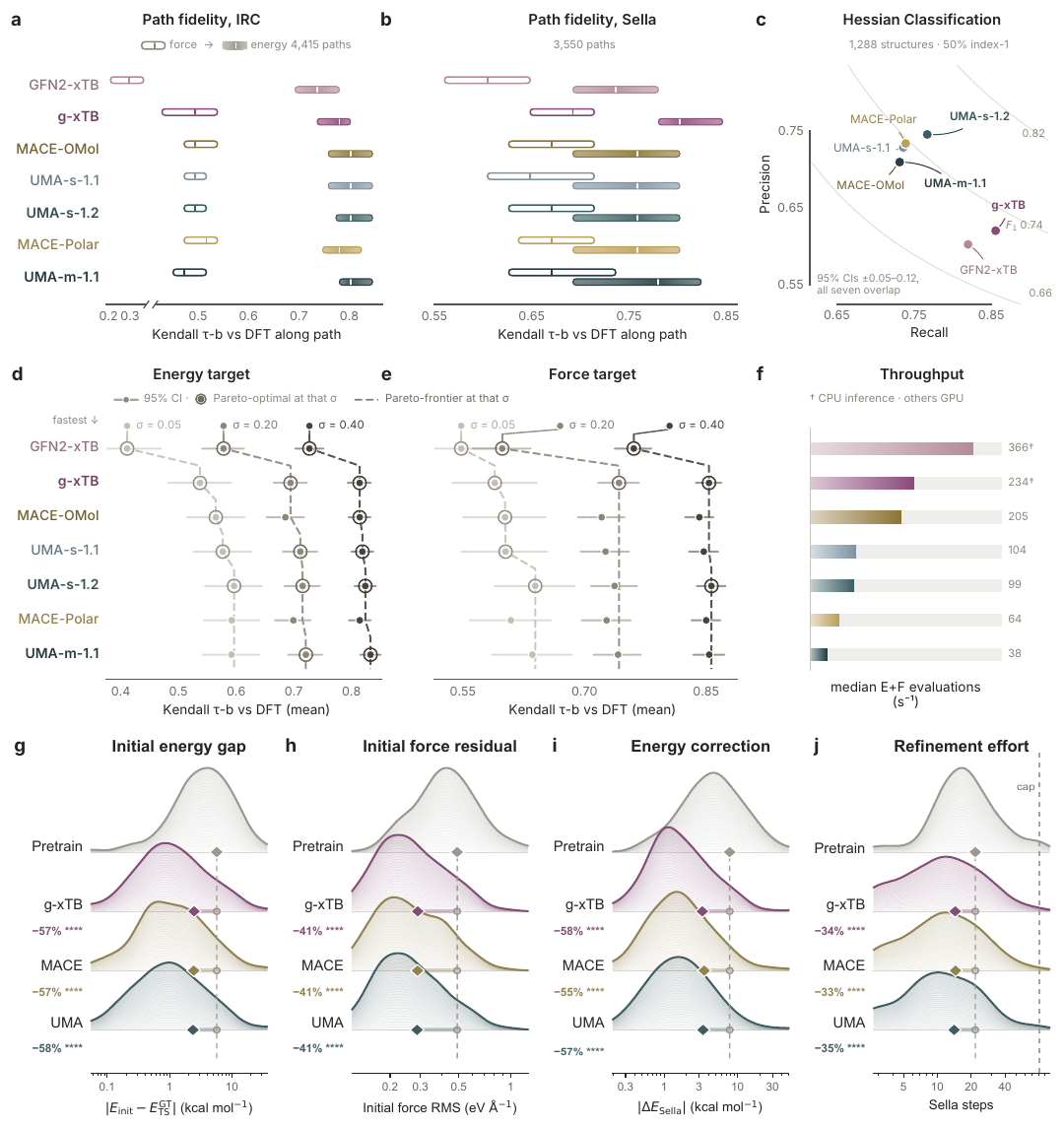}
  \caption{\textbf{ReactOracleBench identifies an accuracy--cost Pareto
  frontier, and each tested frontier oracle improves DFT-verified
  outcomes after post-training.}
  \textbf{a}, Agreement with DFT along Sella (circles) and IRC (squares)
  paths: median Kendall $\tau_b$ for energy (filled) and force (open)
  rankings; bars, interquartile ranges.
  \textbf{b}, Precision and recall for index-one Hessian
  classification; grey, $F_1$ contours.
  \textbf{c},\,\textbf{d}, Mean $\tau_b$ for within-group energy
  (\textbf{c}) and force (\textbf{d}) rankings at three midpoint-prior
  noise scales $\sigma$; bars, $95\%$ confidence intervals.  Right,
  median joint energy--force throughput (structures\,s$^{-1}$, one
  \gpuname\ node).
  GFN2-xTB, g-xTB, MACE-OMol XL and UMA-s-1.2 form the Pareto frontier.
  \textbf{e}--\textbf{h}, DFT and Sella evaluation of the pre-trained
  generator and generators post-trained with g-xTB, MACE or UMA on
  $n=200$ paired Reaction-QM reactions: initial energy gap
  (\textbf{e}), initial force RMS (\textbf{f}), energy correction
  (\textbf{g}) and capped optimizer steps (\textbf{h}); logarithmic
  axes.  Diamonds, means; percentages, mean change from pre-training;
  paired sign-flip tests, Holm-corrected, $P<10^{-4}$ (****).}
  \label{fig:mlff-benchmark}
\end{figure}

All three improve all four measures by similar margins
(Fig.~\ref{fig:mlff-benchmark}e--h): the energy gap to the labelled TS
falls by $57$--$58\%$, the initial DFT force RMS by $41\%$, the
refinement energy correction by $55$--$58\%$ and the capped Sella step
count by $33$--$35\%$ (reaction-paired sign-flip tests, all twelve
Holm-adjusted $P=6.0\times10^{-5}$).
Post-training also slightly raises the refinement success rate, from
$194/200$ reactions for the pre-trained checkpoint to $199/200$ for
every oracle, although this difference is not statistically significant
at this cohort size.  The gains hold for both a state-of-the-art
semiempirical potential and two foundational machine-learned force
fields, so the framework is robust to its reward oracle.

Given their similar performance in this evaluation, we therefore adopt
the most accessible of the three, g-xTB, for the remaining experiments:
it is semiempirical, runs on CPUs alone and matches the machine-learned
force fields on every downstream measure without their GPU requirement
(configuration in Supplementary Table~\ref{tab:hparams}).  The
corresponding MACE-OMol XL training and validation dynamics are given in
Extended Data Fig.~\ref{fig:mace-reward-dynamics}.

\subsection{\method\ transfers to unseen elements and larger molecules}
\label{sec:exp-transfer}

\begin{figure}[!tp]
\centering
\includegraphics[width=1.0\linewidth]{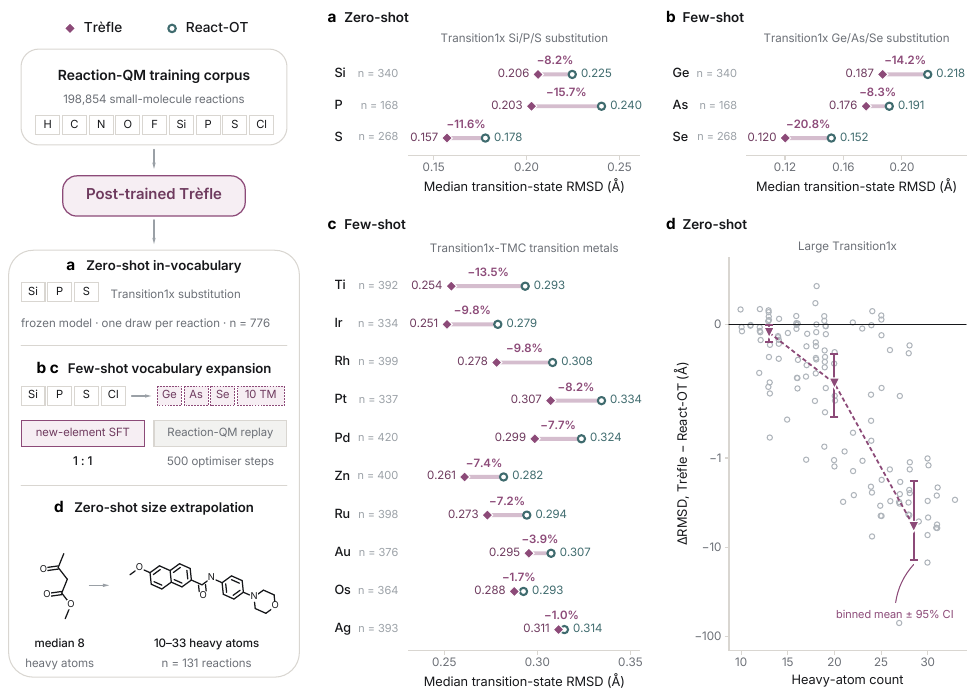}
\caption{\textbf{The post-trained one-step generator improves on
React-OT across all 16 element conditions and widens its margin with
molecular size.}
\textbf{a}, Zero-shot median transition-state RMSD on the Transition1x
Si/P/S substitution benchmark \citep{darouich2026robusttsgen}
(Si $n=340$, P $n=168$, S $n=268$).  Diamonds mark \method, circles
React-OT; labels give \method's relative reduction; lower is better.
\textbf{b}, Few-shot performance on Transition1x Ge/As/Se
(Ge $n=340$, As $n=168$, Se $n=268$).
\textbf{c}, Few-shot performance on the 10 Transition1x-TMC
transition metals ($n=334$--$420$ per metal); no transition metal
appears in the pre-training vocabulary.
\textbf{d}, Zero-shot performance on the large-Transition1x evaluation
set proposed in the FragmentFlow paper \citep{shprints2026fragmentflow}
($n=127$ reactions, 10--33 heavy atoms; up to 69 atoms in total):
RMSD difference, \method\ minus React-OT, plotted against heavy-atom
count.  Four of the 131 published reactions, on which the React-OT
structure diverged numerically ($\mathrm{RMSD}>200$\,\angstrom), are
excluded from the panel and from every statistic derived from it
(Supplementary Note~8).  Small open circles are individual reactions;
large triangles with error bars, connected by the dotted trend line, are
binned means with $95\%$ confidence intervals; negative values favour
\method; on the negative side, the axis is logarithmic in the
magnitude of the RMSD difference.}
\label{fig:transfer}
\end{figure}

Post-training optimizes the generator against reactions drawn from its
own training distribution, so we next ask whether its advantages
survive chemistry the model has never seen.  All transfer evaluations
start from the g-xTB-post-trained Reaction-QM checkpoint, with
React-OT as the baseline (Fig.~\ref{fig:transfer}).
The Transition1x element-substitution benchmark
\citep{darouich2026robusttsgen} replaces C, N and O with heavier
elements from the same groups.  Reaction-QM covers H, C, N, O, F, Si,
P, S and Cl, so the substituted reactions fall into two regimes.  The
third-period substitutions Si, P and S lie within the training
vocabulary and are evaluated zero-shot, whereas the fourth-period
substitutions Ge, As and Se lie outside it, as do the 10 transition
metals of the Transition1x-TMC (transition-metal complex) set.  For these out-of-vocabulary
elements, the model is few-shot fine-tuned on labelled reactions after
extending the element vocabulary (Fig.~\ref{fig:transfer}b,c;
\hyperref[sec:met-transfer]{Methods}).

On the in-vocabulary substitutions, \method\ attains a lower median TS
RMSD than React-OT for every element: $0.206$ versus
$0.225$\,\angstrom\ for Si, $0.203$ versus $0.240$\,\angstrom\ for P
and $0.157$ versus $0.178$\,\angstrom\ for S, reductions of
$8.2$--$15.7\%$ (Fig.~\ref{fig:transfer}a).  The advantage persists
when few-shot adaptation extends the vocabulary to the heavier
main-group elements Ge, As and Se, with $8.3$--$20.8\%$ lower median
RMSD (Fig.~\ref{fig:transfer}b).

Transition metals provide the largest element-vocabulary shift
considered here: none appears during pre-training, and their variable
coordination and oxidation states differ substantially from the
main-group chemistry represented in Reaction-QM.  Few-shot adaptation
nevertheless extends \method\ to all 10 metals of the Transition1x-TMC
set, and \method\ attains a lower median RMSD than React-OT on every
one, by $1.0$--$13.5\%$ (Fig.~\ref{fig:transfer}c).  The margin is
largest for
Ti ($13.5\%$; $0.254$ versus $0.293$\,\angstrom), followed by Rh and
Ir ($9.8\%$ each), both mainstays of homogeneous catalysis.
These
results indicate that, after labelled vocabulary expansion, the
generator can be adapted to these transition-metal test sets with lower
median RMSD than the adapted React-OT baseline.

\label{sec:exp-size}

Having established transfer across new reaction distributions and
element vocabularies, we next ask whether \method's advantage persists
as molecular size extends beyond the range typical of the training
data. Large-Transition1x \citep{shprints2026fragmentflow} provides this
test within the training element vocabulary: its 131 reactions span
10--33 heavy atoms and up to 69 atoms in total, compared with median
heavy-atom counts of 6 in Transition1x and 8 in Reaction-QM IRC. We
evaluate \method{} and React-OT zero-shot on this set.

Four cases with numerically divergent React-OT outputs
($\mathrm{RMSD}>200$\,\angstrom) are excluded from the plotted
paired-difference analysis; their complete values and sensitivity
analyses are reported in Supplementary Note~8. Across the remaining 127
reactions we compare the two methods reaction by reaction through
$\Delta\mathrm{RMSD}=\mathrm{RMSD}(\text{\method})-\mathrm{RMSD}(\text{React-OT})$,
so that negative values favour \method\ (Fig.~\ref{fig:transfer}d).
The advantage grows with molecular size. In the smallest bin (10--16
heavy atoms) the two methods are close, with a median difference of
$-0.06$\,\angstrom\ and \method\ closer to the reference in $72\%$ of
reactions; in the largest bin (24--33 heavy atoms) the median difference
is $-2.4$\,\angstrom\ and \method\ is closer in $94\%$ (Supplementary
Table~\ref{tab:si-sizebins}). The rank correlation between
$\Delta\mathrm{RMSD}$ and heavy-atom count is $\rho=-0.681$
(family-preserving permutation $P<10^{-4}$). The largest differences,
tens of \aa ngstr\"om, arise where React-OT's structures diverge on the
biggest molecules; the binned means in Fig.~\ref{fig:transfer}d are
pulled down by these cases, whereas the medians and win fractions are
not.

\subsection{Reward post-training increases saddle recovery across
ketohydroperoxide decomposition channels}
\label{sec:exp-case}

A transition-state prediction is useful only if refinement reaches an
index-one saddle point whose intrinsic reaction coordinate connects
the intended reactant and product. We therefore evaluate repeated saddle
recovery on 13 decomposition channels of a
$\gamma$-ketohydroperoxide from the public YARP archive
\citep{zhao2023comprehensive}. This bifunctional species carries
carbonyl and hydroperoxide groups on the same backbone.
$\gamma$-Ketohydroperoxides are key chain-branching intermediates in
low-temperature hydrocarbon autoignition, and their competing
decomposition pathways make them a stringent test for automated reaction
discovery. Canonical reactant and product graph matching finds none of
these 13 channels in Reaction-QM, so they are absent from the training
data.

For each KHP channel, the reward post-trained \method{}, the pretrained
generator and React-OT receive the same 20 independently constructed
endpoint pairs. Each prediction is refined for at most 80 Sella steps
using density functional theory energies and gradients; no
machine-learned force field or other low-cost surrogate is used for
Sella refinement. Refinement is followed by a frequency calculation and
a bidirectional intrinsic reaction coordinate calculation. Recovery
requires exactly one significant imaginary mode and the intended
unordered endpoint pair. Percentages below are pooled trial recovery
rates; channel coverage counts channels recovered at least once.
Failures during endpoint construction, inference, refinement, frequency
analysis or intrinsic reaction coordinate tracing remain in the fixed
denominator.

For the ketohydroperoxide, reward post-training recovers the intended
channel in 127 of 260 trials (48.8\%), compared with 66 of 260 (25.4\%)
for the pretrained generator and 64 of 260 (24.6\%) for React-OT
(Fig.~\ref{fig:epoxidation-case-study}b). It recovers all 13 channels
at least once, whereas the pretrained generator recovers 11 and React-OT
recovers 12. React-OT's sole unrecovered KHP pathway is channel 9, the
second-highest-barrier channel (101.7\,kcal\,mol$^{-1}$), which
requires four coordinated heavy-atom bond changes spanning bond cleavage
and formation; reward post-training recovers it in 1 of 20 trials.
Reward post-training also lowers the median Sella step count in every
channel. Across all 260 trials, the median is 25 steps, compared with 55
for pretraining and 45.5 for React-OT
(Fig.~\ref{fig:epoxidation-case-study}a).

As a separate mechanistic illustration, ethene epoxidation by a cyclic
peroxyphosphorane requires four coordinated forming and cleaving bonds.
The post-trained prediction halves the mean absolute error across these
distances (0.122 versus 0.259~\angstrom), refines in 14 rather than 31
accepted Sella steps and reaches an index-one saddle connected to the
intended endpoints; its imaginary mode overlaps the intended
bond-rearrangement coordinate by 0.707 (Supplementary
Fig.~\ref{fig:si-oxygen-transfer};
\hyperref[sec:si-oxygen-transfer]{Supplementary Note~11}).

\subsection{Reward post-training increases saddle recovery across
bimolecular reaction channels}
\label{sec:exp-case-bimolecular}

We next evaluate 85 channels across ten bimolecular systems from the
public YARP archive \citep{zhao2023comprehensive}, including butadiene
and ethene, the parent Diels--Alder cycloaddition. Canonical reactant and
product graph matching finds no overlap with Reaction-QM, so none of
these channels appears in the training data. The same three
models receive the same 20 independently constructed endpoint pairs per
channel and are evaluated with the DFT/Sella protocol above. Each
connected fragment is generated from its mapped molecular graph,
optimized independently and rigidly aligned to the corresponding public
intrinsic-reaction-coordinate terminal. This preserves the reference
docking pose without copying the internal reference conformations.

\begin{figure}[!tp]
  \centering
  \IfFileExists{figures/case-study.pdf}{%
    \includegraphics[width=\textwidth]{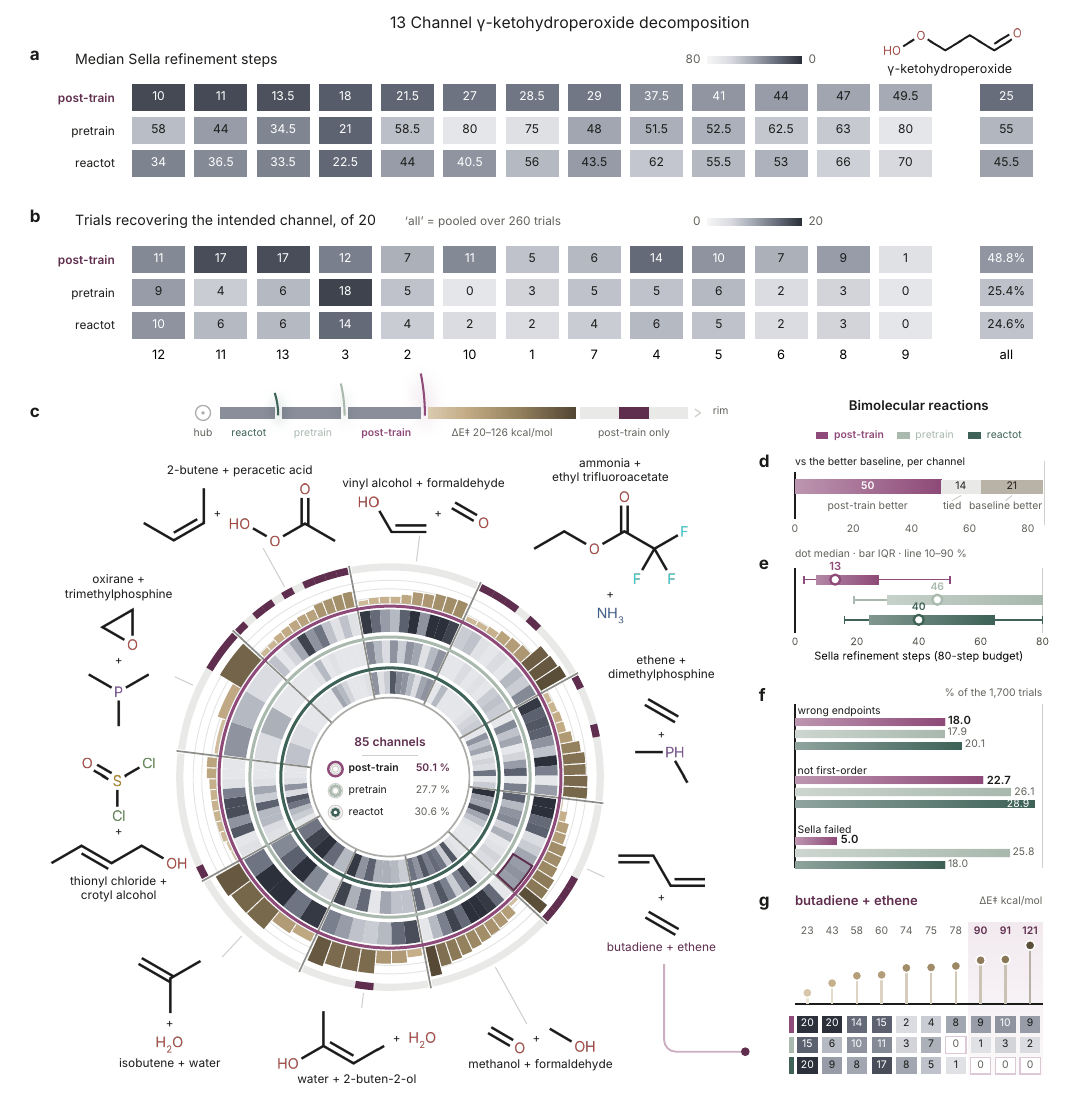}%
  }{%
    \fbox{\parbox[c][38mm][c]{0.94\textwidth}{\centering
      Upload \texttt{figures/case-study.pdf}}}%
  }
  \caption{\textbf{Reward post training improves recovery across
  unimolecular and bimolecular reaction channels.}
  \textbf{a}, Median Sella steps over 20 trials for 13 $\gamma$
  ketohydroperoxide channels; the final column pools 260 trials.
  \textbf{b}, Intended-channel recoveries out of 20. Recovery requires
  one significant imaginary mode and the intended intrinsic reaction
  coordinate endpoints.
  \textbf{c}, Trial recovery counts for 85 channels across ten
  bimolecular systems. Concentric heatmaps compare reward post training,
  pretraining and React-OT; the outer annulus gives the electronic
  activation barrier and centre values give pooled recovery rates over
  1,700 trials.
  \textbf{d}, Per-channel recovery comparison between reward post
  training and the stronger of pretraining and React-OT, classified as
  post training better, tied or baseline better.
  \textbf{e}, Failure-inclusive Sella steps. Dots, medians; bars,
  interquartile ranges; lines, 10th to 90th percentiles.
  \textbf{f}, Fractions ending at wrong endpoints, not first order, or
  Sella failure.
  \textbf{g}, Butadiene plus ethene channels ordered by barrier, with
  recovery counts below; shading marks the three highest-barrier
  channels.}
  \label{fig:epoxidation-case-study}
\end{figure}

Across 1,700 trials per model on the bimolecular set, 50.1\% of
reward-post-trained predictions recover the intended channel, compared
with 27.7\% for pretraining and 30.6\% for React-OT
(Fig.~\ref{fig:epoxidation-case-study}c). The median Sella cost falls to
13 steps, from 46 and 40 steps, respectively
(Fig.~\ref{fig:epoxidation-case-study}e). Wrong-endpoint outcomes remain
similar at 18.0\%, 17.9\% and 20.1\%, whereas non-first-order outcomes
fall to 22.7\% from 26.1\% and 28.9\%, and Sella failures fall to 5.0\%
from 25.8\% and 18.0\%
(Fig.~\ref{fig:epoxidation-case-study}f). The remaining $4.2\%$,
$2.5\%$ and $2.4\%$, respectively, are other pipeline errors retained
in the denominator but not displayed in panel f.

Butadiene--ethene is a useful test because its classical Diels--Alder
cycloaddition and less-favourable competing channels share the same
reactants but require different intermolecular orientations and bond
rearrangements. This system most clearly separates the methods at the
high-barrier tail (Fig.~\ref{fig:epoxidation-case-study}g). Its three
highest-barrier channels, at 90, 91 and 121\,kcal\,mol$^{-1}$, lie far
above the classical [4+2] cycloaddition and are sampled only rarely by
the pretrained generator. Reward post-training recovers them in 9, 10
and 9 of 20 trials; the pretrained generator recovers 1, 3 and 2, and
React-OT recovers none. Group-relative post-training can upweight such
rare samples when they satisfy the stationarity, Hessian-index and
reaction-progress terms of the reward. This result is consistent with
enrichment of rare, reward-aligned transition-state geometries,
including high-barrier channels.  Together, the KHP and bimolecular
evaluations cost approximately 4,465 GPU hours with GPU4PySCF.

\section{Discussion}
\label{sec:discussion}

\method\ combines supervised pre-training on DFT-labelled TS geometries
with online reward post-training against fixed low-cost surrogate-PES
oracles. MeanFlow pre-training learns the one-step map, while likelihood-free
post-training compiles scalar reward preferences into the generator
parameters. Sampling requires no energy, force or Hessian oracle at inference.
The resulting generator produces a TS candidate in one network evaluation,
reduces mean and median RMSD on Reaction-QM by $8.7\%$ and $12.6\%$,
respectively, relative to the 18-evaluation React-OT baseline at $30.3\times$ its sampling
throughput, and, after reward post-training, produces geometries that
independent DFT evaluation shows to be physically better: $41\%$ lower
initial force residuals, more than half the refinement energy correction and
a third fewer refinement steps. The same post-trained generator transfers
few-shot to ten unseen transition metals, generalizes zero-shot to
molecules far larger than any in training, and, seeded into DFT-refined
searches, raises per-attempt recovery of unseen reaction channels from
$28\%$ to $50\%$.

Two design choices may be useful beyond this system.  First, the reward
model is grounded in physics rather than in preference data: cheap surrogates
of the potential energy surface play the role that learned reward models
play in reinforcement learning from human feedback, and
ReactOracleBench provides a task-specific template for evaluating
candidate surrogates on rankings and failure modes relevant to this
transition-state generator. Applying the protocol to other
structure-generation tasks will require task-specific structures,
rewards and validation criteria.  Selection proved forgiving: three
oracles spanning the accuracy--cost frontier, differing in both cost
and ranking fidelity, produced near-identical DFT-verified gains, so
the oracle can be chosen for accessibility rather than accuracy.
Second, the conditional map is deterministic and its pushforward density
is implicit, so standard trajectory-likelihood policy-gradient estimators
\citep{black2024training,fan2023dpok,liu2025flowgrpo} do not directly
provide tractable likelihood ratios for this model. The likelihood-free
regression update \citep{zheng2025diffusionnft} avoids that requirement
and accepts black-box reward terms that cannot be differentiated.  Using scalar surrogate-oracle rewards keeps rollout scoring during
post-training inexpensive, while the one-step map keeps deployment
inexpensive; related post-training of
equivariant diffusion models with force-field rewards
\citep{li2026elign} retains a multi-step sampler.

Several limitations bound the current empirical scope.  Both benchmarks
contain only neutral closed-shell singlets; charged or open-shell
surfaces, though representable in the conditioning, are untested.  The
model is conditioned on a single reactant--product pair and does not
address the multiplicity of TS conformers or their barrier ranking.
Like all endpoint-conditioned generators, \method\ requires an
atom-mapped reactant--product pair; mapping errors propagate to the
candidate.
Geometric accuracy and even low force residuals do not certify a saddle:
refinement and vibrational analysis remain necessary. Post-training
makes them cheaper, not optional. The index-one Hessian reward examines
the full internal surrogate Hessian, but Hessian index alone does not
establish that its unstable mode follows the intended reaction
coordinate. Finally, the reward
inherits the biases of its oracle; agreement among three frontier
oracles bounds this dependence but does not remove it, since all three
approximate the same electronic structure.

Reward post-training is an up-front investment that is repaid at
screening scale: charged against the DFT time it removes from downstream
refinement, the $138.1$ \gpuhour\ post-training cost is offset after roughly
$1.5\times10^{4}$ reactions (Extended Data
Fig.~\ref{fig:model-parameter-counts}d).  The amortization perspective suggests where to go next.  A natural next
step is to deploy \method\ within network-level construction, where
generative frameworks already enumerate and validate elementary steps at
scale \citep{sun2026bytecrn,tuo2025flow}, and every such step needs the
one-step, physics-rewarded seed it provides.  Charge and spin
multiplicity already enter the conditioning, so extending to
electrochemical and radical chemistry is a data problem rather than an
architectural one.  Group rollouts provide a natural interface for
uncertainty estimation and active learning, in which disagreement among
one-step samples triggers selective high-fidelity evaluation.  More
broadly, the same recipe applies to any structure-generation task with a
cheap physical evaluator, including conformer ensembles, adsorbate
placement and molecular crystals: pre-train a one-step map, benchmark candidate
evaluators on the ordering they will actually be asked to produce, and
use group-relative scalar rewards to post-train the generator.

\section{Methods}
\label{sec:methods}

\subsection{Problem formulation}
\label{sec:met-problem}
Given the reactant and product geometries of a chemical reaction, our goal is to
predict the TS geometry.
We consider reactions with $N$ atom-mapped nuclei. The atom mapping provides a
one-to-one correspondence between nuclei in the reactant and product. We write
$\Rmol,\Pmol,\vxs,\vx\in\RR^{3N}$ for the reactant, product, labelled transition
state and candidate geometry in concatenated Cartesian coordinates. We let $Z\in\mathbb{N}^{N}$ collect the atomic numbers, and let
$q_{\mathrm{chg}}\in\mathbb{Z}$ and
$m_{\mathrm{spin}}\in\mathbb{N}$ denote total charge and spin
multiplicity ($m_{\mathrm{spin}}=2S+1$). Both are conserved along an
elementary step on a single adiabatic surface. We condition on
$\cond=(\Rmol,\Pmol,Z,q_{\mathrm{chg}},m_{\mathrm{spin}})$ and
denote the empirical joint distribution over the training set
$\mathcal D$ by $p_{\mathcal D}(\cond,\vxs)$, with marginal
$p_{\mathcal D}(\cond)$.

The tuple $(Z,q_{\mathrm{chg}},m_{\mathrm{spin}})$ fixes the molecular
composition, total charge and spin multiplicity, but does not by itself specify the electronic state. We therefore
let $\Ftrue$ denote the adiabatic potential energy surface selected by the
electronic-structure model and its state convention.

On this surface, a TS is an index-one stationary point. To determine its index
after removing rigid motion, we centre $\vx$ and form
$\bm R_{\mathrm{rigid}}(\vx)$ from three translations and three
infinitesimal rotations. For axis $\bm e_\alpha$, the corresponding atom-$i$
blocks are $\bm e_\alpha$ and $\bm e_\alpha\times\vx_i$, respectively. Let
$\bm B_{\mathrm{int}}(\vx)$ be an orthonormal basis for the null space of
$\bm R_{\mathrm{rigid}}(\vx)^\top$. The internal Hessian is
\begin{equation}
\bm H_{\mathrm{int}}(\vx)
=
\bm B_{\mathrm{int}}(\vx)^\top
\nabla^2\Ftrue(\vx)
\bm B_{\mathrm{int}}(\vx).
\label{eq:internal-hessian}
\end{equation}
Thus, at $\vxs$, the atomic forces vanish,
$\nabla\Ftrue(\vxs)=\vzero$, and $\bm H_{\mathrm{int}}(\vxs)$ has
exactly one negative eigenvalue
\citep{murrell1968symmetries,wilson1955molecular}.

We generate a TS geometry in two steps. First, we draw
$\vx_1\sim p_1(\cdot\mid\cond)$ from a reaction-conditioned stochastic
prior. A deterministic map with trainable parameters $\policy$ then produces
$\widehat{\vx}=T_{\policy}(\vx_1;\cond)$.
The map is equivariant to a common orthogonal transformation and translation
of all three structures, as well as to a common permutation of atom-mapped
indices (Supplementary Note~2). Such symmetry can be implemented with an
equivariant graph neural network. For a fixed reaction, all randomness comes
from the prior draw rather than from the map, so repeated draws can produce
distinct TS candidates. Their distribution is
$\pi_{\policy}(\cdot\mid\cond)
=T_{\policy}(\cdot\,;\cond)_{\#}\,p_1(\cdot\mid\cond)$.
This within-reaction diversity is used during post-training.

Supervised distribution matching conditions the model on reaction identity,
but it does not directly enforce stationarity or Hessian index. One conceptual
way to address this mismatch at inference time is oracle guidance
(Fig.~\ref{fig:overview}a). This approach keeps the pre-trained flow fixed and
modifies the sampling dynamics using local DFT information:
\begin{equation}
\begin{aligned}
\mathcal{O}_{\mathrm{DFT}}(\vx;\cond)
&=
\left(
\Ftrue(\vx;\cond),
-\nabla_{\vx}\Ftrue(\vx;\cond),
\nabla_{\vx}^{2}\Ftrue(\vx;\cond)
\right),\\
\frac{d\vx_t}{dt}
&=
\vv_{\policy_{\mathrm{pre}}}(\vx_t,t;\cond)
+
\gamma(t)\,
\mathcal{G}\!\left(
\mathcal{O}_{\mathrm{DFT}}(\vx_t;\cond),\cond
\right).
\end{aligned}
\label{eq:dft-guidance}
\end{equation}
Here, $\mathcal{G}$ is a saddle-seeking correction. Because the DFT
information is recomputed along the sampling trajectory, this form of guidance
requires repeated oracle calls. We do not implement or benchmark the guided
sampler in this work. Instead, our post-training procedure updates the
generator using a scalar surrogate-PES reward $\reward(\vx;\cond)$.

Post-training against this reward defines a \emph{contextual bandit}. The
reaction condition $\cond$ is the context, a generated geometry is the action,
and the oracle returns a terminal reward. The episode then ends. Because the
horizon is one, there is no temporal credit assignment. The action value is the
reward itself,
$Q^{\pi_{\policy}}(\cond,\vx)=\reward(\vx;\cond)$, whereas the state value is
$V^{\pi_{\policy}}(\cond)=\EE_{\vx\sim\pi_{\policy}(\cdot\mid\cond)}
[\reward(\vx;\cond)]$. The group-relative score in
equation~\eqref{eq:adv} is a standardized Monte Carlo estimate of the one-step
advantage $Q^{\pi_{\policy}}-V^{\pi_{\policy}}$, computed from rollouts of the
same reaction.

The idealized objective for this bandit is the expected return
regularized by the Kullback--Leibler (KL) divergence:
\begin{equation}
\max_{\pi}\quad
\EE_{\cond\sim p_{\mathcal D}(\cond)}
\EE_{\vx\sim\pi(\cdot\mid\cond)}
\left[\reward(\vx;\cond)\right]
-
\lambda_{\mathrm{KL}}\,
\EE_{\cond\sim p_{\mathcal D}(\cond)}
D_{\mathrm{KL}}\!\left(
\pi(\cdot\mid\cond)\,\|\,\pi_{\policy_{\mathrm{pre}}}(\cdot\mid\cond)
\right).
\label{eq:objective-ideal}
\end{equation}
The maximizer is the reward-tilted policy
$\pi^{\star}\propto\pi_{\policy_{\mathrm{pre}}}
\exp(\reward/\lambda_{\mathrm{KL}})$. Because the reward in
equation~\eqref{eq:reward} is nonpositive and bounded below, this tilt is well
defined (Supplementary Note~7).

We cannot optimize equation~\eqref{eq:objective-ideal} directly:
$\pi_{\policy}$ is an implicit policy with no tractable density. Consequently,
neither the KL term nor a likelihood-ratio policy gradient is available. We
instead optimize the tractable surrogate
\begin{equation}
\max_{\policy}\quad
\EE_{\substack{\cond\sim p_{\mathcal D}(\cond)\\
\vx_1\sim p_1(\cdot\mid\cond)}}
\left[
\reward\!\left(T_{\policy}(\vx_1;\cond);\cond\right)
\right]
-
\lambda_{\mathrm{reg}}\,
\mathcal{R}(\policy;\policy_{\mathrm{pre}}),
\label{eq:objective}
\end{equation}
where $\policy_{\mathrm{pre}}$ denotes the pre-trained parameters and
$\lambda_{\mathrm{reg}}\geq0$ controls regularization. The regularizer
penalizes departures of the current average-velocity field, defined in
\hyperref[sec:met-meanflow]{Pre-training}, from its pre-trained counterpart:
\begin{equation}
\mathcal{R}(\policy;\policy_{\mathrm{pre}})
=
\EE_{\substack{
\cond\sim p_{\mathcal D}(\cond),\;
\vx_1\sim p_1(\cdot\mid\cond)\\
\vx_0=T_{\policy^{\mathrm{old}}}(\vx_1;\cond),\;
(s,t)\sim p_{\mathrm{time}}}}
\!\left[
\frac{1}{3N}
\left\|
\vu_{\policy}(\vx_t,s,t;\cond)
-
\vu_{\policy_{\mathrm{pre}}}(\vx_t,s,t;\cond)
\right\|_2^2
\right].
\label{eq:reg}
\end{equation}
Here $\vx_t=(1-t)\vx_0+t\vx_1$ interpolates the rollout and its prior
draw, and $\policy^{\mathrm{old}}$ is a lagged copy of the generator
parameters defined under \hyperref[sec:met-post]{Post-training}. Each velocity-field output lies in $\mathbb{R}^{3N}$ and contains only
the $N$ candidate-geometry atoms, which gives the $3N$ denominator. We
use $\lambda_{\mathrm{reg}}=0.05$.

Because $\mathcal{R}$ constrains a velocity field rather than a policy density,
equation~\eqref{eq:objective} is a surrogate for
equation~\eqref{eq:objective-ideal}, not an instance of it. Our implemented
update, equation~\eqref{eq:nft}, provides a likelihood-free policy-improvement
step for this objective
(\hyperref[sec:met-post]{Post-training: likelihood-free reward fine-tuning}).
The surrogate oracle enters only through a scalar reward: we do not propagate
gradients through it. Sampling after post-training uses no guidance term and
requires no energy, force or Hessian oracle at inference.

\subsection{Datasets}
\label{sec:met-data}
We train and evaluate the model primarily on two benchmarks. Transition1x
\citep{schreiner2022transition1x} contains $10{,}073$ elementary
reactions at the $\omega$B97x/6-31G(d) level and uses the standard
$9{,}000/1{,}073$ split. Reaction-QM \citep{reactionqm2025} is roughly
$20\times$ larger. Its intrinsic-reaction-coordinate
(\texttt{irc\_all}) subset contains $198{,}854$ elementary reactions
involving H, C, N, O, F, Si, P, S and Cl: $159{,}074$ training
reactions and two disjoint held-out partitions containing $19{,}883$
and $19{,}897$ reactions. We call the $19{,}883$-reaction partition the
Reaction-QM test split and use it for coordinate-accuracy, ablation and
throughput comparisons. The separate $19{,}897$-reaction partition
supplies the computationally intensive DFT/Sella evaluation cohorts.
Neither held-out partition enters pre-training, reward post-training,
oracle selection or checkpoint selection.

Coordinate preprocessing is benchmark-specific. For Transition1x, we retain
the reactant as the endpoint reference, use a proper-rotation Kabsch fit
\citep{kabsch1976solution} to superpose the product on the reactant, form the
aligned endpoint midpoint $\bar{\vx}=\tfrac12(\Rmol+\Pmol)$, and separately
Kabsch-fit the labelled TS to $\bar{\vx}$. For Reaction-QM, we determine one
proper rigid transformation by superposing the TS frame of the
intrinsic-reaction-coordinate trajectory on the labelled TS, then apply that
same transformation to every trajectory frame before selecting the reactant
and product endpoints. Thus, Reaction-QM preserves the path-relative pose,
whereas Transition1x uses the pairwise construction above. The data loader then
subtracts the atomic centroid of each structure. No further Kabsch fit is
applied during training or inference. Only when calculating the reported root-mean-square deviations is each
generated structure independently Kabsch-aligned to its labelled TS.

Each reaction includes optimized reactant, TS and
product geometries, together with a complete intrinsic-reaction-coordinate
trajectory at the B3LYP-D3/TZVP level. All reactions in both benchmarks
are neutral closed-shell singlets, so $q_{\mathrm{chg}}=0$ and
$m_{\mathrm{spin}}=1$ are constant across $\mathcal D$. We retain both variables in the
notation because they select the surface on which the saddle conditions
are defined. Dataset composition and structural statistics are summarized
in Extended Data Fig.~\ref{fig:dataset-comparison}.

\subsection{Centroid-projected isotropic Cartesian prior}
\label{sec:met-prior}
One-step sampling should begin near a plausible reaction path, yet
post-training requires distinct candidates for the same reaction. We
therefore place the prior at the midpoint of the preprocessed,
centroid-centred endpoints and give it a small but nonzero width. We set
$\bar{\vx}=\tfrac12(\Rmol+\Pmol)$ and add isotropic Cartesian noise with
the centroid removed, as in equivariant diffusion models
\citep{hoogeboom2022equivariant,xu2023geoldm}. With
$P_0=I_N-\tfrac1N\mathbf{1}_N\mathbf{1}_N^{\top}$ denoting the
atom-index centring projector, we draw
\begin{equation}
\vx_1
=
\bar{\vx}
+
\sigma\,(P_0\otimes I_3)\,\bm{\epsilon},
\qquad
\bm{\epsilon}\sim\mathcal N(\vzero,I_{3N}),
\qquad
\sigma=0.1~\angstrom,
\label{eq:prior}
\end{equation}
so that $\vx_1$ lies in the zero-centroid subspace
$\mathcal S_0=\{\vx\in\RR^{3N}:
(\mathbf{1}_N^\top\!\otimes I_3)\vx=\vzero\}$ and has covariance
$\sigma^2 P_0\otimes I_3$. The midpoint reduces the transport distance,
centroid removal fixes global translation, and the nonzero width
$\sigma$ preserves the within-reaction exploration needed during
post-training.

\subsection{Pre-training: conditional flow matching}
Let $\vx_0=\vxs$ be a labelled TS and $\vx_1$ a prior draw. Each pair
defines transport from the source at $t=1$ to the data at $t=0$. Along
the linear path $\vx_t=(1-t)\vx_0+t\vx_1$, $t\in[0,1]$, the
sample-conditional velocity is
$\vv_t^{\mathrm{cond}}=\vx_1-\vx_0$.

We draw training-time pairs from a fixed distribution
$p_{\mathrm{time}}$ over $0\leq s\leq t\leq1$. With probability
$\tfrac12$, the pair lies on the diagonal:
$s=t=\operatorname{sigmoid}(z-0.4)$ with
$z\sim\mathcal N(0,1)$. Otherwise, we draw two independent samples from
the same logit-normal distribution and sort them into $(s,t)$.
Conditional flow matching
\citep{lipman2023flow,liu2023flow,albergo2023building} then fits
\begin{equation}
\mathcal{L}_{\mathrm{FM}}(\policy)
=
\EE_{\substack{(\cond,\vx_0)\sim p_{\mathcal D},\;
\vx_1\sim p_1(\cdot\mid\cond)\\ t\sim p_t}}
\left\|
\vv_{\policy}(\vx_t,t;\cond)-\vv_t^{\mathrm{cond}}
\right\|_2^2,
\label{eq:fm}
\end{equation}
where $p_t$ is the $t$-marginal of $p_{\mathrm{time}}$. This
conditional objective has the same parameter gradients as the
corresponding marginal flow-matching objective
\citep{lipman2023flow}.

\subsection{Pre-training: distilling transport into one MeanFlow step}
\label{sec:met-meanflow}
MeanFlow \citep{geng2025meanflow} learns the average velocity over a
finite interval $0\leq s<t\leq1$:
\begin{equation}
\vu(\vx_t,s,t)
=
\frac{1}{t-s}\int_s^t \vv(\vx_\tau,\tau)\,d\tau,
\qquad
\Phi_{t\to s}(\vx_t)
=
\vx_t-(t-s)\vu(\vx_t,s,t),
\label{eq:avgvel}
\end{equation}
with the continuous boundary extension
$\vu(\vx_t,t,t):=\vv(\vx_t,t)$. We suppress $\cond$ here for
readability. The average and instantaneous velocities satisfy
\begin{equation}
\vu(\vx_t,s,t)
=
\vv(\vx_t,t)
-
(t-s)\frac{d}{dt}\vu(\vx_t,s,t),
\qquad
\frac{d}{dt}\vu
=
\partial_t\vu+(\vv(\vx_t,t)\cdot\nabla_{\vx})\vu.
\label{eq:mfidentity}
\end{equation}
This identity converts the path integral into a local regression target
(derivation in Supplementary Note~1). Following improved MeanFlow
(iMF) \citep{geng2025improvedmeanflow}, we estimate the marginal
instantaneous velocity inside the total derivative with the detached
boundary prediction
$\widetilde{\vv}_{\policy}(\vx_t,t;\cond)
=\sg[\vu_{\policy}(\vx_t,t,t;\cond)]$, while retaining the
sample-conditional velocity as the regression label. This gives
\begin{equation}
\mathcal{L}_{\mathrm{MF}}(\policy)
=
\EE_{\substack{(\cond,\vx_0)\sim p_{\mathcal D},\;
\vx_1\sim p_1(\cdot\mid\cond)\\
(s,t)\sim p_{\mathrm{time}}}}
\left\|
\vu_{\policy}(\vx_t,s,t;\cond)
-
\sg\!\left[
\vv_t^{\mathrm{cond}}-(t-s)
\left(
\partial_t\vu_{\policy}
+(\widetilde{\vv}_{\policy}\cdot\nabla_{\vx})\vu_{\policy}
\right)
\right]
\right\|_2^2.
\label{eq:mfloss}
\end{equation}
The total derivative requires one Jacobian--vector product with tangent
$(\widetilde{\vv}_{\policy},0,1)$, so training does not require a
numerical rollout of the flow. All reported models use this improved
MeanFlow construction wherever the MeanFlow loss is applied; the change
affects training only and leaves the one-step deployment map unchanged.

We combine the two pre-training objectives as
\begin{equation}
\mathcal{L}_{\mathrm{pre}}
=
\mathcal{L}_{\mathrm{FM}}
+
\lambda_{\mathrm{MF}}\mathcal{L}_{\mathrm{MF}},
\label{eq:pretrain}
\end{equation}
with $\lambda_{\mathrm{MF}}\geq0$. The flow-matching term anchors the
instantaneous boundary $s=t$, where
$\vv_{\policy}(\cdot,t;\cond)=\vu_{\policy}(\cdot,t,t;\cond)$, whereas
the MeanFlow term supervises finite transport intervals. In practice,
the reported models use the AlphaFlow curriculum
\citep{zhang2025alphaflow}, which gradually shifts the objective from
trajectory flow matching to the MeanFlow target. Supplementary Note~1
provides the full objective and schedule.

Pre-training uses Adam with AMSGrad ($\beta_1=0.9$, $\beta_2=0.999$, no
weight decay) with learning rate $10^{-3}$. We apply a 20-epoch warm-up,
then decay the learning rate by a factor of $0.9$ every 20 epochs
starting at epoch 40. The effective batch size is 256 across eight
\gpunames, and gradients are clipped at 1. At deployment, we
evaluate the full boundary map $\Phi_{1\to 0}$ once:
\begin{equation}
\widehat{\vx}
=
T_{\policy}(\vx_1;\cond)
=
\vx_1-\vu_{\policy}(\vx_1,0,1;\cond),
\qquad
\vx_1\sim p_1(\cdot\mid\cond).
\label{eq:onestep}
\end{equation}

\subsection{Architecture}
\label{sec:met-arch}
The velocity field outputs one transport vector per atom, and each vector
must rotate with the molecule. It therefore has the same geometric
structure as an interatomic force field. We parameterize this field with
GeoditE, a GotenNet-derived architecture that combines scalar features
with steerable vector features
\citep{aykent2025gotennet,reschutzegger2026geodite}. The scalar channels
encode atom types and time. They can also encode global conditioning
variables such as $q_{\mathrm{chg}}$ and $m_{\mathrm{spin}}$, although these are constant in our
benchmarks and are therefore omitted. The steerable channels encode the
current geometry and displacements to the reactant and product.

Geometric messages pass only between atom pairs in the same structure
and within the distance cutoff. Reactant and product information instead
reaches transition-state atoms through a geometry-free scalar route.
This separation prevents cross-structure coordinate differences from
being interpreted as physical distances. The readout returns one
zero-centroid velocity per atom.

The reported model uses six interaction layers, hidden width 256, eight
attention heads, 64 radial basis functions, maximum degree $L=1$ and a
$10$~\AA\ cutoff. Supplementary Note~2 gives the full layer-by-layer
specification and equivariance property.

\subsection{Composite heuristic reward}
\label{sec:met-reward}
We use a fixed, low-cost energy--force oracle $\Ehat$ to score generated
geometries. The oracle can be either a semiempirical method or a
foundational machine-learned force field (MLFF). Our main post-training
run uses g-xTB \citep{froitzheim2025gxtb}, which ReactOracleBench places
on the accuracy--cost frontier and which requires no GPU. g-xTB returns
energies in eV and Cartesian forces in
eV\,\angstrom$^{-1}$.

For the reward comparison in Fig.~\ref{fig:mlff-benchmark}e--h, we repeat
the same post-training procedure with the other two frontier oracles:
MACE-OMol XL \citep{batatia2022mace,levine2025omol25} (the
\texttt{extra\_large} MACE-OMol-0 checkpoint) and UMA-s-1.2
(\texttt{uma-s-1p2}, OMol task) \citep{wood2025uma}. The MACE-based arm uses energy--force
batches of 32 and executes the oracle eagerly. Every calculation uses the charge $q_{\mathrm{chg}}$ and spin multiplicity
$m_{\mathrm{spin}}$ supplied by
$\cond$.

The composite heuristic reward combines five nonnegative costs. We cap
their weighted sum at $C_{\max}=5$ and assign the same maximum cost to
invalid samples and failed oracle evaluations. For g-xTB, failures include calls that exceed
a 300~s timeout. With nonnegative coefficients
$\boldsymbol\lambda$, the reward is
\begin{equation}
\reward(\vx;\cond)
=
-\min\!\Bigl\{
C_{\max},
\lambda_F C_F(\vx)
+\lambda_{\mathrm{prog}}C_{\mathrm{prog}}(\vx)
+\lambda_{H}C_{H}(\vx;\cond)
+\lambda_{\mathrm{above}}C_{\mathrm{above}}(\vx)
+\lambda_{\mathrm{geom}}C_{\mathrm{geom}}(\vx)
\Bigr\}.
\label{eq:reward}
\end{equation}
Extended Data Fig.~\ref{fig:reward-components} illustrates how four of
these five costs distinguish reward-aligned candidates from representative
low-reward failures; its $C_{\mathrm{barrier}}$ panel corresponds to
$C_{\mathrm{above}}$ in equation~\eqref{eq:reward}.
For the MACE-OMol XL training-dynamics run in Extended Data
Fig.~\ref{fig:mace-reward-dynamics}, the weights are
$\boldsymbol{\lambda}=(1.5,\,0.5,\,0,\,0.05,\,1)$, ordered as
$(F,\mathrm{prog},H,\mathrm{above},\mathrm{geom})$.

\paragraph{Residual-force cost.}
With $\widehat{\bm F}(\vx)=-\nabla\Ehat(\vx)\in\RR^{3N}$ and atom-wise
blocks $\widehat{\bm F}_i(\vx)\in\RR^3$,
\begin{equation}
F_{\mathrm{rms}}(\vx)
=
\left[
\frac{1}{N}\sum_{i=1}^{N}
\left\|\widehat{\bm F}_{i}(\vx)\right\|_2^2
\right]^{1/2},
\qquad
C_F(\vx)
=
\frac{F_{\mathrm{rms}}(\vx)}
{1~\mathrm{eV}\,\angstrom^{-1}}.
\label{eq:force_reward}
\end{equation}

\paragraph{Endpoint-progress cost.}
For unordered atom pairs $1\leq i<j\leq N$, let $d_{ij}^{\mathrm R}$,
$d_{ij}^{\mathrm P}$ and $d_{ij}(\vx)$ denote the interatomic distances
in the reactant, product and candidate. Among pairs satisfying
$|d_{ij}^{\mathrm P}-d_{ij}^{\mathrm R}|\ge 0.3~\angstrom$, the active
set $\mathcal A$ retains at most the 12 largest endpoint distance
changes. For each $(i,j)\in\mathcal A$, we define
\begin{equation}
\xi_{ij}(\vx)
=
\frac{d_{ij}(\vx)-d_{ij}^{\mathrm R}}
     {d_{ij}^{\mathrm P}-d_{ij}^{\mathrm R}}.
\label{eq:pair_progress}
\end{equation}
Using $\xi_{\mathrm{low}}=0.15$, $\xi_{\mathrm{high}}=0.85$ and
$\lambda_{\mathrm{var}}=0$,
\begin{align}
C_{ij}(\vx)
&=
\bigl[\xi_{\mathrm{low}}-\xi_{ij}(\vx)\bigr]_+^2
+
\bigl[\xi_{ij}(\vx)-\xi_{\mathrm{high}}\bigr]_+^2,
\\
C_{\mathrm{prog}}(\vx)
&=
\frac{1}{|\mathcal A|}
\sum_{(i,j)\in\mathcal A}C_{ij}(\vx)
+
\lambda_{\mathrm{var}}\,
\operatorname{Var}_{(i,j)\sim\operatorname{Unif}(\mathcal A)}
\!\left[
\operatorname{clip}\!\left(\xi_{ij}(\vx),0,1\right)
\right],
\label{eq:progress_reward}
\end{align}
where $[z]_+=\max(0,z)$. A reaction with no eligible active pair receives
the maximum cost $C_{\max}$. This term favours coordinated motion through
the interior of the reaction.

\paragraph{Index-one Hessian cost.}
A TS has exactly one imaginary vibrational mode. Let
$\widehat{\bm H}_{\mathrm{int}}(\vx;\cond)$ be the oracle analogue of
the internal Hessian in equation~\eqref{eq:internal-hessian}, with
$\Ehat$ replacing $\Ftrue$, and let
$\widehat n_-(\vx;\cond)
=\operatorname{ind}(\widehat{\bm H}_{\mathrm{int}})$ count its
negative eigenvalues. We penalize every departure from Hessian index one:
\begin{equation}
C_H(\vx;\cond)
=
\left|\widehat n_-(\vx;\cond)-1\right|.
\label{eq:hessian_reward}
\end{equation}
Thus, $C_H$ vanishes only when the surrogate internal Hessian has exactly
one negative eigenvalue. The full Hessian is computed analytically for
differentiable oracles, including the MLFFs; for g-xTB, it is assembled
from central finite differences of forces with each Cartesian coordinate
displaced by $\pm0.005$~\angstrom, before rigid translations and rotations
are projected out.

\paragraph{Endpoint-above energy and geometry costs.}
Define the endpoint-above energy
$\Delta\Ehat_{\mathrm{above}}(\vx)
=\Ehat(\vx)-\max\{\Ehat(\Rmol),\Ehat(\Pmol)\}$.
Then
\begin{align}
C_{\mathrm{above}}(\vx)
={}&
\operatorname{softplus}\!\left(
\frac{E_{\min}-\Delta\Ehat_{\mathrm{above}}(\vx)}{s_E}
\right)
+
\lambda_{\mathrm{high}}
\operatorname{softplus}\!\left(
\frac{\Delta\Ehat_{\mathrm{above}}(\vx)-E_{\max}}{s_E}
\right),
\label{eq:above_energy_reward}
\\[-0.25em]
&E_{\min}=0.03~\mathrm{eV},
\quad E_{\max}=8.0~\mathrm{eV},
\quad s_E=0.2~\mathrm{eV},
\quad \lambda_{\mathrm{high}}=0.25.
\nonumber
\end{align}
This cost encourages the candidate to lie above both endpoints. We also
guard against invalid atomic contacts. For
$d_{\min}(\vx)=\min_{1\leq i<j\leq N}d_{ij}(\vx)$, with
$d_{\mathrm{cut}}=0.4~\angstrom$ and $\alpha_{\mathrm{geom}}=10$,
\begin{equation}
C_{\mathrm{geom}}(\vx)
=
\begin{cases}
0,
& d_{\min}(\vx)\ge d_{\mathrm{cut}},\\[2pt]
\alpha_{\mathrm{geom}}\,
\dfrac{d_{\mathrm{cut}}-d_{\min}(\vx)}{d_{\mathrm{cut}}},
& d_{\min}(\vx)<d_{\mathrm{cut}}.
\end{cases}
\label{eq:geometry_reward}
\end{equation}
DFT-level vibrational analysis is used only for evaluation.

\subsection{Post-training: likelihood-free reward fine-tuning}
\label{sec:met-post}
Likelihood-based policy gradients such as Flow-GRPO
\citep{liu2025flowgrpo} assume a multi-step stochastic sampler. They
obtain importance ratios by converting the sampling ODE into an SDE
\citep{song2021score}, which provides a Gaussian transition kernel at
each step. A MeanFlow map has neither property: it samples through one
deterministic jump, with all stochasticity supplied by the prior.
The policy $\pi_{\policy}(\cdot\mid\cond)$ defined in
\hyperref[sec:met-problem]{Problem formulation} is therefore stochastic
through its prior draw. Conditional on $\vx_1$, however, the map is
deterministic. Its one-step transition kernel is a Dirac measure, and
the resulting implicit pushforward has no tractable action likelihood or
likelihood ratio. We therefore follow the implicit-guidance perspective
of Diffusion-NFT (negative-aware fine-tuning, NFT) \citep{zheng2025diffusionnft} and absorb the
reward-improving direction into the flow weights.

For each reaction, we generate a group of $K=48$ on-policy pairs,
\[
\mathcal{G}_{\cond}
=\{(\vx_0^{(j)},\vx_1^{(j)})\}_{j=1}^{K}.
\]
Here, $\vx_1^{(j)}\sim p_1(\cdot\mid\cond)$ is a prior draw, and
$\vx_0^{(j)}=T_{\policy^{\mathrm{old}}}(\vx_1^{(j)};\cond)$ is the
corresponding one-step candidate produced by an exponential-moving-average
(EMA) copy of the generator. The oracle scores every candidate. We then
standardize rewards within each reaction using the group-relative
baseline of group-relative policy optimization (GRPO) \citep{shao2024deepseekmath}:
\begin{equation}
\adv^{(j)}
=
\frac{
\reward(\vx_0^{(j)};\cond)-\mu_{\cond}
}{
\varsigma_{\cond}+\epsilon_A
},
\qquad
\mu_{\cond},\varsigma_{\cond}
=
\operatorname{mean/std}_{j\in\mathcal{G}_{\cond}}
\reward(\vx_0^{(j)};\cond).
\label{eq:adv}
\end{equation}
We use the population standard deviation and $\epsilon_A=0.01$.
This per-reaction baseline is important because force and barrier scales
vary widely across chemistries. Standardization makes the update compare
candidates within a reaction rather than compare raw rewards across
different reactions. We convert each standardized advantage into an
optimality weight:
\begin{equation}
q^{(j)}
=
\tfrac12+\tfrac12
\operatorname{clip}\!\left(\adv^{(j)},-1,1\right)
\in[0,1].
\label{eq:optimality}
\end{equation}
A single trainable field represents both attraction and repulsion through
two branches symmetric around $\vu^{\mathrm{old}}$:
\begin{equation}
\vu_{\policy}^{+}
=
(1-\beta)\vu^{\mathrm{old}}+\beta\vu_{\policy},
\qquad
\vu_{\policy}^{-}
=
(1+\beta)\vu^{\mathrm{old}}-\beta\vu_{\policy}.
\label{eq:nftbranches}
\end{equation}

For each pair, we form
$\vx_t^{(j)}=(1-t)\vx_0^{(j)}+t\vx_1^{(j)}$. We evaluate the MeanFlow
target from equation~\eqref{eq:mfloss} on the frozen EMA copy, rather
than on the trainable field:
\begin{equation}
\vu_{\mathrm{target}}^{(j)}
=
\vv_t^{\mathrm{cond}}
-
(t-s)\left[
\partial_t\vu^{\mathrm{old}}
+
\left(\vv_t^{\mathrm{cond}}\cdot\nabla_{\vx}\right)\vu^{\mathrm{old}}
\right].
\label{eq:nft-target}
\end{equation}
All terms are evaluated at $(\vx_t^{(j)},s,t;\cond)$, and the
sample-conditional velocity remains
$\vv_t^{\mathrm{cond}}=\vx_1^{(j)}-\vx_0^{(j)}$. We obtain the total
derivative exactly with one Jacobian--vector product using tangent
$(\vv_t^{\mathrm{cond}},0,1)$. Both branches share this target, and we
minimize
\begin{equation}
\begin{aligned}
\mathcal{L}_{\mathrm{NFT}}(\policy)
=
\EE_{\cond,j,s\leq t}\Big[
&q^{(j)}
\left\|
\vu_{\policy}^{+}(\vx_t^{(j)},s,t;\cond)
-
\sg[\vu_{\mathrm{target}}^{(j)}]
\right\|^2\\
&+
(1-q^{(j)})
\left\|
\vu_{\policy}^{-}(\vx_t^{(j)},s,t;\cond)
-
\sg[\vu_{\mathrm{target}}^{(j)}]
\right\|^2
\Big].
\end{aligned}
\label{eq:nft}
\end{equation}
For a high-reward pair ($q^{(j)}\!\approx\!1$), the positive branch pulls
$\vu_{\policy}$ towards the sample. For a low-reward pair
($q^{(j)}\!\approx\!0$), the negative branch pushes it away. Extended
Data Fig.~\ref{fig:post-training-schematic} summarizes the rollout group
in equation~\eqref{eq:adv}, the optimality weight in
equation~\eqref{eq:optimality}, and the branch update in
equations~\eqref{eq:nftbranches} and~\eqref{eq:nft}.

The complete post-training loss adds the proximity regularizer from
equation~\eqref{eq:objective}:
\begin{equation}
\mathcal{L}_{\mathrm{post}}
=
\mathcal{L}_{\mathrm{NFT}}
+
\lambda_{\mathrm{reg}}
\mathcal{R}(\policy;\policy_{\mathrm{pre}}).
\label{eq:posttrain}
\end{equation}
The field $\vu^{\mathrm{old}}$ is a lagged copy of $\vu_{\policy}$.
We initialize $\policy^{\mathrm{old}}$ from the pre-trained parameters
and update it once per epoch $e$:
\[
\policy^{\mathrm{old}}\leftarrow
\rho_e\policy^{\mathrm{old}}+(1-\rho_e)\policy,
\qquad
\rho_e=\min\{5\times10^{-4}(e+1),\,0.5\}.
\]
Sampling from this lagged copy separates rollout generation from the
latest optimization step and improves stability. Because the advantage
enters through weighted regression rather than through a
likelihood-ratio surrogate, all supervision acts on clean pairs. The
update requires neither trajectory likelihoods nor back-propagation
through the oracle.

\paragraph{Relation to policy improvement.}
Equation~\eqref{eq:nft} is not a likelihood-ratio policy-gradient
estimator, and we do not claim that its fixed point maximizes
equation~\eqref{eq:objective}. It is nevertheless a policy-improvement
step for the contextual bandit defined in
\hyperref[sec:met-problem]{Problem formulation}, rather than an arbitrary
weighting.

To see this, define
$\bar{\adv}=2q-1=\operatorname{clip}(\adv,-1,1)$ as the clipped
advantage, $\vz=\vu_{\policy}-\vu^{\mathrm{old}}$ as the displacement
from the behaviour policy, and
$\bm{d}=\vu_{\mathrm{target}}-\vu^{\mathrm{old}}$. The per-sample term
in equation~\eqref{eq:nft} is exactly
\begin{equation}
q\left\|\vu_{\policy}^{+}-\vu_{\mathrm{target}}\right\|^{2}
+
(1-q)\left\|\vu_{\policy}^{-}-\vu_{\mathrm{target}}\right\|^{2}
=
\beta^{2}
\left\|
\vz-\frac{\bar{\adv}}{\beta}\bm{d}
\right\|^{2}
+
\left(1-\bar{\adv}^{2}\right)\|\bm{d}\|^{2}.
\label{eq:nft-decomp}
\end{equation}
Its minimizer is
$\vz^{\star}=(\bar{\adv}/\beta)\,\bm{d}$. The field therefore moves
towards a sampled candidate in proportion to a positive advantage,
moves away in proportion to a negative advantage, and remains unchanged
for a candidate with exactly average reward. Equivalently, at
$\policy=\policy^{\mathrm{old}}$, the gradient of
equation~\eqref{eq:nft} is $\beta\bar{\adv}$ times the gradient of plain
MeanFlow regression onto the same sample.

At the distribution level, reweighting a rollout group by $q$ produces a
candidate distribution whose expected reward is at least that of the
behaviour policy. The inequality is strict whenever the within-reaction
reward is not constant, and clipping in equation~\eqref{eq:optimality}
preserves this guarantee (Supplementary Note~7).

The implemented update only approximates this idealized operator. Finite
rollout groups, empirical reward standardization, clipping, a fixed
branch scale $\beta$, the EMA behaviour policy and finite model capacity
all separate equation~\eqref{eq:nft} from the exact improvement step.
We therefore do not claim monotonic reward improvement at every optimizer
step. Concurrent work applies the same forward-process update to
average-velocity generators for image and video synthesis
\citep{huang2026meanflownft,flowmapgrpo2026}. Because
equation~\eqref{eq:nft-target} evaluates the total derivative on the
frozen behaviour policy, the branch residuals in
equation~\eqref{eq:nft} are algebraically identical to those obtained by
branching the induced instantaneous velocity in that formulation.

\paragraph{Training procedure.}
Given dataset $\mathcal{D}$, frozen oracle $\Ehat$, time-pair
distribution $p_{\mathrm{time}}$, step counts
$N_{\mathrm{pre}},N_{\mathrm{post}}$, rollout size $K$ and lagged-copy
schedule $\rho_e$:
\begin{enumerate}[leftmargin=1.5em,itemsep=1pt,topsep=2pt]
\item \emph{Pre-training.} For $N_{\mathrm{pre}}$ steps, sample
$(\cond,\vxs)\sim p_{\mathcal D}$, $\vx_1\sim p_1(\cdot\mid\cond)$ and
$(s,t)\sim p_{\mathrm{time}}$. Set
$\vx_t=(1-t)\vxs+t\vx_1$ and update $\policy$ with
$\mathcal{L}_{\mathrm{pre}}$ (equation~\eqref{eq:pretrain}). Then freeze
$\policy_{\mathrm{pre}}\leftarrow\policy$ and initialize
$\policy^{\mathrm{old}}\leftarrow\policy$.
\item \emph{Reward model.} Define $\reward$ using
equation~\eqref{eq:reward}, with $\Ehat$ and all scales frozen.
\item \emph{Post-training.} For $N_{\mathrm{post}}$ steps, draw $K$
prior samples for each $\cond$ in a minibatch. Generate one-step
candidates with the EMA generator, score them with $\reward$, and compute
$\{\adv^{(j)},q^{(j)}\}$ using equations~\eqref{eq:adv} and
\eqref{eq:optimality}. Draw fresh time pairs and take one optimizer step
on $\mathcal{L}_{\mathrm{post}}$ (equation~\eqref{eq:posttrain}). At the
end of each epoch, update the lagged copy.
\end{enumerate}

\subsection{Baselines}
\label{sec:met-baselines}
On Transition1x, baseline results are as published
\citep{duan2023oareactdiff,duan2025reactot,kim2024tsdiff,schreiner2022neuralneb,schlama2026driftreact}.
On Reaction-QM, all methods are implemented in one consistent suite:
React-OT \citep{duan2025reactot}, MeanFlow \citep{geng2025meanflow} with
a LEFTNet encoder \citep{du2023leftnet}, a two-stage latent-flow model,
RitS (midpoint prior), TS-DFM (distance-geometry flow matching with
L-BFGS coordinate reconstruction), TS-drift, and geometry-only
interpolants (linear, image-dependent pair potential (IDPP)
\citep{smidstrup2014idpp} and geodesic \citep{zhu2019geodesic}).
Architectures, conditioning, samplers, neural function evaluations,
training budgets and selected checkpoints for every reimplemented
comparator are given in Supplementary Note~5.

\subsection{Evaluation metrics and throughput protocol}
\label{sec:met-eval}
We measure accuracy by the root-mean-square deviation (RMSD) from the
reference TS. Following \citet{duan2023oareactdiff}, we
normalize over all $3N$ coordinates; these values are a factor of
$\sqrt{3}$ lower than per-atom RMSD. Unless noted otherwise, inference starts from the deterministic
reactant--product midpoint, $\vx_1=\bar{\vx}$ in
equation~\eqref{eq:prior} (equivalently, $\bm\epsilon=\vzero$). We
generate one candidate per reaction and report both the mean and median
over the stated test set. \NFE\ denotes the number of neural function
evaluations per candidate.

\paragraph{Multi-step inference.}
For a budget of $\NFE=S$, we apply the learned finite-interval map
sequentially on the uniform grid $t_i=1-i/S$, $i=0,\dots,S$. The state
advances in the zero-centroid subspace according to
$\vx_{t_{i+1}}=\vx_{t_i}-(t_i-t_{i+1})\,
\vu_{\policy}(\vx_{t_i},t_{i+1},t_i;\cond)$. Thus, $S=1$ recovers the
single jump in equation~\eqref{eq:onestep}, whereas $S=4$ uses the nodes
$\{1,0.75,0.5,0.25,0\}$. Time runs from the
reactant--product-conditioned source at $t=1$ to the transition-state
geometry at $t=0$. Each interval costs one network
evaluation, so an $\NFE=S$ schedule makes exactly $S$ calls without
guidance or auxiliary passes.

React-OT uses the opposite orientation and integrates its instantaneous
velocity field with a fixed-step explicit midpoint rule. Its default
uniform grid has ten nodes, corresponding to nine steps and two velocity
evaluations per step, or 18 calls per structure. In the \NFE-scaling
comparison, we meet the requested budget by coarsening this grid, so
$\NFE=S$ corresponds to $S/2$ midpoint steps. At $\NFE=1$, a midpoint
step would exceed the budget, so we use one explicit Euler step.

We benchmark end-to-end throughput on a single \gpushort\ at batch size
16. Each measurement uses two warm-up training steps, twelve timed
training steps and one test-set sampling batch. In a separate
measurement, compiling the model with PyTorch Inductor using dynamic
shapes increases device-matched one-step sampling throughput from
$421.55$ to $1269.19$ samples/s, a $3.01\times$ increase.

\subsection{Reward-oracle benchmark protocol}
\label{sec:met-oracle}

\textbf{Benchmark construction.}
The path component of ReactOracleBench contains $48{,}497$ observations
from $3{,}550$ complete Sella saddle-point refinement trajectories and
$61{,}803$ observations from $4{,}415$ complete B3LYP
IRC paths. In total, these are
$110{,}300$ observations from $105{,}855$ unique geometries along
$7{,}965$ paths spanning 69 reaction families. We include a path only
when it is complete, allowing ranking over the full trajectory. The
separate preference component contains generator candidates
whose provenance is specified below. Four evaluation sets are derived
from these components.

The \emph{Sella set} contains $3{,}550$ complete refinement
trajectories, each represented by its initial, interior and final
structures. The \emph{IRC set} contains $4{,}415$ complete paths, each
represented by the transition state, interior points and both terminal
structures. Together, these two sets give the $7{,}965$ paths above.

The \emph{natural Hessian set} contains the $3{,}550$ structures with an
exact reference Hessian. Of these, $2{,}906$ are index-one positives and
$644$ are negatives, preserving the class imbalance encountered by an
oracle in deployment. The \emph{balanced Hessian set} is a
$1{,}288$-structure subset containing all $644$ negatives and $644$
positives sampled without replacement. This balanced set prevents
precision and recall from being dominated by the prevalence of positive
examples.

The \emph{preference set} uses 500 reactions from the Reaction-QM
training split. For each reaction and each of three noise scales
$\sigma\in\{0.05,\,0.20,\,0.40\}$, we generate eight candidates
from independent draws of the centroid-projected isotropic Cartesian
prior centred at the reactant--product midpoint
(\hyperref[sec:met-prior]{Centroid-projected isotropic Cartesian prior}).
The generator is the epoch-0 checkpoint from a g-xTB
reward-post-training run initialized from the supervised Reaction-QM
model, and each candidate uses four sampling steps. The full set contains
$500\times3\times8=12{,}000$ candidates. The preference task tests
within-reaction energy and force ranking on grouped generator candidates,
following the same group-relative comparison principle as post-training.
Neither
held-out Reaction-QM partition nor any reaction in the $n=200$ or
$n=500$ DFT evaluation cohorts enters benchmark construction or oracle
selection.

The exact GPU4PySCF reference completed $11{,}999$ of the $12{,}000$
evaluations. The single failure leaves 499 of 500 complete groups at the
largest noise scale and all 500 groups complete at the other two scales.
Every approximate oracle scored all $12{,}000$ candidates, and we compute
reference-dependent statistics using complete groups only. All
structures are evaluated with the charge and spin multiplicity of their
reaction ($q_{\mathrm{chg}}=0$, $m_{\mathrm{spin}}=1$ throughout).

\textbf{Scoring.}
We compare rankings using Kendall rank correlation $\tau_b$. Its tie
correction is needed because oracles can return identical scores for
symmetry-equivalent structures. For each path, we rank structures by
oracle energy and, separately, by oracle force magnitude. We then
correlate each ranking with the corresponding DFT ranking of the same
structures. Figure~\ref{fig:mlff-benchmark}a reports the median
$\tau_b$ over paths and its interquartile range, with Sella and IRC paths
shown separately.

For saddle classification, an oracle labels a structure as index one
when the oracle analogue of the internal Hessian in
equation~\eqref{eq:internal-hessian} has exactly one negative eigenvalue.
We compare this label with the exact GPU4PySCF Hessian and report
precision, recall and $F_1$ (Fig.~\ref{fig:mlff-benchmark}b).

For preference ranking, we independently rank each eight-candidate group
by energy and by force. Figure~\ref{fig:mlff-benchmark}c,d reports the
mean $\tau_b$ across groups at each noise scale, with a $95\%$
confidence interval of the mean. This follows the same group-relative comparison principle as
post-training, but post-training itself ranks the scalar composite reward
(\hyperref[sec:met-post]{Post-training}).

The reference for all three tasks is the density-fitted GPU4PySCF
B3LYP-D3(BJ)/def2-TZVP calculation described under
\hyperref[sec:met-dft]{DFT evaluation and saddle-point refinement}. We
benchmark seven oracles: the semiempirical methods GFN2-xTB
\citep{bannwarth2019gfn2} and g-xTB \citep{froitzheim2025gxtb};
the machine-learned force fields MACE-OMol XL and MACE-Polar; and the
universal models UMA-s-1.1, UMA-s-1.2 and UMA-m-1.1.

\textbf{Throughput.}
We measure cost on a separate common cohort of 32 labelled
transition-state geometries. The cohort is selected from Reaction-QM by
seeded random sampling without replacement after reaction-pair
deduplication. MACE and UMA use an operational batch size of eight. Each
model receives 128 excluded warm-up evaluations, followed by eight
complete timed passes over the cohort. We reshuffle the order
deterministically before every pass.

The semiempirical oracles use CPU batches of 32, with 128 excluded
warm-up evaluations followed by 16 complete timed repeats. The serial
GPU4PySCF reference uses four complete repeat-level passes.
Figure~\ref{fig:mlff-benchmark} reports the median throughput across the
resulting repeat-level estimates, measured as returned energy--force
results per second. We exclude one-time backend initialization. CUDA is
synchronized immediately before and after each GPU batch, and timing uses
\texttt{time.perf\_counter}.

All measurements use one visible \gpufull\ and \cpuname\ on the same
node. MACE-OMol XL and MACE-Polar run
eager, fixed-shape GPU batches of eight, padded to at most 33
atoms and 1,056 directed edges per structure. Both timed paths return
energies and autograd forces. Because MACE computes force as the negative
coordinate gradient of predicted energy, the reported cost includes the
energy forward pass and its coordinate-gradient backward pass. The UMA models use one direct
FairChem GPU call per batch of eight. GFN2-xTB and g-xTB evaluate a CPU batch of 32 through concurrent external
subprocesses, each restricted to one thread, and do not use the GPU.

GPU4PySCF has no multi-molecule batch interface. One process therefore
evaluates structures serially on the single \gpubare, without hidden GPU
multiplicity. Each structure uses one density-fitted restricted
Kohn--Sham calculation followed by its analytic gradient at
B3LYP-D3(BJ)/def2-TZVP, with grid level 3, eight CPU threads and
64\,GB calculator memory. Forces are requested first, after which the
energy is read from the calculator cache. This ensures one
self-consistent-field calculation and one gradient per joint
energy--force evaluation. Per-structure calculator construction is
included in the timing.

Separately, we measured peak allocated and reserved GPU memory for one
complete reinforcement-learning update with 48 rollout members on a
single \gpuname\ (Extended Data
Fig.~\ref{fig:model-parameter-counts}c).

\subsection{DFT evaluation and saddle-point refinement}
\label{sec:met-dft}
All independent DFT energy and force evaluations used density-fitted
GPU4PySCF~1.7.4 \citep{li2025gpu4pyscf,wu2025gpu4pyscf} at the
B3LYP-D3(BJ)/def2-TZVP level \citep{grimme2010d3,grimme2011bj}. We treated every
reaction as a neutral closed-shell singlet, with charge 0 and PySCF spin
0. Calculations used numerical-integration grid level 3, a
self-consistent-field convergence tolerance of $10^{-6}$ Hartree, at
most 200 self-consistent-field cycles, and auxiliary-basis response in
the analytic gradients. The software environment contained
PySCF~2.13.1 \citep{sun2020pyscf}, pyscf-dispersion~1.5.0, the Atomic Simulation Environment (ASE)~3.29.0, NumPy~1.26.4 and
Sella~2.4.2. We defined DFT force RMS as
$\left[(3N)^{-1}\sum_i\lVert\bm F_i\rVert_2^2\right]^{1/2}$, a
per-coordinate quantity that is $\sqrt{3}$ smaller than the per-atom
$F_{\mathrm{rms}}$ of equation~\eqref{eq:force_reward} for the same
forces.

Saddle refinement used Sella~2.4.2 \citep{hermes2022sella} with target order one, a maximum-force
criterion of $0.05$~eV\,\angstrom$^{-1}$ and a limit of 80 optimizer
steps. We did not override the coordinate system or trust radius in the
Sella constructor, so the version-2.4.2 defaults were retained. Reported
refinement counts are Sella optimizer steps.

We assigned a failure-inclusive value of 81 to every non-converged run
and to any run terminated by an electronic-structure or optimizer
exception. This value lies one step beyond the 80-step cap, making every
failure strictly worse than any converged refinement while keeping the
penalty bounded. Because this fixed value raises the mean in proportion
to the failure rate, we also report step-count distributions. Specifically,
Figs.~\ref{fig:nfe-scaling}d and~\ref{fig:mlff-benchmark}h show medians
and interquartile ranges alongside the means, so convergence behaviour is
not hidden by the cap.

For accepted-step trajectory plots, we matched each Sella log row
sequentially to its stored trajectory frame using energy and maximum
force. This procedure excludes rejected trial frames. The plotted
trajectory begins with the step-zero structure, whereas the reported
optimizer-step count includes only subsequent accepted moves.

For the case-study structures in
Fig.~\ref{fig:epoxidation-case-study}, we evaluated the electronic
Hessian analytically using a tighter self-consistent-field tolerance of
$10^{-9}$ Hartree. The D3(BJ) Hessian contribution was obtained by finite
differences of D3(BJ) gradients, with grid and auxiliary-basis responses
included. For the frequency calculation, we constructed the projection
in equation~\eqref{eq:internal-hessian} in mass-weighted coordinates and
applied it before diagonalization. Frequencies below
$-50$~cm$^{-1}$ were classified as significant imaginary modes, and an
index-one assignment required exactly one such mode. To test endpoint
connectivity, we integrated the IRC in
both directions from the refined saddle at the same level of theory. We
then compared the bond connectivity of the relaxed endpoints with that
of the intended reactant and product.

A refinement counts as recovering the intended saddle in
Fig.~\ref{fig:epoxidation-case-study} only if both tests pass: the
refined structure has exactly one significant imaginary mode, and its
bidirectional IRC relaxes to the intended reactant and product. This
criterion is independent of the optimizer-step count. A refinement can
therefore terminate in few steps and still fail. We apply the same
criterion to all three inference recipes across both evaluation cohorts:
the 13 KHP decomposition channels and 85 channels from ten bimolecular
systems.

The KHP benchmark contains the 13 b2f2 channels, with two bonds formed
and two broken, retained after an $80$~kJ\,mol$^{-1}$ enthalpy filter for
$\gamma$-ketohydroperoxide 3-hydroperoxypropanal (InChIKey prefix
\texttt{XSASRUDTFFBDDK}). These channels come from version 7 of the
public Zhao--Savoie YARP archive
(doi:\href{https://doi.org/10.6084/m9.figshare.14766624.v7}%
{10.6084/m9.figshare.14766624.v7}). Two channels required endpoint
remapping:
\texttt{\_21\_0\_3}$\rightarrow$\texttt{\_21\_0\_0} and
\texttt{\_29\_0\_1}$\rightarrow$\texttt{\_29\_0\_3}.
Supplementary Table~\ref{tab:si-khp-channels} lists all 13 archive
channel identifiers and these two endpoint-geometry remappings.

The bimolecular benchmark contains 85 channels across ten systems from
the same archive. For each trial, we generated every connected fragment
from its mapped molecular graph, optimized each fragment independently
and rigidly aligned it to the corresponding public IRC terminal. This
preserves the archived intermolecular docking pose without copying the
internal reference conformations. Each inference recipe received the
same 20 independently constructed endpoint pairs per channel.
Fig.~\ref{fig:epoxidation-case-study}g reports the butadiene--ethene
subset, with channels ordered by their archived electronic activation
barriers and the three highest-barrier channels highlighted.

Independent DFT scoring and Sella refinement were performed on random
subsets of the separate Reaction-QM test split: 500 reactions for
Fig.~\ref{fig:nfe-scaling} and 200 reactions for
Fig.~\ref{fig:mlff-benchmark}e--h. We used these subsets because applying
the DFT/Sella protocol to the complete test split was beyond the
available computational budget.

\subsection{Zero- and few-shot transfer protocol}
\label{sec:met-transfer}
We evaluate transfer using the Transition1x Si/P/S and Ge/As/Se
element-substitution sets and the Transition1x-TMC set from
ref.~\citenum{darouich2026robusttsgen}, together with the
large-Transition1x evaluation set proposed by FragmentFlow
\citep{shprints2026fragmentflow}. We evaluate Si/P/S and
large-Transition1x zero-shot from the Reaction-QM post-trained
checkpoint. Ge/As/Se and the ten transition metals in TMC fall outside
the Reaction-QM element vocabulary, so we evaluate them after few-shot
adaptation.

For Ge/As/Se, we randomly draw 50 labelled training reactions for each
new element, giving 150 reactions in total. For TMC, we randomly draw 50
parent-reaction families and retain all ten metal variants from each
family, giving 500 reactions in total. In both benchmarks, we hold out an
equally sized, parent-disjoint tuning set. The official test split is
never accessed during training or checkpoint selection.

Each adaptation set is mixed 1:1 with Reaction-QM replay. Vocabulary
expansion is append-only: the original H/C/N/O/F/Si/P/S/Cl indices and
weights remain unchanged. We initialize Ge, As and Se from the
corresponding Si, P and S channels, respectively, and initialize each
transition-metal channel from the mean of the Si/P/S/Cl channels.

We supervise the expanded checkpoints for 500 optimizer steps at
learning rate $10^{-5}$ and batch size 16, without warm-up. We use the
epoch-15 checkpoint for Ge/As/Se and the epoch-4 checkpoint for TMC.

\subsection{Statistics}
\label{sec:met-stats}
All model comparisons use matched reaction identifiers and are paired by
reaction. In Fig.~\ref{fig:nfe-scaling}, each point is a
reaction-level mean. Shaded bands show 95\% percentile bootstrap
confidence intervals of the mean from 10,000 reaction-level resamples
(base seed 20260720).

At each neural function evaluation budget, we compare \method{} with
React-OT for each of the four metrics using a two-sided Wilcoxon
signed-rank test. Zero differences are discarded, and the implementation
selects the asymptotic or exact test automatically. Holm's procedure
controls the family-wise error rate across the twelve displayed
metric--budget comparisons.

The Holm-adjusted \(P\) values, listed for one, four and sixteen neural
function evaluations, respectively, are
\(1.88355\times10^{-26}\), \(2.68204\times10^{-22}\) and
\(3.36881\times10^{-22}\) for initial RMSD;
\(5.61999\times10^{-75}\), \(7.13162\times10^{-41}\) and
\(1.30009\times10^{-36}\) for the initial energy gap;
\(8.63604\times10^{-78}\), \(4.82015\times10^{-36}\) and
\(2.26097\times10^{-42}\) for initial force RMS; and
\(6.61421\times10^{-36}\), \(1.33406\times10^{-26}\) and
\(1.11156\times10^{-21}\) for capped Sella steps. Every comparison uses
the same \(n=500\) paired reactions.

In Fig.~\ref{fig:mlff-benchmark}e--h, each reward-trained recipe is
compared with the shared Pretrain reference on the same \(n=200\)
reactions. The plotted distributions are over reactions, and diamonds
mark arithmetic means. We use the mean within-reaction difference as the
test statistic. Two-sided Monte Carlo paired sign-flip tests use 200,000
randomizations and deterministic metric--arm seeds based on 568701. We
apply the finite-sample correction
\(P=(b+1)/(200{,}000+1)\), where \(b\) is the number of randomized
absolute mean differences at least as large as the observed value.

All twelve reward-versus-Pretrain comparisons reach the resolution floor
of this correction: \(b=0\) and
\(P=1/200{,}001=4.999975\times10^{-6}\). These comparisons cover the
initial energy gap, initial force RMS, absolute Sella energy correction
and capped Sella steps for each of g-xTB, MACE and UMA. Because all twelve
raw values are tied, Holm's step-down procedure gives the same adjusted
value for each comparison,
\(12/200{,}001=5.99997\times10^{-5}\). Thus, all twelve comparisons are
significant at \(P<10^{-4}\) (****). This floor reflects the resolution
of the randomization test, not the effect size.

Effect-size intervals use 20,000 paired reaction-level bootstrap
resamples (base seed 20260719). Sella failed to converge for \(6/200\)
Pretrain reactions and \(1/200\) reactions for each of g-xTB, MACE and
UMA. These runs enter the step-count mean at the 81-step cap. In both
figures, \(\mathrm{ns}\) denotes \(P\geq0.05\), and one to four
asterisks denote \(P<0.05\), \(P<0.01\), \(P<0.001\) and
\(P<0.0001\), respectively.

The ReactOracleBench panels are descriptive, and no hypothesis test is
applied to Fig.~\ref{fig:mlff-benchmark}a--d. Panel \textbf{a} reports
the median Kendall \(\tau_b\) over complete paths with the interquartile
range across paths. Results are separated for the 3,550 Sella
trajectories and 4,415 intrinsic-reaction-coordinate paths, and for
energy and force rankings. Panel \textbf{b} reports precision, recall
and \(F_1\) for index-one classification against the exact reference
Hessian.
Panels \textbf{c} and \textbf{d} report the mean \(\tau_b\) over
eight-candidate groups at each midpoint-prior noise scale. Their
normal-approximation 95\% confidence intervals are the mean plus or
minus 1.96 standard errors across groups. Throughput is summarized by the
median over repeat-level estimates.

\section*{Data availability}
The Transition1x dataset used in this study is publicly available on
Figshare (version 4) under DOI
\url{https://doi.org/10.6084/m9.figshare.19614657.v4}
\citep{schreiner2022transition1xdata}. The Reaction-QM dataset used in
this study, including the B3LYP-IRC data, is publicly available on
Zenodo (version 2) under DOI
\url{https://doi.org/10.5281/zenodo.18551029}
\citep{lee2026reactionqmdata}. The LargeT1x files used for the
size-extrapolation evaluation (\texttt{fragmented\_dataset.pkl} and
\texttt{filtered\_reactions\_data.csv}) are publicly available on Zenodo
under the version-specific DOI
\url{https://doi.org/10.5281/zenodo.18612166}. The exact
Transition1x-2p3p4p element-substitution and Transition1x-TMC files used
for the out-of-domain evaluations are available in the
\texttt{ReactOT.zip} and \texttt{XYZ\_files.zip} archives of the
RobustTSGen release on Zenodo under the version-specific DOI
\url{https://doi.org/10.5281/zenodo.20569500}.

\section*{Code availability}
The code required to reproduce this study, including data preparation, the
generator and its training and reward post-training, one-step and
multi-step sampling, the g-xTB and machine-learned force-field reward
oracles, and the GPU4PySCF/Sella evaluation and analysis pipelines,
together with the pre-trained and post-trained model checkpoints used for
the reported results, will be released publicly upon publication. The
release pins the software environment used here (PySCF~2.13.1,
pyscf-dispersion~1.5.0, GPU4PySCF~1.7.4, ASE~3.29.0, NumPy~1.26.4 and
Sella~2.4.2).

\section*{Acknowledgements}
The authors acknowledge the ByteDance Seed AI for Science teams for their
invaluable support.

\section*{Author contributions}
Y.L. conceived the method, implemented the generator, the MeanFlow
pre-training and the reward post-training, prepared the datasets, carried
out the training, sampling and DFT validation experiments, analysed the
results and drafted the manuscript.
Z.S. contributed to the early model codebase and to the preliminary
studies.
K.Y. and W.Y. contributed insight and discussion on model development and
on the physical validation of the generated transition states.
M.G. supervised Y.L., contributed insight and discussion, and revised the
manuscript.
H.Q.P. conceived the overall project, shaped its direction and numerical
examples, contributed to the writing, and oversaw the work.
All authors contributed to the discussion of this manuscript.

\section*{Competing interests}
The authors declare no competing interests.

\FloatBarrier
\clearpage
\bibliographystyle{unsrtnat}   %
\bibliography{references}

\clearpage
\begingroup
\setcounter{figure}{0}
\setcounter{table}{0}
\renewcommand{\figurename}{Extended Data Fig.}
\renewcommand{\tablename}{Extended Data Table}
\renewcommand{\thefigure}{\arabic{figure}}
\renewcommand{\thetable}{\arabic{table}}
\renewcommand{\theHfigure}{ED\arabic{figure}}
\renewcommand{\theHtable}{ED\arabic{table}}
\section*{Extended Data}

\begin{table}[p]
\centering
\small
\renewcommand{\arraystretch}{1.08}
\caption{\textbf{Transition-state generation accuracy on Transition1x.}
Standard $9{,}000/1{,}073$ split at the $\omega$B97x/6-31G(d) level.
\method\ is reported both after supervised pre-training and after
reward post-training; baseline results are as published. RMSD uses the $3N$
normalization of ref.~\citenum{duan2023oareactdiff} and is lower than
per-atom RMSD by a factor of $\sqrt{3}$. RGD1-xTB denotes pre-training on
the RGD1 reaction set \citep{zhao2023comprehensive}. Unless noted, each
row uses one sample per reaction. Bold and underlining mark the best and
second-best results; shading marks our method.
$^{\dagger}$Different train/test seed \citep{duan2025reactot}; TSDiff
conditions on the 2D reaction graph rather than 3D endpoints.
$^{\ddagger}$Ranks 40 samples per reaction.}
\label{tab:t1x-rmsd}
\begin{tabularx}{0.85\textwidth}{l>{\centering\arraybackslash}X>{\centering\arraybackslash}Xc}
\toprule
Method & Mean RMSD (\AA) $\downarrow$
& Median RMSD (\AA) $\downarrow$ & \NFE\ $\downarrow$ \\
\midrule
TSDiff$^{\dagger}$ \citep{kim2024tsdiff}
& 0.2526 & 0.2206 & ${\sim}5\!\times\!10^{3}$ \\
NeuralNEB$^{\dagger}$ \citep{schreiner2022neuralneb}
& 0.1358 & 0.0959 & --- \\
OA-ReactDiff \citep{duan2023oareactdiff}
& 0.1800 & 0.0752 & ${\sim}5\!\times\!10^{3}$ \\
\quad + recommender$^{\ddagger}$
& 0.1297 & 0.0582 & ${\sim}5\!\times\!10^{4}$ \\
Drift-React \citep{schlama2026driftreact}
& 0.168 & --- & 1 \\
React-OT \citep{duan2025reactot}
& 0.1029 & 0.0527 & 50 \\
\quad + RGD1-xTB pre-training
& 0.0981 & \underline{0.0441} & 50 \\
\midrule
\rowcolor{trefleRow}
\method\ (pre-trained) & \underline{0.0761} & 0.0538 & 1 \\
\rowcolor{trefleRow}
\method\ (reward post-trained) & \textbf{0.0752} & \textbf{0.0432} & 1 \\
\bottomrule
\end{tabularx}
\end{table}

\begin{table}[p]
\centering
\small
\renewcommand{\arraystretch}{1.08}
\caption{\textbf{Transition-state generation accuracy on the Reaction-QM
\texttt{irc\_all} test split ($n=19{,}883$).} All methods are implemented,
trained and evaluated in one consistent suite (Methods). The comparators are evaluated
after supervised pre-training and before post-training; \method\ is
reported both as the matched pre-trained checkpoint (GeoditE one-step
MeanFlow with the AlphaFlow curriculum) and as the full
reward-post-trained checkpoint (Extended Data Table~\ref{tab:ablation}
and Supplementary Note~6).
Lower values are better.
Bold and underlining mark the best and second-best results; shading marks
our method.}
\label{tab:ts-rmsd}
\begin{tabularx}{\textwidth}{ll>{\centering\arraybackslash}X>{\centering\arraybackslash}Xc}
\toprule
Method & Encoder & Mean RMSD (\AA) $\downarrow$
& Median RMSD (\AA) $\downarrow$ & \NFE\ $\downarrow$ \\
\midrule
React-OT \citep{duan2025reactot} & LEFTNet
& \underline{0.1214} & 0.1056 & 18 \\
\rowcolor{trefleRow}
\method\ (reward post-trained) & GeoditE & \textbf{0.1108} & \textbf{0.0923} & 1 \\
\rowcolor{trefleRow}
\method\ (pre-trained) & GeoditE & 0.1222 & \underline{0.1054} & 1 \\
MeanFlow \citep{geng2025meanflow} & LEFTNet & 0.1269 & 0.1107 & 1 \\
Latent flow & LEFTNet latent & 0.1255 & 0.1110 & 20 \\
RitS (midpoint prior) & Pairformer + DiT & 0.1355 & 0.1203 & 25 \\
TS-DFM (distance geometry) & Pairformer-style & 0.1383 & 0.1227 & 40 \\
TS-drift & LEFTNet & 0.1659 & 0.1519 & 1 \\
\midrule
Geodesic interpolant & --- & 0.322 & 0.309 & --- \\
IDPP & --- & 0.382 & --- & --- \\
Linear interpolant & --- & 0.391 & 0.382 & --- \\
\bottomrule
\end{tabularx}
\end{table}

\begin{table}[p]
\centering
\small
\renewcommand{\arraystretch}{1.15}
\caption{\textbf{Encoder, generator and post-training ablation on the
Reaction-QM \texttt{irc\_all} test split ($n=19{,}883$).} All rows are
trained and evaluated in the same consistent suite with one sample per reaction;
RMSD uses the $3N$ normalization of ref.~\citenum{duan2023oareactdiff}.
``$+$ AlphaFlow'' denotes the AlphaFlow curriculum of Supplementary
Note~1.
The first row corresponds to the React-OT recipe and the last to the
full \method{} model (GeoditE encoder, one-step MeanFlow with the
AlphaFlow curriculum, reward post-training); the penultimate row is the
matched pre-trained control for that final row. Shading marks \method.
Lower is better; the best mean and median are in bold. Supplementary
Note~6 discusses the comparisons. SMILES, simplified molecular-input
line-entry system.}
\label{tab:ablation}
\begin{tabularx}{\textwidth}{%
  l%
  >{\raggedright\arraybackslash}X%
  >{\centering\arraybackslash}p{2.15cm}%
  >{\centering\arraybackslash}p{2.30cm}%
  >{\centering\arraybackslash}p{2.30cm}}
\toprule
Encoder & Generator & Reward post-training
& Mean RMSD (\AA)~$\downarrow$ & Median RMSD (\AA)~$\downarrow$ \\
\midrule
LEFTNet & Multi-step conditional flow matching & --- & 0.1214 & 0.1056 \\
LEFTNet & One-step MeanFlow & --- & 0.1269 & 0.1107 \\
LEFTNet & One-step MeanFlow $+$ AlphaFlow & --- & 0.1233 & 0.1059 \\
GeoditE & Multi-step conditional flow matching & --- & 0.1202 & 0.1032 \\
GeoditE & One-step MeanFlow & --- & 0.1262 & 0.1094 \\
GeoditE & One-step MeanFlow $+$ AlphaFlow & --- & 0.1222 & 0.1054 \\
\rowcolor{trefleRow}
GeoditE & One-step MeanFlow $+$ AlphaFlow (\method) & \checkmark
& \textbf{0.1108} & \textbf{0.0923} \\
\bottomrule
\end{tabularx}
\end{table}

\begin{table}[p]
\centering
\small
\renewcommand{\arraystretch}{1.08}
\caption{\textbf{Single-GPU throughput on the Reaction-QM
\texttt{irc\_all} test split ($n=19{,}883$) at batch size 16.} Higher values are
better; shading marks our method. \NFE\ counts neural function evaluations
per sample. TS-DFM additionally performs L-BFGS distance-to-coordinate
reconstruction. For \method, the arrow shows training throughput before
and after the curriculum activates the MeanFlow term (Methods).}
\label{tab:throughput}
\begin{tabularx}{\textwidth}{ll>{\centering\arraybackslash}X>{\centering\arraybackslash}Xc}
\toprule
Method & Encoder & Training samples/s $\uparrow$
& Sampling samples/s $\uparrow$ & \NFE\ $\downarrow$ \\
\midrule
React-OT & LEFTNet & 114.05 & 13.90 & 18 \\
\rowcolor{trefleRow}
\method & GeoditE
& $196.45\!\rightarrow\!86.35$ & \textbf{421.55} & 1 \\
\rowcolor{trefleRow}
\method{} + Inductor compilation & GeoditE
& $294.67\!\rightarrow\!90.99$ & 1269.19 & 1 \\
MeanFlow & LEFTNet & 56.85 & 199.54 & 1 \\
TS-DFM & Pairformer-style & 90.06 & 1.71 & 40 \\
TS-drift & LEFTNet & 24.15 & 150.94 & 1 \\
RitS (SMILES midpoint) & Pairformer + DiT & 89.77 & 12.52 & 25 \\
Latent flow, stage 2 & LEFTNet latent & 98.67 & 12.37 & 20 \\
\bottomrule
\end{tabularx}
\end{table}

\begin{figure}[p]
\centering
\includegraphics[width=1.0\textwidth]{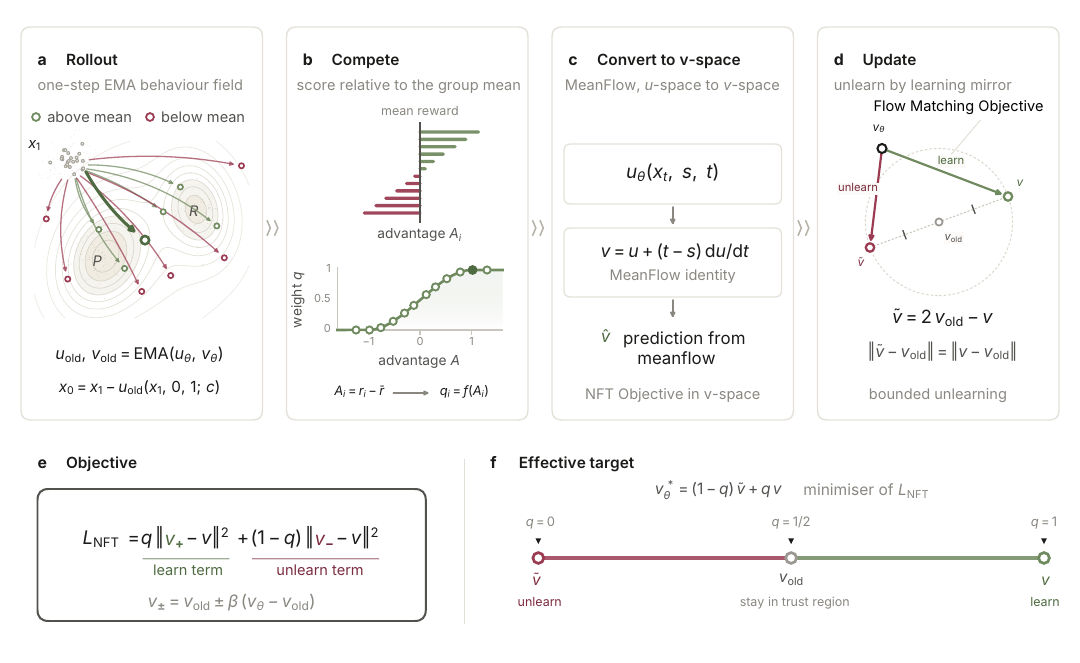}
\caption{\textbf{Group-relative reward post-training of the one-step
MeanFlow map.}
Schematic of the update in \hyperref[sec:met-post]{Methods}.
\textbf{a}, A lagged exponential-moving-average behaviour field maps
prior draws $\vx_1$ to one-step candidate transition-state geometries
$\vx_0$; oracle rewards distinguish candidates above (green) and below
(magenta) the within-reaction group mean.
\textbf{b}, Rewards produce group-relative advantages $\adv_i$ and
clipped optimality weights $q_i$; above- and below-mean candidates receive
$q_i>1/2$ and $q_i<1/2$, respectively.
\textbf{c}, The MeanFlow identity converts the average-velocity field
$\vu_{\policy}$ into the induced instantaneous-velocity field
$\vv_{\policy}$.
\textbf{d}, In instantaneous-velocity space, the sample target $\vv$ and
its mirror $\widetilde{\vv}=2\vv^{\mathrm{old}}-\vv$ lie at equal
distance on opposite sides of the lagged behaviour velocity
$\vv^{\mathrm{old}}$; learning towards the mirror therefore implements
bounded unlearning from a low-reward sample.
\textbf{e}, Symmetric branches
$\vv_{\pm}=\vv^{\mathrm{old}}\pm\beta
(\vv_{\policy}-\vv^{\mathrm{old}})$ define the weighted objective
$q\lVert\vv_{+}-\vv\rVert^2+(1-q)\lVert\vv_{-}-\vv\rVert^2$.
\textbf{f}, For the implemented branch scale $\beta=1$, its minimizer
$\vv_{\policy}^{\star}=(1-q)\widetilde{\vv}+q\vv$ interpolates from the
mirror target at $q=0$ (unlearn), through the behaviour field at $q=1/2$
(no displacement), to the sample target at $q=1$ (learn). This
instantaneous-velocity view is algebraically equivalent to the
average-velocity update in Methods.}
\label{fig:post-training-schematic}
\end{figure}

\begin{figure}[p]
\centering
\includegraphics[width=0.86\textwidth]{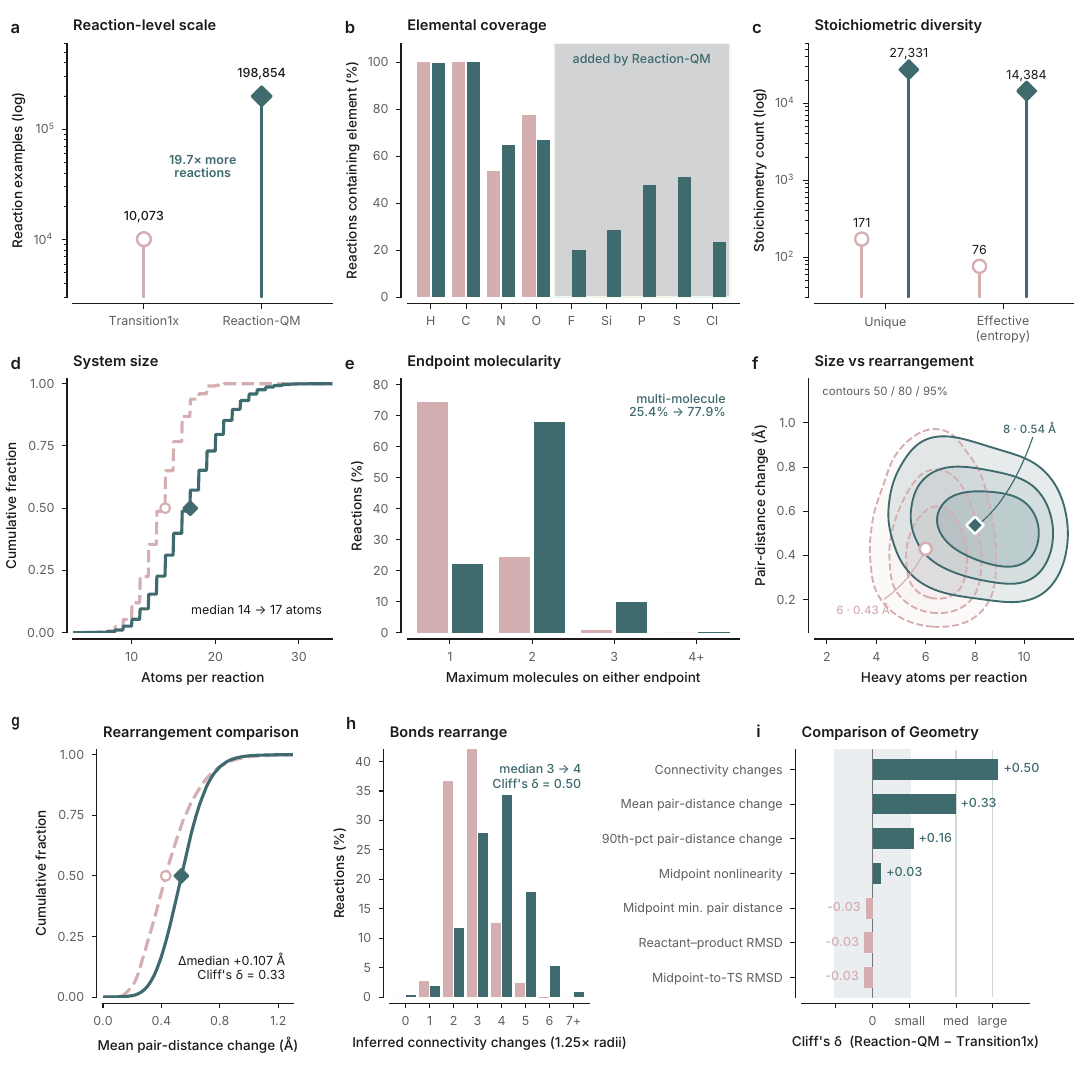}
\caption{\textbf{Reaction-QM is larger, chemically broader and more
strongly rearranging than Transition1x.}
Transition1x is shown in pink and Reaction-QM in dark green throughout.
\textbf{a}, Reaction count: $10{,}073$ and $198{,}854$, a $19.7\times$
increase.
\textbf{b}, Fraction of reactions containing each element; the shaded
band marks the five elements absent from Transition1x, and $93.0\%$ of
Reaction-QM reactions contain an element outside H, C, N and O.
\textbf{c}, Distinct stoichiometries ($171$ and $27{,}331$) and
entropy-effective stoichiometries ($76$ and $14{,}384$).
\textbf{d}, Cumulative distribution of atoms per reaction, with medians
of $14$ and $17$.
\textbf{e}, Maximum number of molecules on either endpoint;
multi-molecule endpoints occur in $25.4\%$ and $77.9\%$ of reactions.
\textbf{f}, Joint distribution of heavy-atom count and mean absolute
reactant--product pair-distance change. Contours enclose $50\%$, $80\%$
and $95\%$ of each dataset; markers give the medians, $6$ heavy atoms
and $0.43$~\AA\ against $8$ and $0.54$~\AA.
\textbf{g}, Cumulative distribution of the mean absolute pair-distance
change; the median rises by $0.107$~\AA\ (Cliff's $\delta=0.33$).
\textbf{h}, Geometry-inferred connectivity changes under a
$1.25\times$ covalent-radius cutoff, with medians of $3$ and $4$
(Cliff's $\delta=0.50$).
\textbf{i}, Cliff's $\delta$ between the datasets for seven structural
descriptors; shading marks the negligible range. Only connectivity
change ($0.50$) and the two rearrangement-magnitude descriptors ($0.33$
and $0.16$) leave it, whereas the endpoint- and midpoint-geometry
descriptors do not ($|\delta|\leq0.03$): the datasets differ in how much
bonding reorganizes, not in how far the reactant--product midpoint sits
from the transition state.}
\label{fig:dataset-comparison}
\end{figure}

\begin{figure}[p]
  \centering
  \IfFileExists{figures/model-parameter-counts.pdf}{%
    \includegraphics[width=\textwidth]{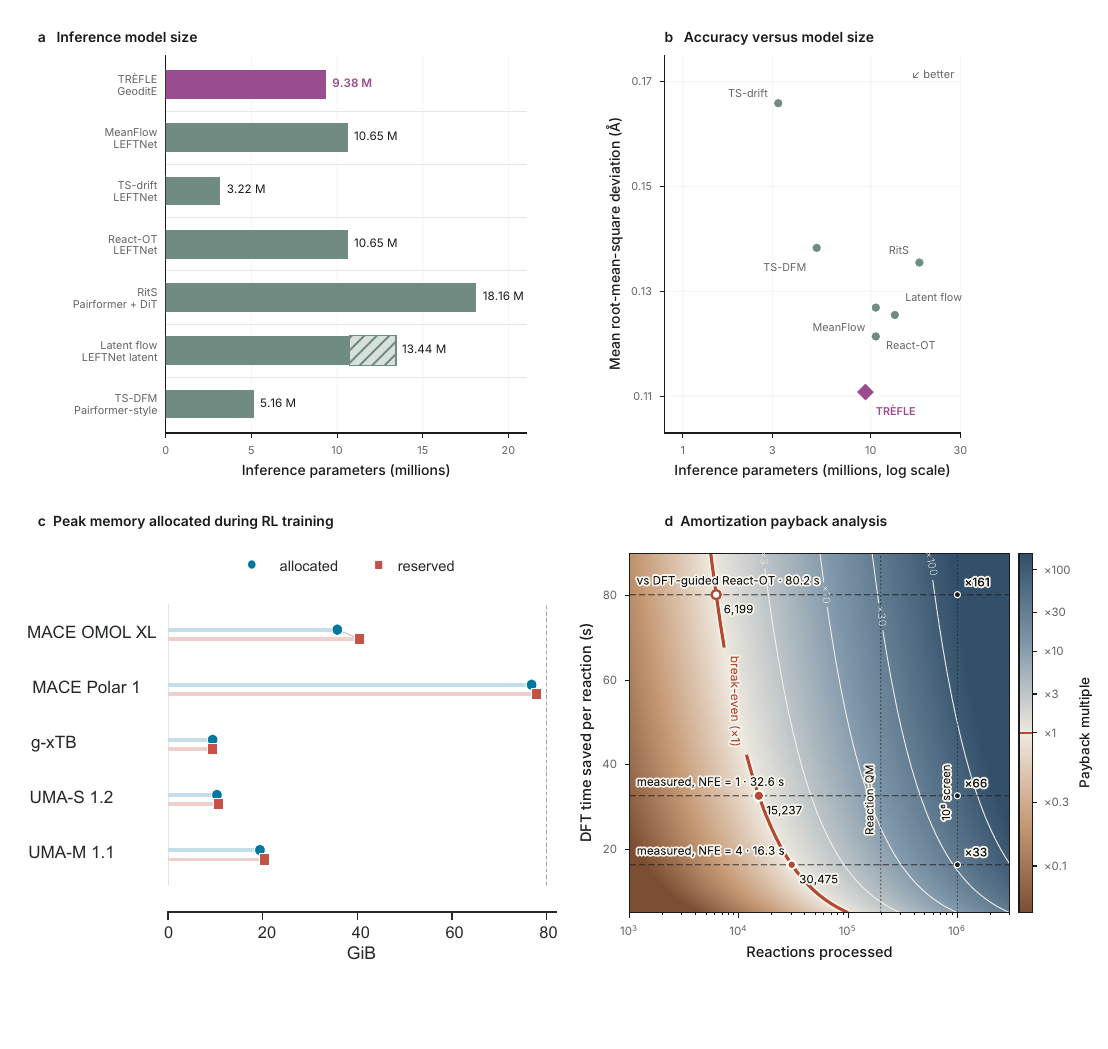}%
  }{%
    \fbox{\parbox[c][38mm][c]{0.94\textwidth}{\centering
      Upload \texttt{figures/model-parameter-counts.pdf}}}%
  }
  \caption{\textbf{Model size, training memory and the amortization payback
  of reward post-training.}
  \textbf{a}, Learned parameters used by each method at inference; the
  hatched segment is the frozen stage-one autoencoder of Latent flow.
  \textbf{b}, Reaction-QM mean root-mean-square deviation against
  inference-parameter count; lower and farther left are better. \method\
  uses $9.38$ million parameters, $12\%$ fewer than the matched LEFTNet
  React-OT and MeanFlow models and $48\%$ fewer than RitS, while attaining
  the lowest mean deviation, so accuracy here is not explained by model
  size.
  \textbf{c}, Peak allocated (circles) and reserved (squares) GPU memory for
  one complete reinforcement-learning update with 48 rollout members on a
  single \gpuname; the dashed line marks the 80-GiB device capacity.
  \textbf{d}, Illustrative break-even map for the post-training investment:
  $\eta_{\mathrm{pay}}=Ns/C_{\mathrm{post}}$ over a deployment of $N$
  reactions, where $s$ is the DFT-side wall-clock saving per reaction from
  the downstream Sella campaign and $C_{\mathrm{post}}=138.1$
  \gpuhours. The red curve is break-even; dashed lines mark the two
  measured operating points ($s=32.6$ and $16.3$\,s), which repay at
  $N\approx15{,}237$ and $30{,}475$ reactions and return
  $\eta_{\mathrm{pay}}\approx66$ and $33$ at the $10^{6}$-reaction screen.
  Methods gives the accounting.}
  \label{fig:model-parameter-counts}
\end{figure}

\begin{figure}[p]
\centering
\includegraphics[width=\textwidth]{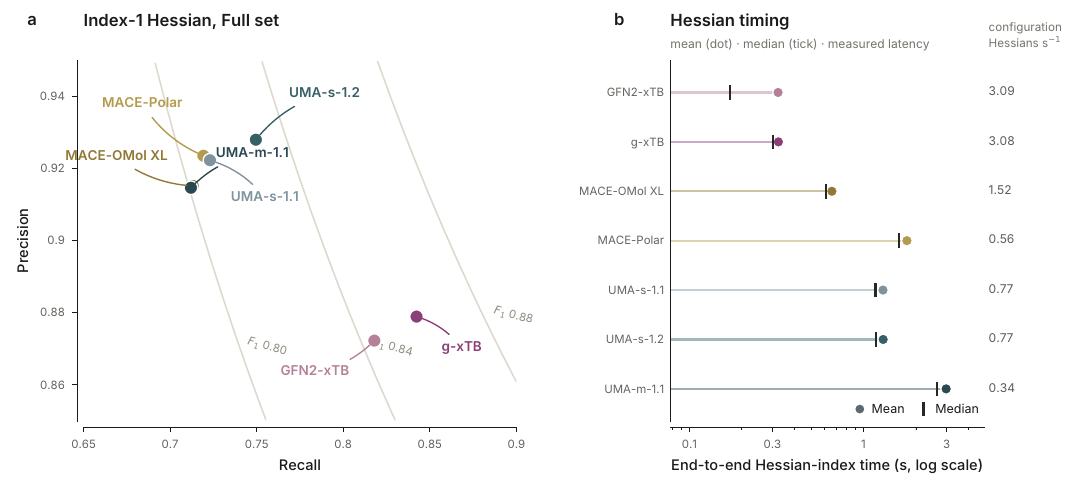}
\caption{\textbf{Natural-prevalence index-one Hessian classification and
complete Hessian-index timing across ReactOracleBench oracles.}
\textbf{a}, Precision and recall on the full $3{,}550$-structure saddle
cohort, comprising $2{,}906$ strict DFT index-one positives and $644$
negatives from 77 reaction families; grey curves are iso-$F_1$ contours.
A strict positive has exactly one projected negative mode below
$-50$\,cm$^{-1}$ and no additional soft-negative modes. The five
machine-learned force fields classify all $3{,}550$ structures;
GFN2-xTB and g-xTB classify $3{,}546$ and $3{,}547$, respectively, and
precision and recall are calculated over those successfully classified
structures.
\textbf{b}, Mean (filled circle) and median (vertical tick) wall time per
successful structure for the complete Hessian-index operation; values at
right are the reciprocal of the mean latency under the stated benchmark
configuration. Timing includes all force calls for $0.005$\,\AA\ central
differences, Hessian assembly, mass weighting, translation--rotation
projection, eigendecomposition, frequency conversion and strict
classification, but excludes model loading and result serialization.
All machine-learned force fields use force batch size 8 on \agpufull;
GFN2-xTB and g-xTB use 32 persistent one-thread CPU
processes pinned to CPU IDs 1--31 and 33.}
\label{fig:oracle-hessian-fullset-timing}
\end{figure}

\begin{figure}[p]
\centering
\includegraphics[width=\textwidth]{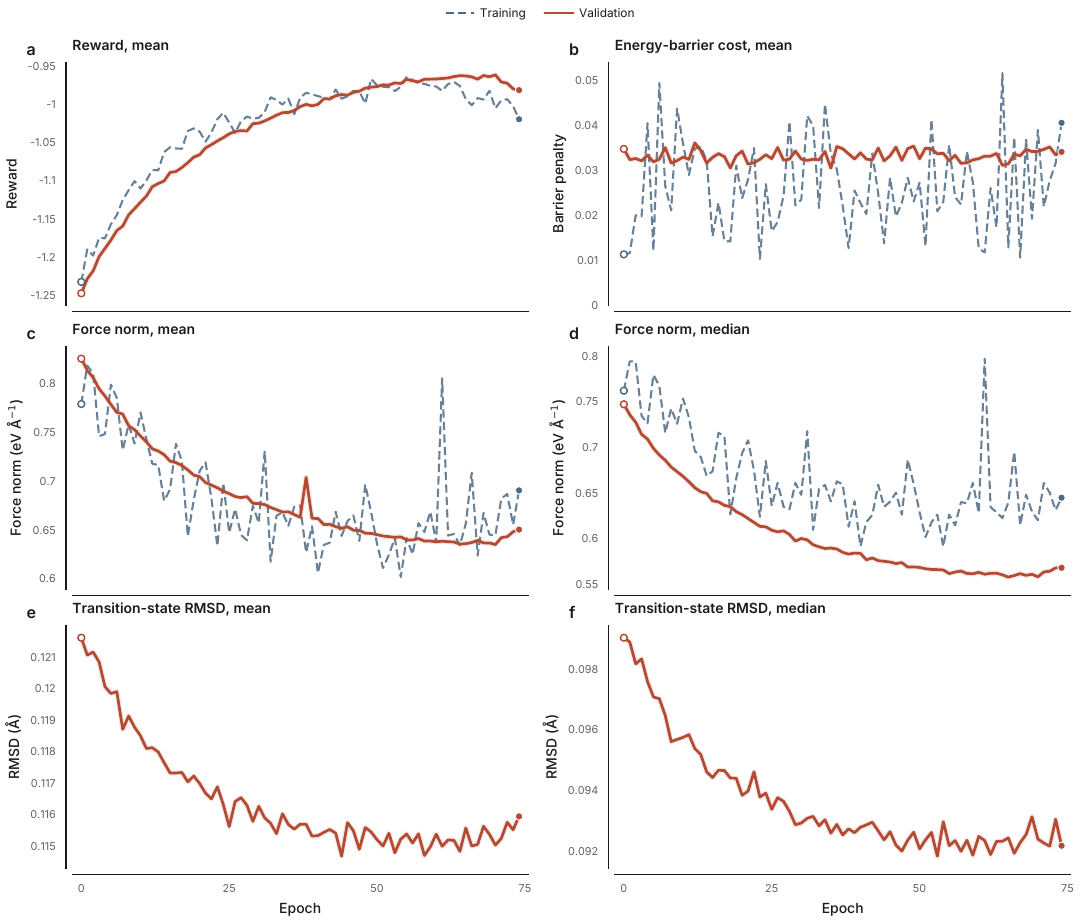}
\caption{\textbf{Training and validation dynamics under MACE-OMol XL
reward post-training.}
Blue dashed curves denote training and clay solid curves validation;
open and filled circles mark the first and final epochs, respectively.
\textbf{a}, Mean reward, for which higher values are better.
\textbf{b}, Mean energy-barrier penalty.
\textbf{c,d}, Mean and median surrogate-oracle force RMS ($F_{\mathrm{rms}}$ of equation~\eqref{eq:force_reward}), respectively.
\textbf{e,f}, Mean and median validation transition-state RMSD,
respectively. Lower values are better in panels \textbf{b--f}.
Lines show the exact, unsmoothed observations over post-training epochs
0--74 from one completed run. Training curves are epoch means over 128
logged batches, except that panel \textbf{a} uses the logged epoch reward;
validation was evaluated once per epoch on a fixed set. No between-seed
uncertainty is shown.}
\label{fig:mace-reward-dynamics}
\end{figure}

\begin{figure}[p]
\centering
\includegraphics[width=\textwidth]{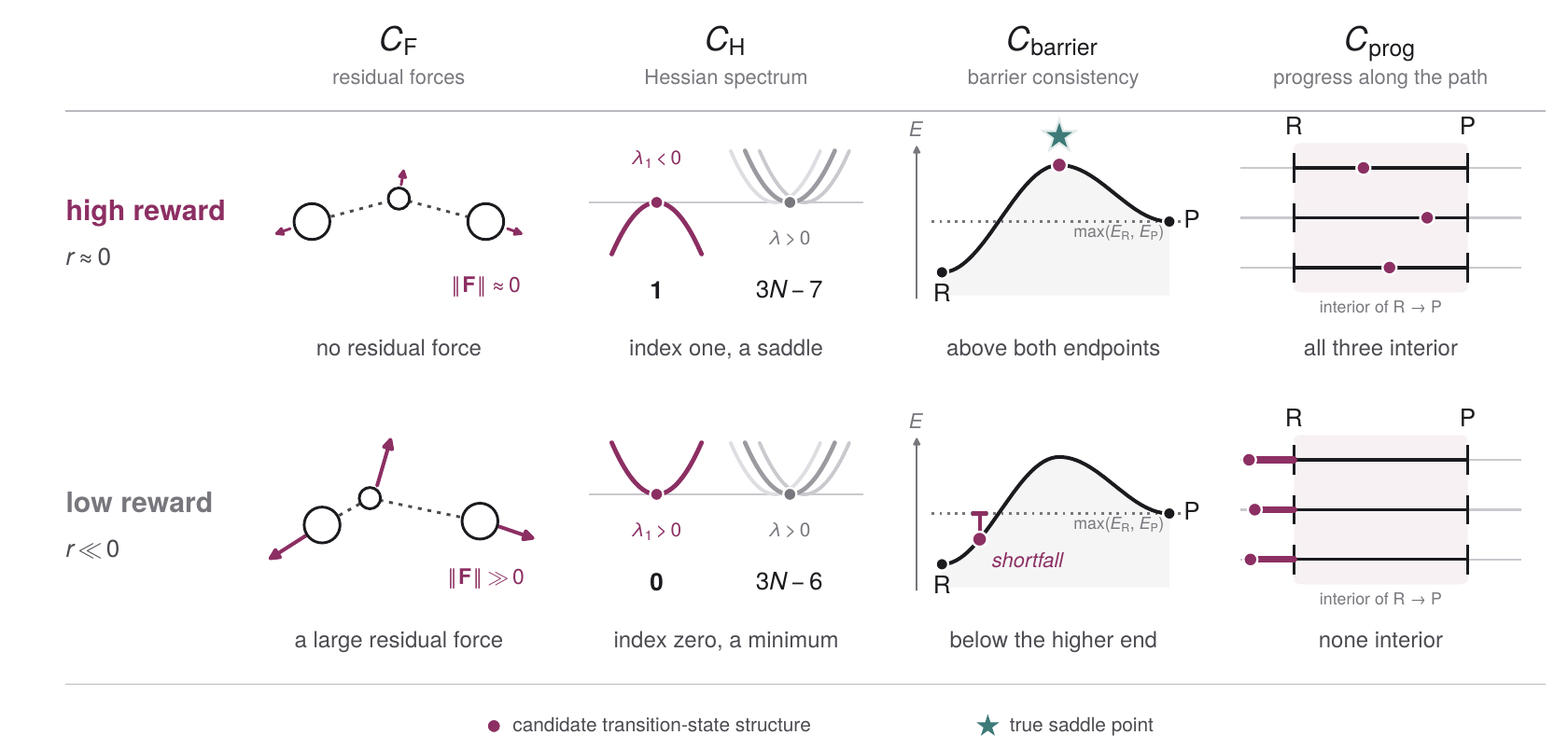}
\caption{\textbf{Physical interpretation of four components of the
composite heuristic reward.}
Top, configurations that receive high reward: a vanishing residual force
($C_F$), exactly one negative internal-Hessian eigenvalue ($C_H$), a
candidate energy above both endpoints ($C_{\mathrm{above}}$, labelled
$C_{\mathrm{barrier}}$ in the schematic), and active-pair progress values
inside the reactant-to-product interval ($C_{\mathrm{prog}}$).
Bottom, representative low-reward failures: a large residual force,
Hessian index zero, energy below the higher endpoint and progress outside
the path interior. Magenta circles denote candidate transition-state
structures and the teal star the true saddle point. The geometric
validity cost $C_{\mathrm{geom}}$, the fifth term in
equation~\eqref{eq:reward}, is not shown.}
\label{fig:reward-components}
\end{figure}

\endgroup

\clearpage
\begingroup
\setcounter{figure}{0}
\setcounter{table}{0}
\setcounter{section}{0}
\renewcommand{\figurename}{Supplementary Figure}
\renewcommand{\tablename}{Supplementary Table}
\renewcommand{\thefigure}{S\arabic{figure}}
\renewcommand{\thetable}{S\arabic{table}}
\renewcommand{\theHfigure}{S\arabic{figure}}
\renewcommand{\theHtable}{S\arabic{table}}

\section*{Supplementary Information}
\subsection*{Table of contents}
\begingroup
\small
\setlength{\tabcolsep}{4pt}
\renewcommand{\arraystretch}{1.08}
\begin{tabularx}{\textwidth}{@{}p{0.25\textwidth}Xr@{}}
\toprule
\textbf{Item} & \textbf{Title} & \textbf{Page} \\
\midrule
\multicolumn{3}{@{}l}{\textbf{Supplementary Notes}} \\
\hyperref[app:proofs]{Supplementary Note 1}
& Flow matching and the MeanFlow identity; AlphaFlow curriculum
& \pageref{app:proofs} \\
\hyperref[app:geodite]{Supplementary Note 2}
& GeoditE velocity field
& \pageref{app:geodite} \\
\hyperref[app:reward-implementation]{Supplementary Note 3}
& Composite heuristic reward implementation
& \pageref{app:reward-implementation} \\
\hyperref[app:nft]{Supplementary Note 4}
& Post-training details
& \pageref{app:nft} \\
\hyperref[sec:si-baselines]{Supplementary Note 5}
& Baseline implementations
& \pageref{sec:si-baselines} \\
\hyperref[sec:si-ablation]{Supplementary Note 6}
& Architecture and training-scheme ablation
& \pageref{sec:si-ablation} \\
\hyperref[app:rl]{Supplementary Note 7}
& Contextual-bandit formulation and policy improvement
& \pageref{app:rl} \\
\hyperref[sec:si-size-stats]{Supplementary Note 8}
& Size-extrapolation statistics for large-Transition1x
& \pageref{sec:si-size-stats} \\
\hyperref[sec:si-test200]{Supplementary Note 9}
& The 200-reaction Reaction-QM test cohort
& \pageref{sec:si-test200} \\
\hyperref[sec:si-khp-channels]{Supplementary Note 10}
& Ketohydroperoxide channel identifiers
& \pageref{sec:si-khp-channels} \\
\hyperref[sec:si-oxygen-transfer]{Supplementary Note 11}
& Oxygen-transfer epoxidation case study
& \pageref{sec:si-oxygen-transfer} \\
\hyperref[sec:si-gxtb-guidance-budget]{Supplementary Note 12}
& Matched inference-time g-xTB saddle guidance
& \pageref{sec:si-gxtb-guidance-budget} \\
\addlinespace[4pt]
\multicolumn{3}{@{}l}{\textbf{Supplementary Tables and Figures}} \\
\hyperref[tab:hparams]{Supplementary Table S1}
& Model, training and reward configuration
& \pageref{tab:hparams} \\
\hyperref[tab:si-sizebins]{Supplementary Table S2}
& Size-extrapolation statistics by heavy-atom bin
& \pageref{tab:si-sizebins} \\
\hyperref[tab:si-test200]{Supplementary Table S3}
& The 200-reaction Reaction-QM test cohort
& \pageref{tab:si-test200} \\
\hyperref[tab:si-khp-channels]{Supplementary Table S4}
& The 13 ketohydroperoxide decomposition channels
& \pageref{tab:si-khp-channels} \\
\hyperref[fig:si-oxygen-transfer]{Supplementary Figure S1}
& Oxygen-transfer epoxidation case study
& \pageref{fig:si-oxygen-transfer} \\
\hyperref[fig:si-gxtb-guidance-budget]{Supplementary Figure S2}
& Matched inference-time g-xTB saddle guidance
& \pageref{fig:si-gxtb-guidance-budget} \\
\bottomrule
\end{tabularx}
\endgroup
\clearpage

\section*{Supplementary Note 1: Flow matching and the MeanFlow identity;
AlphaFlow curriculum}
\label{app:proofs}
\paragraph{Derivation of the MeanFlow identity \eqref{eq:mfidentity}.}
By definition of the average velocity \eqref{eq:avgvel},
$(t-s)\,\vu(\vx_t,s,t)=\int_s^t\vv(\vx_\tau,\tau)\,d\tau$. Differentiate both
sides with respect to $t$ at fixed $s$. By the fundamental theorem of calculus
the right-hand side gives $\vv(\vx_t,t)$, while the left-hand side gives
$\vu(\vx_t,s,t)+(t-s)\tfrac{d}{dt}\vu(\vx_t,s,t)$, where
$\tfrac{d}{dt}\vu=\partial_t\vu+
(\dot{\vx}_t\!\cdot\!\nabla_{\vx})\vu
=\partial_t\vu+(\vv(\vx_t,t)\!\cdot\!\nabla_{\vx})\vu$ is the total derivative along the
flow $\dot{\vx}_t=\vv(\vx_t,t)$. Equating the two yields
$\vu=\vv-(t-s)(\partial_t\vu+(\vv\!\cdot\!\nabla_{\vx})\vu)$, which is
\eqref{eq:mfidentity}.

\paragraph{Improved MeanFlow objective.}
For a sampled training pair, the conditional path velocity is
$\vv_t^{\mathrm{cond}}=\vx_1-\vx_0$. At a fixed interpolated state,
however, the instantaneous velocity in the MeanFlow identity is the
marginal quantity
\begin{equation}
\vv(\vx_t,t;\cond)
=
\EE\!\left[
\vv_t^{\mathrm{cond}}
\,\middle|\,
\vx_t,t,\cond
\right].
\label{eq:si-marginal-velocity}
\end{equation}
Original MeanFlow substitutes the sampled conditional velocity for this
marginal velocity inside the total derivative. The same conditional draw
therefore acts both as the regression label and as the spatial direction
of the Jacobian--vector product.

We instead use the parameter-free boundary construction of improved
MeanFlow \citep{geng2025improvedmeanflow}. The boundary identity
$\vu(\vx_t,t,t;\cond)=\vv(\vx_t,t;\cond)$ provides the detached marginal
estimate
\begin{equation}
\widetilde{\vv}_{\policy}(\vx_t,t;\cond)
=
\sg\!\left[
\vu_{\policy}(\vx_t,t,t;\cond)
\right].
\label{eq:si-imf-velocity}
\end{equation}
The resulting total derivative of the finite-interval field is
\begin{equation}
\widetilde{\mathcal D}_t
\vu_{\policy}(\vx_t,s,t;\cond)
=
\partial_t\vu_{\policy}
+
\left(
\widetilde{\vv}_{\policy}\!\cdot\!\nabla_{\vx}
\right)\vu_{\policy}.
\label{eq:si-imf-derivative}
\end{equation}
This derivative is evaluated by one Jacobian--vector product at
$(\vx_t,s,t)$ with tangent
$(\widetilde{\vv}_{\policy},0,1)$. The improved MeanFlow target is
therefore
\begin{equation}
\bm Y_{\mathrm{iMF}}
=
\vv_t^{\mathrm{cond}}
-
(t-s)\,
\widetilde{\mathcal D}_t
\vu_{\policy}(\vx_t,s,t;\cond),
\qquad
\mathcal L_{\mathrm{MF}}
=
\EE\!\left[
\left\|
\vu_{\policy}(\vx_t,s,t;\cond)
-
\sg[\bm Y_{\mathrm{iMF}}]
\right\|_2^2
\right],
\label{eq:si-imf-objective}
\end{equation}
which is the objective in equation~\eqref{eq:mfloss}. Thus, the
sample-conditional velocity remains the supervised regression label,
whereas the detached diagonal prediction supplies the trajectory direction.
When $s=t$, the derivative term vanishes and the objective reduces to
ordinary flow matching without a Jacobian--vector product.

All reported models use this improved construction wherever the MeanFlow
loss is applied. We use the boundary prediction in
equation~\eqref{eq:si-imf-velocity} directly and do not add an auxiliary
instantaneous-velocity head. The modification is confined to training:
sampling still evaluates only
$\vu_{\policy}(\vx_1,0,1;\cond)$ in the one-step map
\eqref{eq:onestep}.

\subsection*{AlphaFlow curriculum for MeanFlow pre-training}
\label{app:alphaflow}
AlphaFlow \citep{zhang2025alphaflow} changes the training objective but not
the velocity-field architecture or the sampling rule.  Its motivation is
that the MeanFlow objective couples trajectory flow matching to trajectory
consistency, whose gradients can conflict when both are learned from scratch.
The curriculum first learns the lower-variance trajectory velocity and then
moves smoothly to the finite-interval MeanFlow target, improving convergence
without adding a network module.  Consequently, our GeoditE field and the
one-step deployment map in Eq.~\eqref{eq:onestep} remain unchanged.

For a sampled pair $0\leq s\leq t\leq1$, write
$\delta=t-s$ and let $\alpha_k\in[0,1]$ be the curriculum value at optimizer
step $k$.  AlphaFlow places an intermediate time and state at
\begin{equation}
m_k=\alpha_k s+(1-\alpha_k)t
=t-\alpha_k\delta,
\qquad
\vx_{m_k}
=
\vx_t-(t-m_k)\vv_t^{\mathrm{cond}}.
\label{eq:alphaflow_intermediate}
\end{equation}
Thus $t-m_k=\alpha_k\delta$ is supervised by the known trajectory velocity,
whereas the remaining interval $m_k-s=(1-\alpha_k)\delta$ is represented by
a detached prediction of the same average-velocity field.  With
$\widetilde{\vu}_{\policy}(\vx_{m_k},s,m_k;\cond)
=\sg[\vu_{\policy}(\vx_{m_k},s,m_k;\cond)]$, the regression target is
\begin{equation}
\bm{Y}_{\alpha_k}
=
\begin{cases}
\vv_t^{\mathrm{cond}},
& \delta=0,\\[3pt]
\alpha_k\vv_t^{\mathrm{cond}}
+(1-\alpha_k)
\widetilde{\vu}_{\policy}(\vx_{m_k},s,m_k;\cond),
& \delta>0,\ \alpha_k>0,\\[3pt]
\vv_t^{\mathrm{cond}}
-\delta\!\left[
\partial_t\vu_{\policy}
+
(\vv_t^{\mathrm{cond}}\!\cdot\nabla_{\vx})\vu_{\policy}
\right](\vx_t,s,t;\cond),
& \delta>0,\ \alpha_k=0.
\end{cases}
\label{eq:alphaflow_target}
\end{equation}
The middle branch is equivalently the interval-length-weighted target
\[
\bm{Y}_{\alpha_k}
=
\frac{(t-m_k)\vv_t^{\mathrm{cond}}
+(m_k-s)\widetilde{\vu}_{\policy}(\vx_{m_k},s,m_k;\cond)}
{t-s}.
\]
At $\alpha_k=1$, $m_k=s$ and Eq.~\eqref{eq:alphaflow_target} reduces exactly
to trajectory flow matching.  At $\alpha_k=0$, we evaluate the continuous
MeanFlow endpoint with the Jacobian--vector product in
Eq.~\eqref{eq:mfloss}.  Intermediate values therefore interpolate the
supervision assigned to the known trajectory and the learned
finite-interval consistency target.

Let $\ell_i(\policy;\alpha_k)$ denote the atom-normalized squared coordinate
error between
$\vu_{\policy}(\vx_t,s,t;\cond)$ and
$\sg[\bm{Y}_{\alpha_k}]$ for sample $i$.  The implemented objective is
\begin{equation}
\mathcal{L}_{\alpha}(\policy;k)
=
\EE_i\!\left[
\frac{q_i(\alpha_k)}
{\sg[\ell_i(\policy;\alpha_k)]+\epsilon_{\alpha}}\,
\ell_i(\policy;\alpha_k)
\right],
\qquad
q_i(\alpha_k)=
\begin{cases}
\alpha_k, & s_i<t_i,\ 0<\alpha_k<1,\\
1, & s_i=t_i\ \text{or}\ \alpha_k\in\{0,1\},
\end{cases}
\label{eq:alphaflow_loss}
\end{equation}
with $\epsilon_{\alpha}=10^{-3}$.  The denominator is detached and acts only
as per-sample gradient normalization.  We draw half of each batch on the
diagonal $s=t$ to retain direct flow-matching supervision; off-diagonal
time pairs follow the logit-normal distribution used for MeanFlow.  For
non-diagonal samples with $\alpha_k>0$, coordinate targets are clipped
elementwise to $[-4,4]$ before evaluating the loss.

We anneal $\alpha_k$ with a clamped sigmoid.  For schedule endpoints
$k_{\mathrm{s}}<k_{\mathrm{e}}$, midpoint
$k_{\mathrm{m}}=(k_{\mathrm{s}}+k_{\mathrm{e}})/2$, temperature $\gamma$,
and clamp $\eta$, define
\begin{align}
\widetilde{\alpha}_k
&=
1-\operatorname{sigmoid}\!\left(
\gamma\,\frac{k-k_{\mathrm{m}}}{k_{\mathrm{e}}-k_{\mathrm{s}}}
\right),\\
\alpha_k
&=
\begin{cases}
1, & k<k_{\mathrm{s}}\ \text{or}\
\widetilde{\alpha}_k>1-\eta,\\
0, & k>k_{\mathrm{e}}\ \text{or}\
\widetilde{\alpha}_k<\eta,\\
\widetilde{\alpha}_k, & \text{otherwise}.
\end{cases}
\label{eq:alphaflow_schedule}
\end{align}
The reported pre-training uses
$(k_{\mathrm{s}},k_{\mathrm{e}},\gamma,\eta)
=(0,50{,}000,25,0.005)$.  This yields three stages: exact trajectory flow
matching early in training, a smooth AlphaFlow transition, and exact
MeanFlow supervision after the schedule reaches zero.

\section*{Supplementary Note 2: GeoditE velocity field}
\label{app:geodite}
GeoditE parameterizes the conditional velocity field used by both flow
matching and MeanFlow. It adapts the tensor-product-free message passing of
GotenNet and GeoditE \citep{aykent2025gotennet,reschutzegger2026geodite} to
direct vector prediction: the original interatomic-potential energy readout is
replaced by an equivariant velocity readout. The flow-matching model is
\(\vv_{\policy}(\vx_t,t;\cond)\); the MeanFlow model uses the same architecture
with the interval endpoint as an additional scalar input,
\(\vu_{\policy}(\vx_t,s,t;\cond)\).

\paragraph{Reaction graph and input states.}
For one reaction, let
\(\mathcal V=\mathcal V_{\mathrm R}\cup\mathcal V_{\mathrm T}\cup
\mathcal V_{\mathrm P}\) contain the aligned reactant, current transition-state
candidate, and product atoms. We use the directed complete graph on
\(\mathcal V\). For atom \(i\), \(f(i)\in\{\mathrm R,\mathrm T,\mathrm P\}\)
denotes its structure, \(\vx_i\in\RR^3\) its coordinate, and \(\bm{a}_i\) its
invariant atom attributes. Separate fragment encoders are followed by the
shared GeoditE projection:
\begin{equation}
\bm{h}_i^{\mathrm{in}}
=
\left[
\operatorname{Enc}_{f(i)}(\bm{a}_i),\;
t,\;
\Delta,\;
\cond_{\mathrm{inv}}
\right],
\qquad
\Delta=t-s.
\label{eq:geodite-input}
\end{equation}
Here \(\cond_{\mathrm{inv}}\) denotes any global invariant conditioning.
Charge and spin are constant in our experiments and are therefore omitted.
The instantaneous flow field drops \(\Delta\), whereas MeanFlow supplies both
\(t\) and \(\Delta\).

\paragraph{Local geometric basis and complete-graph conditioning.}
Only same-structure pairs inside the cutoff carry geometry:
\begin{equation}
\chi_{ij}
=
\mathbf 1\!\left[f(i)=f(j)\right]
\mathbf 1\!\left[d_{ij}<d_{\mathrm c}\right],
\qquad
d_{ij}=\|\vx_i-\vx_j\|_2,
\qquad
\widehat{\bm{r}}_{ij}=\frac{\vx_i-\vx_j}{d_{ij}+\varepsilon}.
\label{eq:geodite-local-mask}
\end{equation}
For \(k=1,\ldots,K\), the effective radial embedding is
\begin{equation}
g_{ij,k}
=
\chi_{ij}\,
w_{Z_iZ_j}(d_{ij})\,
c_6\!\left(\frac{d_{ij}}{d_{\mathrm c}}\right)
\sqrt{\frac{2}{d_{\mathrm c}}}\,
\frac{\sin(k\pi d_{ij}/d_{\mathrm c})}{d_{ij}+\varepsilon},
\label{eq:geodite-radial}
\end{equation}
with the smooth sixth-order envelope
\begin{equation}
c_6(z)
=
\begin{cases}
1-28z^6+48z^7-21z^8, & 0\leq z<1,\\
0, & z\geq1.
\end{cases}
\end{equation}
The trainable covalent-radius factor uses
\(s_{ij}=d_{ij}/(\rho_{Z_i}+\rho_{Z_j})\) and
\begin{equation}
w_{Z_iZ_j}(d_{ij})
=
\frac{a\,s_{ij}^{\,q}}
{1+s_{ij}^{\,q-p}+a\,s_{ij}^{\,q}},
\qquad
a
=
-\frac{2(p+q-2pq)}{p^2+p+q^2+q},
\qquad p,q>1.
\label{eq:geodite-covalent}
\end{equation}
Direction is represented by real spherical harmonics
\(\bm{Y}_{ij}^{(\ell)}=Y_{\ell}(\widehat{\bm{r}}_{ij})\),
\(1\leq\ell\leq L\). In the reported \(L=1\) model,
\(\bm{Y}_{ij}^{(1)}=\widehat{\bm{r}}_{ij}\). Cross-structure and out-of-cutoff
edges have \(g_{ij,k}=0\) and \(\bm{Y}_{ij}^{(\ell)}=\vzero\), but remain in a
separate geometry-free scalar route. Thus reactant and product information
reaches transition-state atoms without treating cross-structure coordinate
differences as physical distances.

\paragraph{State initialization.}
Let \(\bar{\bm{h}}_i=W_0\bm{h}_i^{\mathrm{in}}\). Local radial messages initialize the
invariant node and edge states,
\begin{align}
\bm{m}_{ij}^{0}
&=
W_n\bar{\bm{h}}_j\odot W_r\vg_{ij},
&
\delta_i^{0}
&=
1+\sum_j
\tanh\!\left[
\operatorname{SiLU}\!\left(D_0(\vg_{ij})\right)^2
\right],
\\
\bm{h}_i^{0}
&=
\phi_0\!\left(
\left[
\bar{\bm{h}}_i,\;
\frac{1}{\delta_i^{0}}\sum_j\bm{m}_{ij}^{0}
\right]\right),
&
\bm{e}_{ij}^{0}
&=
(\bm{h}_i^{0}+\bm{h}_j^{0})\odot W_e\vg_{ij},
\qquad
\bm{X}_i^{0}=\vzero.
\label{eq:geodite-init}
\end{align}
Here \(\odot\) is channel-wise multiplication,
\(\bm{h}_i^\ell\in\RR^d\) is invariant, and
\(\bm{X}_i^{\ell,(\ell')}\in\RR^{(2\ell'+1)\times d}\) is steerable.

\paragraph{One GeoditE interaction layer.}
Each layer first applies a geometry-free complete-graph update. With
\(\widetilde{\bm{h}}_i=\mathcal N_s(\bm{h}_i^\ell)\),
\begin{align}
\bm{m}_{ij}^{\mathrm{all}}
&=
\phi_m^\ell([\widetilde{\bm{h}}_i,\widetilde{\bm{h}}_j])\,
\phi_a^\ell\!\left(
\phi_m^\ell([\widetilde{\bm{h}}_i,\widetilde{\bm{h}}_j])
\right),
\\
\bar{\bm{h}}_i^\ell
&=
\widetilde{\bm{h}}_i+
\phi_c^\ell\!\left(
\left[
\widetilde{\bm{h}}_i,\;
\frac{1}{|\mathcal V|-1}\sum_{j\neq i}\bm{m}_{ij}^{\mathrm{all}}
\right]\right).
\label{eq:geodite-complete}
\end{align}
The scalar gate \(\phi_a^\ell\) has one output, so it rescales the entire
message. The invariant and steerable states are then normalized together.
Suppressing learned per-degree scales, the shared statistic is
\begin{equation}
\sigma_i^2
=
\frac{1}{d}\sum_{c=1}^{d}
\left[
(\bar h_{ic}^{\ell})^2
+
\sum_{\ell'=1}^{L}
\frac{1}{2\ell'+1}
\sum_{m=-\ell'}^{\ell'}
(X_{i m c}^{\ell,(\ell')})^2
\right],
\qquad
(\widehat{\bm{h}}_i^\ell,\widehat{\bm{X}}_i^\ell)
=
\frac{(\bar{\bm{h}}_i^\ell,\bm{X}_i^\ell)}
{\sqrt{\sigma_i^2+\varepsilon}}.
\label{eq:geodite-coupled-norm}
\end{equation}

For attention head \(a\), the edge-conditioned score and additive SiLU weight
are
\begin{equation}
s_{ij,a}^{\ell}
=
\sum_c
(W_q\widehat{\bm{h}}_i^\ell)_{ac}
(W_k\widehat{\bm{h}}_j^\ell)_{ac}
(\phi_e(\bm{e}_{ij}^{\ell}))_{ac},
\qquad
\alpha_{ij,a}^{\ell}
=
\frac{\operatorname{SiLU}(s_{ij,a}^{\ell})}{\sqrt d}.
\label{eq:geodite-attention}
\end{equation}
The attention branch
\(\boldsymbol\alpha_{ij}^{\ell}\odot W_v\widehat{\bm{h}}_j^\ell\)
and spatial branch
\(W_f\bm{e}_{ij}^{\ell}\odot W_s\widehat{\bm{h}}_j^\ell\)
are split into a scalar message \(\bm{m}_{ij}^{h}\), directional coefficients
\(\bm{b}_{ij}^{(\ell')}\), and transported-state coefficients
\(\bm{c}_{ij}^{(\ell')}\). The resulting update is
\begin{align}
\bm{m}_{ij}^{X,(\ell')}
&=
\chi_{ij}\left[
\bm{Y}_{ij}^{(\ell')}\otimes\bm{b}_{ij}^{(\ell')}
+
\widehat{\bm{X}}_j^{\ell,(\ell')}\odot
\bm{c}_{ij}^{(\ell')}
\right],
\\
\delta_i^\ell
&=
1+\sum_j
\tanh\!\left[
\operatorname{SiLU}\!\left(D_\ell(\vg_{ij}/K)\right)^2
\right],
\\
\bm{h}_i^{+}
&=
\widehat{\bm{h}}_i^\ell
+
\frac{1}{\delta_i^\ell}\sum_j\bm{m}_{ij}^{h},
\qquad
\bm{X}_i^{+,(\ell')}
=
\widehat{\bm{X}}_i^{\ell,(\ell')}
+
\frac{1}{\delta_i^\ell}\sum_j\bm{m}_{ij}^{X,(\ell')}.
\label{eq:geodite-message}
\end{align}
This additive attention avoids cutoff-dependent softmax renormalization.
The equivariant feed-forward block then couples the two streams without a
Clebsch--Gordan tensor product:
\begin{align}
\bm{P}_i &= W_X\bm{X}_i^{+},
&
[\bm{b}_i,\bm{c}_i]
&=
\phi_{\mathrm{ff}}\!\left(
\left[
\bm{h}_i^{+},\;
\sum_m(\bm{P}_{im})^2
\right]\right),
\\
\bm{h}_i^{\ell+1}
&=
\bm{h}_i^{+}+\bm{b}_i,
&
\bm{X}_i^{\ell+1}
&=
\bm{X}_i^{+}+\bm{c}_i\odot\bm{P}_i.
\label{eq:geodite-eqff}
\end{align}

Between interaction layers, the local edge state is refined by projected
inner products. For a degree-\(\ell'\) direction \(\bm{Y}\), define rejection
from that direction as
\begin{equation}
\mathcal R(\bm{A},\bm{Y})
=
\bm{A}-\bm{Y}(\bm{Y}^\top\bm{A}).
\end{equation}
With \(\bm{Q}_i^{(\ell')}=W_Q^{(\ell')}\bm{X}_i^{\ell+1,(\ell')}\) and
\(\bm{K}_j^{(\ell')}=W_K^{(\ell')}\bm{X}_j^{\ell+1,(\ell')}\),
\begin{align}
\vz_{ij}
&=
\sum_{\ell'=1}^{L}
\left\langle
\mathcal R(\bm{Q}_i^{(\ell')},\bm{Y}_{ij}^{(\ell')}),
\mathcal R(\bm{K}_j^{(\ell')},-\bm{Y}_{ij}^{(\ell')})
\right\rangle_{\ell'},
\\
\bm{e}_{ij}^{\ell+1}
&=
\bm{e}_{ij}^{\ell}
+
\chi_{ij}\,
\phi_{\mathrm{edge}}(\bm{e}_{ij}^{\ell})
\odot
W_z\mathcal N_e(\vz_{ij}).
\label{eq:geodite-edge-update}
\end{align}
The inner product contracts only the spherical-harmonic index and therefore
produces an invariant channel vector.

\paragraph{Direct velocity readout.}
After \(D\) layers, two gated equivariant blocks map the
\(\ell=1\) state to a Cartesian vector,
\begin{equation}
(\bm{s}_i,\bm{V}_i)
=
\operatorname{GEB}_1(\bm{h}_i^D,\bm{X}_i^{D,(1)}),
\qquad
\Delta\widetilde{\vv}_i
=
\operatorname{GEB}_2(\bm{s}_i,\bm{V}_i),
\qquad
\widetilde{\vv}_i
=
\vx_i+\Delta\widetilde{\vv}_i.
\end{equation}
The last equality is the backbone's residual coordinate parameterization; the
flow wrapper interprets \(\widetilde{\vv}_i\) as the velocity prediction. It
retains the transition-state block and projects it to the zero-centroid
subspace:
\begin{equation}
\vv_{\policy,i}
=
\widetilde{\vv}_i
-
\frac{1}{|\mathcal V_{\mathrm T}|}
\sum_{j\in\mathcal V_{\mathrm T}}\widetilde{\vv}_j,
\qquad i\in\mathcal V_{\mathrm T}.
\label{eq:geodite-velocity}
\end{equation}
For MeanFlow, this is \(\vu_{\policy}(\vx_t,s,t;\cond)\) and the finite-time
map is
\begin{equation}
\Phi_{t\rightarrow s}(\vx_t)
=
\vx_t-(t-s)\vu_{\policy}(\vx_t,s,t;\cond).
\end{equation}
Let \(\mathcal X=\{\vx_i:i\in\mathcal V\}\) collect all three
structures. For any orthogonal matrix \(Q\) and translation \(\bm{b}\), the
construction satisfies
\begin{equation}
\vv_{\policy}(Q\mathcal X+\bm{b},t;\cond_{\mathrm{inv}})
=
Q\,\vv_{\policy}(\mathcal X,t;\cond_{\mathrm{inv}}),
\qquad
\sum_{i\in\mathcal V_{\mathrm T}}\vv_{\policy,i}=\vzero.
\label{eq:geodite-equivariance}
\end{equation}

The reported model uses \(D=6\) interaction layers, hidden width \(d=256\),
eight attention heads, \(K=64\) radial functions, \(L=1\), a
\(10~\text{\AA}\) cutoff, zero attention dropout, additive SiLU attention,
density-normalized aggregation, and trainable covalent-radius weighting.

\section*{Supplementary Note 3: Composite heuristic reward implementation}
\label{app:reward-implementation}
The g-xTB oracle is executed in analytical-gradient mode. Oracle calls
from a rollout batch are dispatched one molecule per task to a persistent
CPU worker pool, and their original sample order is restored before the
per-sample costs are assembled. Each result is checked for a finite energy
and a finite Cartesian force array with the expected atom count. Invalid
coordinates, timeouts, nonzero process exits, malformed outputs and other
oracle failures return a failure sentinel and receive the configured capped
failure cost while the rollout-group shape is retained. The Hessian-index
term diagonalizes the full oracle internal Hessian: it is computed
analytically for the differentiable MLFF oracles and assembled from
finite differences of forces for g-xTB, with rigid translations and
rotations projected out before all negative eigenvalues are counted.

\section*{Supplementary Note 4: Post-training details}
\label{app:nft}
We use the negative-sample formulation of Diffusion-NFT
\citep{zheng2025diffusionnft}. For each reaction, $K=48$ candidate
geometries are sampled with a lagged copy of the generator. The lagged
parameters are initialized from the pre-trained model and updated once
per epoch $e$ as
$\policy^{\mathrm{old}}\!\leftarrow\!
\rho_e\policy^{\mathrm{old}}+(1-\rho_e)\policy$ with
$\rho_e=\min\{5\times10^{-4}(e+1),\,0.5\}$. Sampling from this lagged
copy separates rollouts from the latest optimization step and improves
training stability.

For the recent g-xTB experiment, the branch scale was $\beta=1$. We
computed $\varsigma_{\cond}$ as the population standard deviation of the
$K$ rewards and implemented the denominator of
equation~\eqref{eq:adv} as
$\varsigma_{\cond}+\epsilon_A$, with $\epsilon_A=10^{-2}$; standardized
advantages were then clipped to $[-1,1]$. The separate $10^{-8}$
numerical tolerance in the implementation only identifies exactly
zero-spread groups and is not the advantage stabilizer. The fixed
post-training AlphaFlow target-mixture coefficient was $0.5$.

A pre-training epoch is one complete traversal of the distributed
training loader. The supervised lineage used by the g-xTB run comprised
$549{,}226$ optimizer updates through base checkpoint epoch 882
($883\times622$) and a further $377{,}608$ updates through
optimized-endpoint checkpoint epoch 307 ($308\times1{,}226$), giving
$N_{\mathrm{pre}}=926{,}834$. Post-training instead defined one epoch as
64 distributed minibatches. Each minibatch contained two reactions per
GPU on 16 GPUs, or 32 reactions globally, and therefore
$32\times48=1{,}536$ g-xTB candidates. Checkpoint epochs are zero
indexed: the selected epoch-30 checkpoint contains
$N_{\mathrm{post}}=1{,}984$ optimizer updates. Validation was performed
after every epoch, and checkpoints were ranked by the minimum
validation-set mean $F_{\mathrm{rms}}$ (equation~\eqref{eq:force_reward}). Epoch 30 was best
($0.78299$~eV\,\AA$^{-1}$); training stopped after epoch 45 when this
quantity had not improved for 15 validations. The completed run
processed $46\times64=2{,}944$ minibatches and recorded 2,943 completed
optimizer updates, while all reported parameters came from the selected
epoch-30 checkpoint.

The complete model, training and reward configuration used for this
experiment is collected in Supplementary Table~\ref{tab:hparams}.

\begin{table}[p]
\centering
\caption{\textbf{Model, training, and reward configuration.}}
\label{tab:hparams}
\small
\begin{tabularx}{\textwidth}{@{}p{0.31\textwidth}X@{}}
\toprule
\textbf{Component} & \textbf{Setting} \\
\midrule

\multicolumn{2}{@{}l}{\textit{Model architecture}} \\
\addlinespace[2pt]
Encoder
& GeoditE; 6 layers, width 256, 8 attention heads, $L=1$, 10~\AA{} cutoff \\
Radial functions
& 64 \\
Attention dropout
& 0 \\

\addlinespace[5pt]
\multicolumn{2}{@{}l}{\textit{Flow and sampling}} \\
\addlinespace[2pt]
Prior
& Centroid-projected isotropic Cartesian, $\sigma=0.1$~\AA{} \\
AlphaFlow normalization
& $\epsilon_{\alpha}=10^{-3}$ \\
Base pre-training time pairs
& $P(s=t)=0.75$; otherwise sorted logit-normal$(-0.4,1)$ \\
Post-training time pairs
& $P(s=t)=0.75$; otherwise sorted logit-normal$(-0.4,1)$ \\
AlphaFlow schedule
& $(k_{\mathrm{s}},k_{\mathrm{e}},\gamma,\eta_{\alpha})
   =(0,50{,}000,25,0.005)$ \\

\addlinespace[5pt]
\multicolumn{2}{@{}l}{\textit{Post-training}} \\
\addlinespace[2pt]
Branch / advantage
& $\beta=1$; $\epsilon_A=10^{-2}$; standardized advantages clipped to $[-1,1]$ \\
Minibatch
& 2 reactions per GPU $\times$ 16 GPUs = 32 reactions per global step;
1,536 oracle calls per global step \\
Optimizer
& AdamW; constant learning rate $5\times10^{-5}$;
$(\beta_1,\beta_2)=(0.9,0.999)$; weight decay $10^{-4}$;
no AMSGrad; gradient clipping at 1.0 \\

\addlinespace[5pt]
\multicolumn{2}{@{}l}{\textit{Data loading and oracle parallelism}} \\
\addlinespace[2pt]
Data-loader workers
& 16 per distributed training process; prefetch factor 4 \\
g-xTB oracle
& 32 persistent single-thread processes per training process;
up to 512 concurrent oracle calls across 16 training processes \\

\addlinespace[5pt]
\multicolumn{2}{@{}l}{\textit{Reward}} \\
\addlinespace[2pt]
Reward weights
& $(F,\mathrm{prog},H,\mathrm{above},\mathrm{geom})
   =(1,1,0.01,0.2,1)$ \\

\bottomrule
\end{tabularx}
\end{table}

\section*{Supplementary Note 5: Baseline implementations}
\label{sec:si-baselines}

\paragraph{Shared protocol and provenance.}
All learned Reaction-QM comparators use the same
training and evaluation splits,
Reaction-QM atom mapping, explicit hydrogen atoms and transition-state
coordinate target.  Unless stated otherwise, training uses seed 43,
full-precision arithmetic and an exponential moving average of the model
weights.  

\paragraph{React-OT.}
React-OT leverages the conditional flow-matching run without any stochasticity.
Its LEFTNet has six message-passing layers, 196 hidden channels, 96 radial
functions and a 10-\AA{} cutoff, with object-aware, reflection-equivariant
updates.  It was trained on eight accelerators with configured batch size
32, learning rate \(5\times10^{-4}\) and exponential-moving-average decay
0.999.  The sampler uses ten time-grid
points, hence nine fixed midpoint intervals and two network calls per
interval, for 18 neural function evaluations per sample.

\paragraph{ReactOT MeanFlow}
The MeanFlow baseline
uses the same six-layer, 196-channel LEFTNet and was warm-started from
the React-OT LEFTNet.  It was trained on
eight accelerators with configured batch size 8, learning rate
\(10^{-3}\) and exponential-moving-average decay 0.999. 

\paragraph{Latent flow.}
The latent model is trained in two stages.  The autoencoder
has four 128-channel layers, a 32-dimensional latent feature and a
16-dimensional edge-type representation; its is frozen for the second stage.  The
latent vector field
uses the six-layer, 196-channel LEFTNet.  Both stages use configured batch
size 16, learning rate \(5\times10^{-4}\), 16 accelerators and
full-precision training; the latent-flow sampler uses 20 neural
function evaluations.

\paragraph{RitS.}
The RitS
conditions on SMILES-derived bonds and starts from the Cartesian midpoint,
rather than the Gaussian prior.  Its Pairformer--DiT has ten layers,
256-dimensional invariant node features, 64-dimensional invariant edge
and vector features, four attention heads, 16 distance features, 61 atom
classes and nine edge classes.  It was trained in full precision on
16 accelerators with configured batch size 16, learning rate
\(5\times10^{-4}\), weight decay 0.01 and exponential-moving-average decay
0.999.  The sampler uses 25 explicit Euler
steps and therefore 25 neural function evaluations.

\paragraph{TS-DFM.} TS-DFM is implemented using a six-layer Pairformer-style encoder with a hidden dimension of 128, eight attention heads, 32 radial basis functions, and a 20-\AA{} cutoff.
  Distance-space flow matching uses coordinate-noise
standard deviation \(0.1\) and the constant-velocity parameterization.
Training uses seed 42, eight accelerators, configured batch size 8,
learning rate \(5\times10^{-4}\), full precision and
exponential-moving-average decay 0.999.  Inference uses 20 fixed midpoint distance-flow steps.  Each step calls
the network at the interval start and midpoint, giving 40 neural function
evaluations.  This is followed by one multidimensional-scaling
initialization and L-BFGS distance-to-coordinate reconstruction (at most
100 iterations, learning rate 0.1 and tolerance 0.01).  Reconstruction
iterations are excluded from the neural-function-evaluation count but
included in the end-to-end throughput measurement.

\paragraph{TS-drift.}
The drifting-model baseline
uses a four-layer LEFTNet with 128 hidden channels, 96 radial functions
and a 10-\AA{} cutoff.  Its prior is the endpoint interpolation plus
Gaussian noise of standard deviation \(0.05\) modulated by a sine envelope;
the drift objective uses a 16-dimensional noise embedding, eight negative
trajectories and temperature 0.1.  It was trained on 16 accelerators with
configured batch size 16, learning rate \(10^{-3}\), Adam coefficients
\((0.9,0.95)\), weight decay 0.01, five warm-up epochs, minimum learning
rate \(10^{-6}\) and exponential-moving-average decay 0.9999.  The
sampling applies the displacement generator once, giving one neural
function evaluation.

\paragraph{Geometry-only interpolants.}
These comparators have no learned parameters or training budget.  All
construct a three-frame reactant--transition-state--product path and use
the middle frame as the prediction.  Linear interpolation uses the
Cartesian midpoint.  IDPP uses ASE's image-dependent pair-potential NEB
interpolation.  Geodesic interpolation first redistributes the endpoint
path and then smooths it with scaling 1.7, distance cutoff \(3.0\),
friction \(10^{-2}\), tolerance \(2\times10^{-3}\), at most 15 outer
iterations and 20 micro-iterations for the sweep used on systems with
more than 35 atoms.  If geodesic smoothing fails, the redistributed path
is retained; a baseline-level exception falls back to linear
interpolation.  

\section*{Supplementary Note 6: Architecture and training-scheme ablation}
\label{sec:si-ablation}
To attribute \method's Reaction-QM accuracy to its individual design
choices rather than to their combination, we vary one factor at a time:
the encoder (LEFTNet versus GeoditE), the generator (a multi-step
conditional-flow bridge versus a one-step MeanFlow map, with and without
the AlphaFlow curriculum of Supplementary Note~1), and reward
post-training.  All variants are trained and evaluated in the single
consistent Reaction-QM suite on the \texttt{irc\_all} test split
($n=19{,}883$), with one candidate per reaction from the deterministic
reactant--product midpoint and the $3N$-normalized RMSD used throughout
(Extended Data Table~\ref{tab:ablation}).

Four comparisons isolate the contributions.  At a matched LEFTNet
encoder, replacing the multi-step conditional-flow bridge (mean/median
$0.1214/0.1056$~\angstrom) with a one-step MeanFlow map
($0.1269/0.1107$~\angstrom) costs under $5\%$ in mean RMSD, and the
AlphaFlow curriculum recovers most of that gap
($0.1233/0.1059$~\angstrom); one-step generation is therefore nearly
free relative to multi-step integration.  At a matched one-step MeanFlow
scheme, swapping LEFTNet for GeoditE gives a small further improvement
($0.1262/0.1094$~\angstrom).  Adding the AlphaFlow curriculum to the
GeoditE one-step model, still without any reward post-training, gives
$0.1222/0.1054$~\angstrom.  Relative to that matched pre-trained
control, reward post-training then provides the remaining gain,
reducing the GeoditE one-step MeanFlow model to
$0.1108/0.0923$~\angstrom, which carries \method\ past the
multi-step React-OT recipe.

\section*{Supplementary Note 7: Contextual-bandit formulation and
policy improvement}
\label{app:rl}
This note supports two statements made without proof in Methods: that
the physical reward defines a well-posed optimality tilt of the
pre-trained generator, and that the group reweighting of
equation~\eqref{eq:optimality} improves expected reward.  Throughout,
$\pi_{\policy}(\cdot\mid\cond)
=T_{\policy}(\cdot\,;\cond)_{\#}p_1(\cdot\mid\cond)$ is the policy
induced by the one-step map, $\cond$ is the context, a generated
geometry is the action and the oracle supplies the terminal reward, so
the horizon is one.

\paragraph{The reward is nonpositive and bounded below.}
Every cost entering equation~\eqref{eq:reward} is nonnegative:
$C_F$ is a root-mean-square; $C_{\mathrm{prog}}$ is a mean of squared
hinge terms plus a variance; $C_{\kappa}$ is a positive part;
$C_{\mathrm{barrier}}$ is a sum of softplus terms; and
$C_{\mathrm{geom}}$ is a clipped positive ramp.  Hence
$\reward(\vx;\cond)=-\min\{C_{\max},C(\vx;\cond)\}\in[-C_{\max},0]$, so
$\exp(\reward(\vx;\cond)/\lambda_{\mathrm{KL}})\in(0,1]$ is a valid
likelihood for a binary optimality variable at any temperature
$\lambda_{\mathrm{KL}}>0$.  This is the hypothesis that makes the tilt
below well defined; it would fail for an unbounded or sign-indefinite
reward.

\paragraph{The tilted policy maximizes the idealized objective.}
Fix $\cond$ and write $\pi_{\mathrm{pre}}=\pi_{\policy_{\mathrm{pre}}}
(\cdot\mid\cond)$.  Let
$Z(\cond)=\EE_{\vx\sim\pi_{\mathrm{pre}}}
[\exp(\reward(\vx;\cond)/\lambda_{\mathrm{KL}})]$ and
$\pi^{\star}(\vx\mid\cond)=\pi_{\mathrm{pre}}(\vx\mid\cond)
\exp(\reward(\vx;\cond)/\lambda_{\mathrm{KL}})/Z(\cond)$.  For any
policy $\pi$ absolutely continuous with respect to $\pi_{\mathrm{pre}}$,
a direct substitution gives
\begin{equation}
\EE_{\pi}[\reward]
-\lambda_{\mathrm{KL}}
D_{\mathrm{KL}}\!\left(\pi\,\|\,\pi_{\mathrm{pre}}\right)
=
\lambda_{\mathrm{KL}}\log Z(\cond)
-\lambda_{\mathrm{KL}}
D_{\mathrm{KL}}\!\left(\pi\,\|\,\pi^{\star}\right).
\label{eq:si-elbo}
\end{equation}
Because the final divergence is nonnegative and vanishes only at
$\pi=\pi^{\star}$, the objective of
equation~\eqref{eq:objective-ideal} is maximized uniquely by
$\pi^{\star}$, with optimum $\lambda_{\mathrm{KL}}\log Z(\cond)$.
Maximizing KL-regularized expected reward is therefore the same problem
as variationally approximating the pre-trained generator conditioned on
optimality under the physical reward.

\paragraph{The clipped group reweighting improves expected reward.}
Fix a reaction and let $\pi_{\mathrm{old}}$ be the EMA behaviour policy
that generated the rollouts.  With $\mu$ the mean reward under
$\pi_{\mathrm{old}}$, $\adv=\reward-\mu$,
$\psi=\operatorname{clip}(\adv/\varsigma,-1,1)$ and
$q=\tfrac12(1+\psi)$ as in equation~\eqref{eq:optimality}, define the
positively reweighted policy $\pi_{+}\propto q\,\pi_{\mathrm{old}}$.
Writing $J(\pi)=\EE_{\pi}[\reward]$ and substituting
$q=\tfrac12(1+\psi)$,
\begin{equation}
J(\pi_{+})-J(\pi_{\mathrm{old}})
=
\frac{\EE_{\pi_{\mathrm{old}}}[\adv\,\psi]}
{1+\EE_{\pi_{\mathrm{old}}}[\psi]}.
\label{eq:si-improve}
\end{equation}
Pointwise,
$\adv\,\psi=|\adv|\min\{|\adv|/\varsigma,1\}
=\min\{\adv^{2}/\varsigma,\,|\adv|\}\geq0$, and the denominator is
positive because $\psi\geq-1$.  Hence
$J(\pi_{+})\geq J(\pi_{\mathrm{old}})$, with equality only if
$\reward$ is $\pi_{\mathrm{old}}$-almost surely constant.  Clipping
does not weaken the conclusion; it is what keeps $q\in[0,1]$, so that
$\pi_{+}$ is a probability distribution at all.  The same computation
holds for a finite rollout group: with weights
$p^{+}_{j}=(1+\psi_{j})/\sum_{\ell}(1+\psi_{\ell})$,
$\sum_{j}p^{+}_{j}\reward_{j}-\bar{\reward}
=\sum_{j}\adv_{j}\psi_{j}/\sum_{j}(1+\psi_{j})\geq0$, so every
nondegenerate group defines an empirically reward-improved candidate
distribution before any neural approximation enters.

\paragraph{Scope.}
These statements concern the reweighted \emph{distribution}
$\pi_{+}$, which the training procedure never realizes exactly.  The
parametric update projects that target onto the model class by the
regression of equation~\eqref{eq:nft}, and we do not bound the
resulting projection error.  The improvement is also measured relative
to the EMA behaviour policy rather than to the current parameters, and
equation~\eqref{eq:adv} is a finite-$K$ estimator whose baseline and
scale are computed from the same $K$ samples, so it is biased at small
group size and estimates $\adv/\varsigma$ rather than $\adv$.  These
are the reasons Methods claims a policy-improvement step rather than
monotonic reward improvement.

\section*{Supplementary Note 8: Size-extrapolation statistics for
large-Transition1x}
\label{sec:si-size-stats}

Large-Transition1x contains 131 reactions, of which four are excluded as
anomalous: on these the React-OT structure diverges numerically
($\mathrm{RMSD}>200$\,\angstrom; values and justification below).  The
analysed sample is therefore $n=127$, and every statistic in this note, and
in Fig.~\ref{fig:transfer}d, is computed on it.  For each analysed reaction
we take the paired difference
$\Delta=\mathrm{RMSD}(\method)-\mathrm{RMSD}(\text{React-OT})$
after Kabsch alignment, so $\Delta<0$ favours \method.  The reactions form
33 substitution families sharing a parent scaffold, and the estimator
plotted in Fig.~\ref{fig:transfer}d is the family-equal mean: $\Delta$ is
averaged within each family and the family means are then averaged without
weighting, so that no single large family dominates a bin.  All intervals
below are $95\%$ clustered percentile bootstraps that resample whole
families, with $10{,}000$ draws.

\paragraph{Bins, counts and per-bin estimates.}
The bins are heavy-atom terciles, closed on both ends and covering
$10$--$33$ heavy atoms without gaps or overlap.  The four excluded reactions
are held out of the display and of the binned estimates
(Supplementary Table~\ref{tab:si-sizebins}).

\begin{table}[!ht]
\centering
\caption{Per-bin statistics for the RMSD difference $\Delta$ plotted in
Fig.~\ref{fig:transfer}d, analysed sample ($n=127$).  Brackets give
$95\%$ clustered percentile bootstrap intervals over families.}
\label{tab:si-sizebins}
\small
\begin{tabular}{lccccc}
\toprule
Heavy atoms & Reactions & Families &
Family-equal mean $\Delta$ (\angstrom) & Median $\Delta$ (\angstrom) &
$\Pr(\Delta<0)$ \\
\midrule
$10$--$16$ & $39$ & $13$ & $-0.061$ $[-0.132,-0.004]$ &
$-0.059$ $[-0.257,+0.008]$ & $0.72$ \\
$17$--$23$ & $41$ & $19$ & $-0.441$ $[-0.693,-0.225]$ &
$-0.280$ $[-0.552,-0.073]$ & $0.88$ \\
$24$--$33$ & $47$ & $20$ & $-5.892$ $[-13.994,-1.793]$ &
$-2.441$ $[-3.120,-1.116]$ & $0.94$ \\
\bottomrule
\end{tabular}
\end{table}

Median RMSD is flat across the three bins for \method\ ($0.277$, $0.267$ and
$0.322$\,\angstrom) and rises $7.4$-fold for React-OT ($0.373$, $0.598$ and
$2.742$\,\angstrom), so the widening margin reflects degradation of the
baseline rather than improvement in \method.  Because reactions within a
family are not independent, we test on family means: two-sided Wilcoxon
signed-rank tests give $P=0.073$, $6.3\times10^{-5}$ and $9.5\times10^{-7}$
for the three bins, with $8/13$, $17/19$ and $20/20$ families favouring
\method.  The smallest-size bin is therefore not resolved from zero once
clustering is respected, and we claim no effect there.

\paragraph{The four excluded reactions, and the tail that remains.}
The two error distributions differ qualitatively in their tails.  Over the
127 analysed reactions \method\ never exceeds $0.852$\,\angstrom, whereas
React-OT exceeds $1$\,\angstrom\ on 48 of them and $10$\,\angstrom\ on two.
The four excluded reactions are the extreme end of that same asymmetry:
React-OT returns $4.29\times10^{2}$, $3.13\times10^{2}$,
$2.16\times10^{16}$ and $3.13\times10^{26}$\,\angstrom\ on them, with
\method\ below $0.51$\,\angstrom\ on all four.  A single value of that size
determines any mean it enters: retaining the four would make the largest
bin's family-equal mean $-7.8\times10^{24}$\,\angstrom, which is why they
are excluded.  The exclusion is conservative for our comparison, since
every one is a reaction \method\ solved and React-OT did not.

Within the analysed sample the $24$--$33$ estimate still rests heavily on one
reaction, for which React-OT returns $71.40$\,\angstrom\ against \method's
$0.312$\,\angstrom.  Removing it moves that bin from $-5.892$ to
$-2.461$\,\angstrom\ (a $58\%$ change), whereas no other single reaction
moves it by more than $5.9\%$ and no other whole family by more than
$5.1\%$.  Estimators that downweight the tail agree closely with one
another: $20\%$ trimmed mean $-2.362$\,\angstrom, $5\%$ winsorized mean
$-2.702$\,\angstrom, mean over reactions with React-OT below
$10$\,\angstrom\ $-2.487$\,\angstrom, and median $-2.441$\,\angstrom\
$[-3.120,-1.116]$.  We therefore quote the median for this bin.  A
scale-free summary immune to the divergences is the per-reaction RMSD
ratio, whose median is $0.80$, $0.47$ and $0.11$ across the three bins,
i.e.\ an $8.8$-fold typical reduction on the largest systems.

\paragraph{Size association.}
The primary test is a Spearman rank correlation between heavy-atom count and
$\Delta$, chosen for its insensitivity to the unbounded tail:
$\rho=-0.681$, $95\%$ CI $[-0.795,-0.481]$, $P<10^{-4}$ ($n=127$).  The $P$
value comes from a family-preserving permutation test in which heavy-atom
count is permuted across whole families, holding family membership and
within-family structure fixed; none of $10{,}000$ permutations reached the
observed $|\rho|$.  The correlation is stable to dropping the most extreme
retained reaction ($\rho=-0.677$), to reinstating the four excluded
reactions ($\rho=-0.702$) and to replacing $\Delta$ by the log RMSD ratio
($\rho=-0.649$).  Decomposing by method shows that the trend is driven by
the baseline: heavy-atom count correlates with React-OT's error at
$\rho=+0.680$ $[+0.503,+0.781]$ but with \method's at only $\rho=+0.245$
$[+0.002,+0.431]$.  As a secondary, cluster-robust model, regressing the log
RMSD ratio on heavy-atom count gives $-0.0502$ dex per heavy atom
$[-0.0643,-0.0360]$, $P=3.4\times10^{-12}$, that is, roughly $11\%$ per
heavy atom across the observed $23$-heavy-atom span.

The bin boundaries, the estimator and the clustering unit were fixed when
Fig.~\ref{fig:transfer}d was constructed.  The designation of the Spearman
correlation as the primary test of the size association was not
prespecified: it was made for this revision, on the stated grounds of tail
robustness rather than by selecting the strongest of several candidates,
and every alternative we computed is reported above.

\section*{Supplementary Note 9: The 200-reaction Reaction-QM test cohort}
\label{sec:si-test200}

The oracle comparison in Fig.~\ref{fig:mlff-benchmark}e--h is evaluated on
a fixed 200-reaction subset of the Reaction-QM intrinsic-reaction-coordinate
test split ($19{,}897$ reactions). Every arm is evaluated on exactly these
200 reactions, so all reported comparisons are paired.
Supplementary Table~\ref{tab:si-test200} lists the selected zero-based
test-split indices with their Reaction-QM reaction identifiers.

\begin{table}[!ht]
\centering
\caption{The 200 Reaction-QM test-split reactions used for the
GPU4PySCF/Sella oracle comparison, given as zero-based indices into the
intrinsic-reaction-coordinate test split with the corresponding reaction
identifiers. The additional 100 reactions are appended after the original
100; within each block, reading order is down each column.}
\label{tab:si-test200}
\tiny
\setlength{\tabcolsep}{2.5pt}
\renewcommand{\arraystretch}{0.92}
\begin{tabular}{@{}rlrlrlrl@{}}
\toprule
Index & Reaction ID & Index & Reaction ID &
Index & Reaction ID & Index & Reaction ID \\
\midrule
256 & \texttt{RXN\_0000002361} & 4044 & \texttt{RXN\_0000040867} & 9844 & \texttt{RXN\_0000100095} & 15051 & \texttt{RXN\_0000151883} \\
378 & \texttt{RXN\_0000003675} & 4121 & \texttt{RXN\_0000041591} & 10524 & \texttt{RXN\_0000107277} & 15143 & \texttt{RXN\_0000152774} \\
409 & \texttt{RXN\_0000003949} & 4579 & \texttt{RXN\_0000046101} & 10543 & \texttt{RXN\_0000107429} & 15399 & \texttt{RXN\_0000155179} \\
551 & \texttt{RXN\_0000005344} & 4809 & \texttt{RXN\_0000048661} & 10718 & \texttt{RXN\_0000109031} & 15709 & \texttt{RXN\_0000158467} \\
642 & \texttt{RXN\_0000006292} & 5007 & \texttt{RXN\_0000050684} & 11051 & \texttt{RXN\_0000112555} & 15978 & \texttt{RXN\_0000161286} \\
698 & \texttt{RXN\_0000006879} & 5342 & \texttt{RXN\_0000054016} & 11303 & \texttt{RXN\_0000114981} & 16098 & \texttt{RXN\_0000162426} \\
873 & \texttt{RXN\_0000008658} & 5365 & \texttt{RXN\_0000054201} & 12097 & \texttt{RXN\_0000123078} & 16155 & \texttt{RXN\_0000163054} \\
972 & \texttt{RXN\_0000009537} & 5703 & \texttt{RXN\_0000057829} & 12233 & \texttt{RXN\_0000124381} & 16160 & \texttt{RXN\_0000163193} \\
1369 & \texttt{RXN\_0000013676} & 5974 & \texttt{RXN\_0000060811} & 12564 & \texttt{RXN\_0000127800} & 16297 & \texttt{RXN\_0000164495} \\
1413 & \texttt{RXN\_0000014101} & 6190 & \texttt{RXN\_0000062954} & 12696 & \texttt{RXN\_0000129100} & 16329 & \texttt{RXN\_0000164701} \\
1440 & \texttt{RXN\_0000014409} & 6271 & \texttt{RXN\_0000063748} & 12706 & \texttt{RXN\_0000129185} & 16604 & \texttt{RXN\_0000167575} \\
1600 & \texttt{RXN\_0000015942} & 6679 & \texttt{RXN\_0000068185} & 12885 & \texttt{RXN\_0000130992} & 16653 & \texttt{RXN\_0000168039} \\
1675 & \texttt{RXN\_0000016593} & 6747 & \texttt{RXN\_0000069029} & 12920 & \texttt{RXN\_0000131344} & 17232 & \texttt{RXN\_0000173597} \\
1750 & \texttt{RXN\_0000017319} & 6768 & \texttt{RXN\_0000069206} & 13048 & \texttt{RXN\_0000132649} & 17301 & \texttt{RXN\_0000174217} \\
1839 & \texttt{RXN\_0000018314} & 6995 & \texttt{RXN\_0000071415} & 13146 & \texttt{RXN\_0000133610} & 17489 & \texttt{RXN\_0000175977} \\
1993 & \texttt{RXN\_0000019861} & 7480 & \texttt{RXN\_0000076233} & 13157 & \texttt{RXN\_0000133685} & 17587 & \texttt{RXN\_0000176988} \\
2009 & \texttt{RXN\_0000020020} & 7953 & \texttt{RXN\_0000080846} & 13287 & \texttt{RXN\_0000135122} & 17979 & \texttt{RXN\_0000180985} \\
2054 & \texttt{RXN\_0000020490} & 7980 & \texttt{RXN\_0000081113} & 13345 & \texttt{RXN\_0000135577} & 17997 & \texttt{RXN\_0000181113} \\
2415 & \texttt{RXN\_0000024146} & 8058 & \texttt{RXN\_0000081873} & 13544 & \texttt{RXN\_0000137552} & 18001 & \texttt{RXN\_0000181134} \\
2488 & \texttt{RXN\_0000024790} & 8300 & \texttt{RXN\_0000084361} & 13638 & \texttt{RXN\_0000138481} & 18380 & \texttt{RXN\_0000184765} \\
2584 & \texttt{RXN\_0000025721} & 8555 & \texttt{RXN\_0000087128} & 14031 & \texttt{RXN\_0000142199} & 18456 & \texttt{RXN\_0000185399} \\
2605 & \texttt{RXN\_0000025937} & 9060 & \texttt{RXN\_0000092304} & 14096 & \texttt{RXN\_0000142857} & 18865 & \texttt{RXN\_0000189339} \\
2741 & \texttt{RXN\_0000027342} & 9095 & \texttt{RXN\_0000092668} & 14375 & \texttt{RXN\_0000145400} & 19392 & \texttt{RXN\_0000194834} \\
3203 & \texttt{RXN\_0000031987} & 9317 & \texttt{RXN\_0000094991} & 14596 & \texttt{RXN\_0000147485} & 19495 & \texttt{RXN\_0000195965} \\
3677 & \texttt{RXN\_0000036873} & 9360 & \texttt{RXN\_0000095363} & 15006 & \texttt{RXN\_0000151421} & 19521 & \texttt{RXN\_0000196183} \\
\midrule
0 & \texttt{RXN\_0000000001} & 5303 & \texttt{RXN\_0000053470} & 9448 & \texttt{RXN\_0000096218} & 15155 & \texttt{RXN\_0000152865} \\
287 & \texttt{RXN\_0000002658} & 5329 & \texttt{RXN\_0000053756} & 10083 & \texttt{RXN\_0000102607} & 15534 & \texttt{RXN\_0000156521} \\
473 & \texttt{RXN\_0000004584} & 5383 & \texttt{RXN\_0000054423} & 10210 & \texttt{RXN\_0000104142} & 15566 & \texttt{RXN\_0000156850} \\
529 & \texttt{RXN\_0000005155} & 5785 & \texttt{RXN\_0000058830} & 10440 & \texttt{RXN\_0000106441} & 15794 & \texttt{RXN\_0000159308} \\
912 & \texttt{RXN\_0000008954} & 5999 & \texttt{RXN\_0000061047} & 11258 & \texttt{RXN\_0000114561} & 16043 & \texttt{RXN\_0000161911} \\
1086 & \texttt{RXN\_0000010745} & 6015 & \texttt{RXN\_0000061182} & 11388 & \texttt{RXN\_0000115848} & 16086 & \texttt{RXN\_0000162319} \\
1111 & \texttt{RXN\_0000010970} & 6046 & \texttt{RXN\_0000061505} & 11541 & \texttt{RXN\_0000117300} & 16194 & \texttt{RXN\_0000163482} \\
1321 & \texttt{RXN\_0000013183} & 6197 & \texttt{RXN\_0000063046} & 11584 & \texttt{RXN\_0000117720} & 16510 & \texttt{RXN\_0000166628} \\
1455 & \texttt{RXN\_0000014545} & 6461 & \texttt{RXN\_0000065929} & 12104 & \texttt{RXN\_0000123133} & 16701 & \texttt{RXN\_0000168635} \\
2221 & \texttt{RXN\_0000022100} & 6592 & \texttt{RXN\_0000067429} & 12123 & \texttt{RXN\_0000123262} & 16718 & \texttt{RXN\_0000168850} \\
2252 & \texttt{RXN\_0000022384} & 6719 & \texttt{RXN\_0000068693} & 12139 & \texttt{RXN\_0000123391} & 17192 & \texttt{RXN\_0000173232} \\
2285 & \texttt{RXN\_0000022707} & 7308 & \texttt{RXN\_0000074665} & 12228 & \texttt{RXN\_0000124343} & 17455 & \texttt{RXN\_0000175662} \\
2352 & \texttt{RXN\_0000023370} & 7346 & \texttt{RXN\_0000075022} & 12437 & \texttt{RXN\_0000126451} & 17507 & \texttt{RXN\_0000176151} \\
3010 & \texttt{RXN\_0000030077} & 7406 & \texttt{RXN\_0000075562} & 12446 & \texttt{RXN\_0000126509} & 17747 & \texttt{RXN\_0000178655} \\
3020 & \texttt{RXN\_0000030168} & 7575 & \texttt{RXN\_0000077105} & 12537 & \texttt{RXN\_0000127537} & 17792 & \texttt{RXN\_0000179095} \\
3169 & \texttt{RXN\_0000031684} & 7646 & \texttt{RXN\_0000077911} & 13024 & \texttt{RXN\_0000132364} & 17848 & \texttt{RXN\_0000179861} \\
3222 & \texttt{RXN\_0000032152} & 7825 & \texttt{RXN\_0000079512} & 13127 & \texttt{RXN\_0000133445} & 17860 & \texttt{RXN\_0000180001} \\
3268 & \texttt{RXN\_0000032627} & 7868 & \texttt{RXN\_0000079903} & 13216 & \texttt{RXN\_0000134488} & 18110 & \texttt{RXN\_0000182111} \\
4184 & \texttt{RXN\_0000042275} & 8019 & \texttt{RXN\_0000081539} & 13351 & \texttt{RXN\_0000135653} & 18207 & \texttt{RXN\_0000183085} \\
4193 & \texttt{RXN\_0000042340} & 8367 & \texttt{RXN\_0000085004} & 13811 & \texttt{RXN\_0000139971} & 18267 & \texttt{RXN\_0000183746} \\
4480 & \texttt{RXN\_0000045075} & 8906 & \texttt{RXN\_0000090775} & 13874 & \texttt{RXN\_0000140626} & 18452 & \texttt{RXN\_0000185388} \\
4576 & \texttt{RXN\_0000046044} & 8926 & \texttt{RXN\_0000090935} & 14217 & \texttt{RXN\_0000143884} & 18655 & \texttt{RXN\_0000187267} \\
4601 & \texttt{RXN\_0000046295} & 9073 & \texttt{RXN\_0000092396} & 14225 & \texttt{RXN\_0000143942} & 19334 & \texttt{RXN\_0000194286} \\
5089 & \texttt{RXN\_0000051320} & 9219 & \texttt{RXN\_0000094036} & 14299 & \texttt{RXN\_0000144659} & 19883 & \texttt{RXN\_0000199735} \\
5292 & \texttt{RXN\_0000053339} & 9329 & \texttt{RXN\_0000095084} & 14527 & \texttt{RXN\_0000146880} & 19890 & \texttt{RXN\_0000199826} \\
\bottomrule
\end{tabular}
\end{table}

\section*{Supplementary Note 10: Ketohydroperoxide channel identifiers}
\label{sec:si-khp-channels}

\begin{table}[!ht]
\centering
\caption{\textbf{The 13 ketohydroperoxide decomposition channels.}
The KHP labels follow the order used in
Fig.~\ref{fig:epoxidation-case-study}b. The source channel identifier is
the identifier in the public Zhao--Savoie YARP cohort, and the endpoint
geometry identifier is the archived coordinate entry used for evaluation.
The two identifiers differ only for KHP-07 and KHP-10, which required the
endpoint remapping described in Methods.}
\label{tab:si-khp-channels}
\scriptsize
\setlength{\tabcolsep}{4pt}
\renewcommand{\arraystretch}{1.08}
\begin{tabularx}{\textwidth}{c>{\raggedright\arraybackslash}X>{\raggedright\arraybackslash}X}
\toprule
Label & Source channel identifier & Endpoint geometry identifier \\
\midrule
KHP-01 & \texttt{XSASRUDTFFBDDK\_5\_0\_3}  & \texttt{XSASRUDTFFBDDK\_5\_0\_3} \\
KHP-02 & \texttt{XSASRUDTFFBDDK\_6\_0\_1}  & \texttt{XSASRUDTFFBDDK\_6\_0\_1} \\
KHP-03 & \texttt{XSASRUDTFFBDDK\_9\_0\_1}  & \texttt{XSASRUDTFFBDDK\_9\_0\_1} \\
KHP-04 & \texttt{XSASRUDTFFBDDK\_10\_0\_1} & \texttt{XSASRUDTFFBDDK\_10\_0\_1} \\
KHP-05 & \texttt{XSASRUDTFFBDDK\_11\_0\_1} & \texttt{XSASRUDTFFBDDK\_11\_0\_1} \\
KHP-06 & \texttt{XSASRUDTFFBDDK\_13\_2\_0} & \texttt{XSASRUDTFFBDDK\_13\_2\_0} \\
KHP-07 & \texttt{XSASRUDTFFBDDK\_21\_0\_3} & \texttt{XSASRUDTFFBDDK\_21\_0\_0} \\
KHP-08 & \texttt{XSASRUDTFFBDDK\_24\_0\_2} & \texttt{XSASRUDTFFBDDK\_24\_0\_2} \\
KHP-09 & \texttt{XSASRUDTFFBDDK\_26\_0\_2} & \texttt{XSASRUDTFFBDDK\_26\_0\_2} \\
KHP-10 & \texttt{XSASRUDTFFBDDK\_29\_0\_1} & \texttt{XSASRUDTFFBDDK\_29\_0\_3} \\
KHP-11 & \texttt{XSASRUDTFFBDDK\_34\_1\_2} & \texttt{XSASRUDTFFBDDK\_34\_1\_2} \\
KHP-12 & \texttt{XSASRUDTFFBDDK\_36\_0\_2} & \texttt{XSASRUDTFFBDDK\_36\_0\_2} \\
KHP-13 & \texttt{XSASRUDTFFBDDK\_38\_0\_3} & \texttt{XSASRUDTFFBDDK\_38\_0\_3} \\
\bottomrule
\end{tabularx}
\end{table}

\section*{Supplementary Note 11: Oxygen-transfer epoxidation case study}
\label{sec:si-oxygen-transfer}

Ethene epoxidation by a cyclic peroxyphosphorane provides a stringent
reaction-centre test because its transition state is defined by a
concerted reorganization rather than by any single bond length. The
transferred O5 atom must be positioned between the two alkene carbons as
the C3--O5 and C4--O5 bonds form and the C3$=$C4 bond loses
$\pi$ character, while O5--P6 and O5--O8 cleave and P6$=$O8 forms on
the leaving phosphorus fragment (Supplementary
Fig.~\ref{fig:si-oxygen-transfer}a).

We applied the same pretrained and reward-post-trained inference recipes
used in the main experiments. Relative to the pretrained prediction, the
post-trained geometry reduces the mean absolute error over the four
explicitly monitored forming and cleaving distances from 0.259 to
0.122~\angstrom\ (52.9\%) and lowers the initial DFT force RMS from
2.767 to 0.465~eV\,\angstrom$^{-1}$ (83.2\%). Sella reaches the refined
saddle in 14 accepted steps from the post-trained prediction, compared
with 31 from the pretrained prediction. The post-trained refinement
connects the intended endpoints, whereas the pretrained refinement does
not.

To relate the unstable mode to the intended reaction, we construct a
reaction direction from the four bonds that form or break. Let
$\bm G\in\mathbb R^{4\times3N}$ be the Wilson bond-stretch matrix at the
refined saddle. For bond $(i,j)$, its row contains
$\hat{\bm u}_{ij}=(\vx_i-\vx_j)/\lVert\vx_i-\vx_j\rVert$ in the
coordinates of atom $i$ and $-\hat{\bm u}_{ij}$ in those of atom $j$.
With
$\bm M_{\mathrm{mass}}=\operatorname{diag}(m_1,m_1,m_1,\ldots,m_N,m_N,m_N)$,
we set $\bm A=\bm G\bm M_{\mathrm{mass}}^{-1/2}$ and let
$\Delta\bm r=\bm r(\Pmol)-\bm r(\Rmol)$ collect the signed endpoint
bond-length changes. The normalized minimum-norm mass-weighted
displacement that reproduces these changes to first order is
\begin{equation}
\hat{\bm w}
=
\frac{\bm A^{+}\Delta\bm r}
     {\lVert\bm A^{+}\Delta\bm r\rVert},
\qquad
\bm A^{+}
=
\bm A^{\top}(\bm A\bm A^{\top})^{+},
\label{eq:si-oxygen-reaction-direction}
\end{equation}
where $(\cdot)^{+}$ is the Moore--Penrose pseudoinverse with cutoff
$10^{-10}$. We mass-weight and normalize the Cartesian imaginary mode
$\bm c$ as
$\hat{\bm q}=\bm M_{\mathrm{mass}}^{1/2}\bm c/
\lVert\bm M_{\mathrm{mass}}^{1/2}\bm c\rVert$ and report
$\lvert\hat{\bm q}^{\top}\hat{\bm w}\rvert$. This overlap is invariant
to translation, rotation and the arbitrary sign of a normal mode.

The saddle refined from the post-trained prediction has exactly one
significant imaginary frequency, $-364.09$~cm$^{-1}$, with
$\lvert\hat{\bm q}^{\top}\hat{\bm w}\rvert=0.707$. A fraction 0.837 of
the mode's mass-weighted norm lies in the four-dimensional subspace
spanned by the monitored bond coordinates. The unstable mode is
therefore associated with the coupled oxygen transfer rather than an
unrelated molecular distortion.

\begin{figure}[p]
  \centering
  \IfFileExists{figures/case-study-suppl.pdf}{%
    \includegraphics[width=\textwidth]{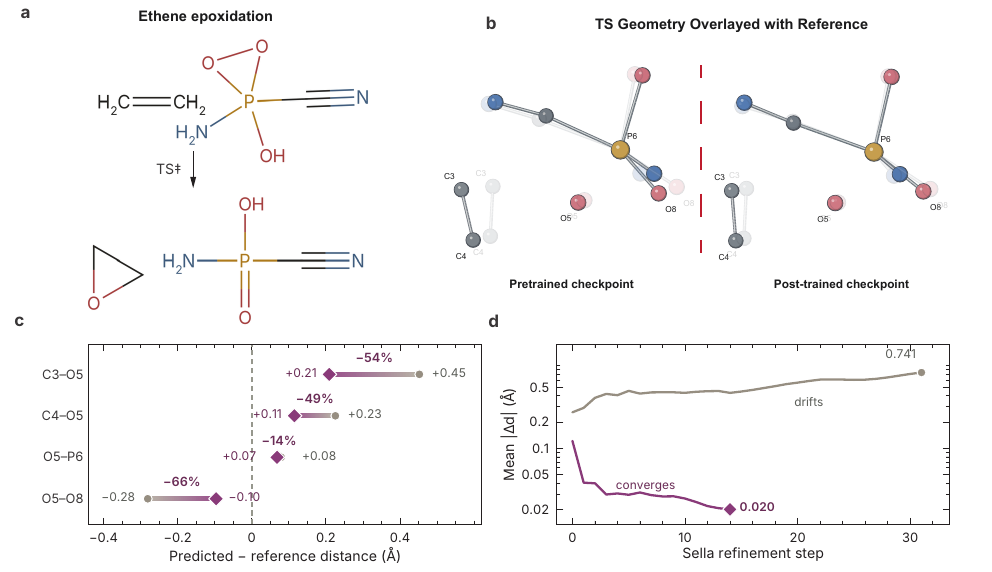}%
  }{%
    \fbox{\parbox[c][38mm][c]{0.94\textwidth}{\centering
      Upload \texttt{figures/case-study-suppl.pdf}}}%
  }
  \caption{\textbf{Post-training better resolves the coupled bond
  reorganization of cyclic-peroxyphosphorane epoxidation.}
  \textbf{a}, Atom-mapped oxygen transfer: O5 leaves the
  peroxyphosphorane motif, forms the two C--O bonds of the epoxide and
  leaves a P6$=$O8 bond on the phosphorus-containing fragment.
  \textbf{b}, Pretrained and post-trained predictions beside the
  labelled transition-state geometry, displayed with a common alignment
  and camera so that reaction-centre displacements are comparable.
  \textbf{c}, Signed errors relative to the reference for the two
  forming C--O distances and the two cleaving donor--oxygen distances.
  \textbf{d}, Mean absolute error over these four reaction-centre
  distances along the accepted Sella refinement steps; the shorter
  post-trained trajectory begins closer to the refined saddle.}
  \label{fig:si-oxygen-transfer}
\end{figure}

\section*{Supplementary Note 12: Matched inference-time g-xTB saddle guidance}
\label{sec:si-gxtb-guidance-budget}

To test whether additional inference-time surrogate optimization recovers the
post-trained result, we applied restricted-step partitioned
rational-function saddle optimization with the g-xTB surrogate to frozen
Pre-train candidates. The 100-reaction cohort, one deterministic
candidate per reaction and four generator function evaluations were fixed
across arms. Guidance budgets of 2, 4, 8 and 16 optimization cycles required
3, 5, 9 and 17 g-xTB energy/force evaluations, respectively, plus one Hessian
evaluation per candidate.

Let $\vx^{(0)}=\widehat{\vx}$ denote a Pre-train output, using the candidate
and surrogate-energy notation of Methods. At guidance cycle $k$, define the
surrogate gradient and the eigendecomposition of the optimizer's current
Cartesian surrogate Hessian by
\begin{equation}
\bm g^{(k)}
=\nabla_{\vx}\Ehat\!\left(\vx^{(k)}\right)
=-\widehat{\bm F}\!\left(\vx^{(k)}\right),
\qquad
\widehat{\bm H}^{(k)}
=\bm V^{(k)}\bm\Lambda^{(k)}\bm V^{(k)\top},
\qquad
\widetilde{\bm g}^{(k)}
=\bm V^{(k)\top}\bm g^{(k)},
\label{eq:si-gxtb-guidance-decomposition}
\end{equation}
where $\widehat{\bm H}^{(0)}=\nabla_{\vx}^{2}\Ehat(\vx^{(0)})$. The
initially lowest-curvature eigenvector is propagated between cycles by maximum
absolute eigenvector overlap. Its current representative $\bm v_1^{(k)}$
defines the one-dimensional uphill subspace $\mathcal U_k$, and the retained
orthogonal modes define the downhill subspace $\mathcal S_k$. At the level of
minimum-mode following, the guidance force is
\begin{equation}
\widehat{\bm F}_{\mathrm{saddle}}^{(k)}
=
\left(\bm I-2\bm v_1^{(k)}\bm v_1^{(k)\top}\right)
\widehat{\bm F}\!\left(\vx^{(k)}\right),
\label{eq:si-gxtb-minimum-mode-force}
\end{equation}
which reverses the force along the followed mode while retaining downhill
motion in its orthogonal complement.

The implemented restricted-step partitioned rational-function update
regularizes this direction. For each block
$\mathcal B\in\{\mathcal U_k,\mathcal S_k\}$, it solves
\begin{equation}
\begin{bmatrix}
\bm\Lambda_{\mathcal B}^{(k)}/\alpha_k
& \widetilde{\bm g}_{\mathcal B}^{(k)}/\alpha_k\\
\widetilde{\bm g}_{\mathcal B}^{(k)\top} & 0
\end{bmatrix}
\begin{bmatrix}\bm p_{\mathcal B}^{(k)}\\1\end{bmatrix}
=
\nu_{\mathcal B}^{(k)}
\begin{bmatrix}\bm p_{\mathcal B}^{(k)}\\1\end{bmatrix},
\qquad
\mathcal B\in\{\mathcal U_k,\mathcal S_k\}.
\label{eq:si-gxtb-rsprfo}
\end{equation}
The largest augmented eigenvalue is selected in $\mathcal U_k$ and the
smallest in $\mathcal S_k$. The restriction parameter $\alpha_k$ is adjusted
until the combined step lies within the current trust radius $\Delta_k$:
\begin{equation}
\bm p^{(k)}
=
\bm V_{\mathcal U_k}^{(k)}\bm p_{\mathcal U_k}^{(k)}
+\bm V_{\mathcal S_k}^{(k)}\bm p_{\mathcal S_k}^{(k)},
\qquad
\vx^{(k+1)}=\vx^{(k)}+\bm p^{(k)},
\qquad
\left\|\bm p^{(k)}\right\|_2\leq\Delta_k.
\label{eq:si-gxtb-guidance-update}
\end{equation}

Only $\widehat{\bm H}^{(0)}$ is evaluated explicitly. For subsequent cycles,
let $\bm s^{(k)}=\vx^{(k+1)}-\vx^{(k)}$,
$\bm y^{(k)}=\bm g^{(k+1)}-\bm g^{(k)}$ and
$\bm z^{(k)}=\bm y^{(k)}-\widehat{\bm H}^{(k)}\bm s^{(k)}$. The Hessian
is propagated by the Bofill update
\begin{equation}
\begin{aligned}
\widehat{\bm H}^{(k+1)}
&=
\widehat{\bm H}^{(k)}
+\phi_k\frac{\bm z^{(k)}\bm z^{(k)\top}}
                 {\bm z^{(k)\top}\bm s^{(k)}}
+(1-\phi_k)\left[
\frac{\bm s^{(k)}\bm z^{(k)\top}+\bm z^{(k)}\bm s^{(k)\top}}
     {\bm s^{(k)\top}\bm s^{(k)}}
-\frac{\bm z^{(k)\top}\bm s^{(k)}}
       {(\bm s^{(k)\top}\bm s^{(k)})^2}
 \bm s^{(k)}\bm s^{(k)\top}
\right],\\
\phi_k
&=
\frac{(\bm z^{(k)\top}\bm s^{(k)})^2}
     {(\bm z^{(k)\top}\bm z^{(k)})(\bm s^{(k)\top}\bm s^{(k)})}.
\end{aligned}
\label{eq:si-gxtb-bofill}
\end{equation}
Near-zero rigid modes are excluded from the step. Each candidate was then
refined with the same
B3LYP-D3(BJ)/def2-TZVP GPU4PySCF/Sella protocol. A search that did not
converge within 80 Sella steps was assigned 81 steps, retaining failures in
the effort endpoint.

Guidance monotonically reduced its own g-xTB force RMS, but downstream
DFT/Sella effort did not improve monotonically (Supplementary
Fig.~\ref{fig:si-gxtb-guidance-budget}). The lowest point estimate occurred at
eight cycles: 18.90 failure-aware steps, compared with 20.63 for Pre-train
(paired difference, $-1.73$; 95\% confidence interval, $-3.79$ to $0.58$).
Sixteen cycles did not improve this endpoint (18.96 steps) and reduced Sella
convergence from 99\% to 97\%. Post-train required 14.74 steps; the
eight-cycle guidance arm remained 4.16 steps worse (95\% confidence interval,
1.63 to 7.05). Thus, increasing inference-time g-xTB guidance did not
reproduce the Post-train result under this control.

\begin{figure}[p]
  \centering
  \includegraphics[width=\textwidth]{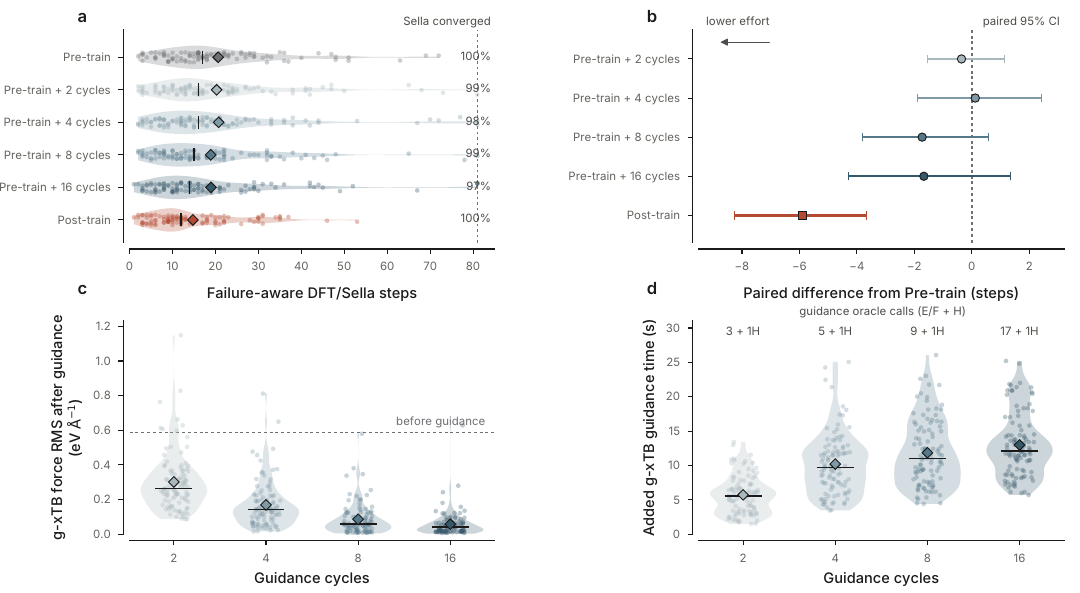}
  \caption{\textbf{Inference-time g-xTB saddle guidance does not
  outperform post-training.}
  \textbf{a}, Per-reaction failure-aware DFT/Sella effort for the frozen
  100-reaction cohort. Each point is one reaction, diamonds denote means,
  short black lines denote medians, and non-converged refinements are assigned
  81 steps (dashed line). Percentages give the Sella convergence rate.
  \textbf{b}, Paired mean differences in failure-aware Sella steps relative
  to direct Pre-train sampling; whiskers are paired percentile-bootstrap 95\%
  confidence intervals from 10,000 reaction-level resamples. Negative values
  indicate lower downstream effort.
  \textbf{c}, g-xTB force RMS after the capped saddle-guidance update. The
  dashed line is the common mean before guidance.
  \textbf{d}, Measured additional g-xTB guidance batch time for batch size
  one. Annotations give the energy/force evaluations plus the single Hessian
  evaluation per candidate. Generator sampling time was not recorded and is
  not included in this panel. All arms use one deterministic candidate per
  reaction and four generator function evaluations. DFT refinements use
  B3LYP-D3(BJ)/def2-TZVP GPU4PySCF/Sella with a maximum of 80 steps and a
  force threshold of 0.05~eV\,\angstrom$^{-1}$.}
  \label{fig:si-gxtb-guidance-budget}
\end{figure}
\endgroup

\end{document}